\documentclass[12pt,oneside]{article}

\usepackage[margin=1.25in]{geometry}
 
\usepackage{amssymb,amsmath,amsthm,bbm,bm,booktabs,multirow,subcaption,setspace,graphicx, float, mathtools, threeparttable}
\def \bbB {\mathbb{B}}                
                
\def \bbD {\mathbb{D}}                
\def \bbE {\mathbb{E}}

    \def \calK {\mathcal{K}}

\def \bbP {\mathbb{P}}                
                
\def \bbR {\mathbb{R}}

    \def \calY {\mathcal{Y}}

\def \Var {\text{Var}}
\def \Cov {\text{Cov}}

\def \conP {\overset{\bbP}\longrightarrow}
\def \conD {\overset{D}\longrightarrow}

\newcommand{\norm}[1]{\left\Vert#1\right\Vert}

\DeclareMathOperator*{\argmin}{arg\,min}

\usepackage[longnamesfirst]{natbib}

\newcommand{\E}{\mathbb{E}}
\newcommand{\inde}{\perp\!\!\!\!\perp} 

\newtheorem{theorem}{Theorem}[section]
\newtheorem{proposition}{Proposition}[section]

\newtheorem{assumption}{Assumption}

\newtheorem{lemma}{Lemma}[section]
\newtheorem{remark}{Remark}[section]

\usepackage[colorlinks,citecolor=blue,urlcolor=blue,hypertexnames=false]{hyperref} 

\usepackage{xr}
\newcommand{\ind}{\mathbbm{1}}
\newcommand{\dottau}{\dot{\tau}}

\newcommand{\dotQ}{\dot{Q}}

\title{Distributionally Robust Treatment Effect}
\author{Ruonan Xu\thanks{ruonan.xu@rutgers.edu}\\
Rutgers University \and 
Xiye Yang\thanks{xiyeyang@economics.rutgers.edu}\\
Rutgers University}
\date{}

\begin{document}
\maketitle

\begin{abstract}
\singlespacing
% 150 words

Using only retrospective data, we study the problem of predicting treatment effects for the same treatment/policy implemented in a different location or time period. We propose a distributionally robust estimator that minimizes the worst-case mean squared error for the prediction of treatment effect over a class of distributions defined by a Wasserstein neighborhood around the source distribution. Because the joint distribution of potential outcomes is unidentified, the problem is inherently one of partial identification. We characterize the sharp upper and lower bounds of the minimax optimizer by exploiting the Fr\'echet class of distributions consistent with the marginal distributions of potential outcomes. The resulting predictor preserves the sign of the average treatment effect under the source distribution but is shrunk toward zero, with the degree of shrinkage depending on the extent of treatment effect heterogeneity. We establish consistency and asymptotic normality of the bound estimators, develop a two-step inference procedure, and discuss the choice of the robustness parameter.

\bigskip


\bigskip 


\end{abstract}

\section{Introduction}
In empirical economics, causal analysis serves two distinct but related objectives. Retrospective studies primarily address internal validity, focusing on the identification of causal effects within a given sample. In contrast, prospective policy analysis concerns external validity, requiring extrapolation of treatment effects to new populations, locations, or time periods. The latter problem is inherently more challenging, as the counterfactual distribution of outcomes under alternative environments is unobserved.

Most empirical work evaluates retrospective causal effects and implicitly treats these estimates as informative for future policy decisions. Such extrapolation, however, relies on strong assumptions about the stability or exchangeability of the underlying population distribution. In many settings, these assumptions are difficult to justify: populations may differ systematically across locations or evolve over time, rendering the original sample unrepresentative of the policy-relevant target population. As a result, external validity remains a central and unresolved challenge.

This paper studies out-of-population prediction of treatment effects that is robust to distributional shift between the source population and an unobserved target population. Our approach contributes to the growing literature on transfer learning, but departs from existing methods by requiring minimal information about the target population. Essentially, our estimator can be used to formulate a prediction of the individual treatment effect for the same treatment/policy implemented in another location or time period with only retrospective data from a single sample. We consider this setting as a common scenario in practice. 

In this paper, we consider the case where there are only outcome variables and treatment status but no covariates.\footnote{Extensions incorporating covariates are left for future work.} In other words, we study how to generalize the findings from randomized experiments. In some cases, there is no universally agreed-upon target population, or the baseline covariates that can be collected from the target population are very limited. As a result, we find ourselves in a situation with limited information, where little is observed from the target population. This scenario arises when we consider expanding the same program to a different location without collecting the necessary data or conducting a cost-benefit analysis to determine whether the policy should be continued in the near future. Instead, we construct an ambiguity/uncertainty set centered around the source/reference distribution—the population distribution from which our sample is drawn. By choosing the radius of the ambiguity set wisely, our hope is that the target population is included in the class of distributions within the neighborhood of the source distribution.

Formally, we consider a distributionally robust optimization (DRO) problem that minimizes the worst-case mean squared error (MSE) of treatment effect prediction over all distributions within a Wasserstein ball centered on the source distribution. Our objective function is conservative here, as we have limited information on the target distribution. The distribution within the ambiguity set that leads to the largest MSE is considered the least favorable distribution. Within the DRO literature, there are many ways to measure the distance between distributions in the ambiguity set. Our use of the Wasserstein distance is motivated by its flexibility: unlike $\phi$-divergences with the Kullback–Leibler divergence serving as a leading example, it does not require the source and target distributions to share common support and admits a natural metric interpretation. See Section S.3.1 in \cite{gu2024wasserstein} for further comparisons between Wasserstein distance and $\phi$-divergences. To maintain tractability while defining the Wasserstein neighborhood, we focus on continuous outcome variables in this paper.

Our choice of the quadratic loss function involves the joint distribution of two potential outcomes. Nevertheless, due to the fundamental missing data problem in the potential outcomes framework, the joint distribution of the potential outcomes remains unidentified but lies within the Fr\'echet class of distributions consistent with the observed marginals. By Sklar’s theorem, this class can be equivalently represented by the set of all copulas linking the marginal distributions of the two potential outcomes. This leads to partial identification. As a result, we select the worst and best distributions within the copula set as our optimistic and pessimistic optimization objects. We pick the current pair of loss function and ambiguity set to balance the goals of generality, tractability, and non-trivial solutions. 

We need to solve a modified three-layer minimax optimization problem coupled with partial identification. The inner maximization primal problem with respect to the ambiguity set can be transformed into a dual problem by minimizing a penalized MSE. Due to the partial identification of the joint distribution of potential outcomes, we derive sharp upper and lower bounds for the minimax optimizer using the Fr\'echet-Hoeffding inequality. 

Our analysis yields several economically interpretable results. The robust predictor preserves the sign of the conventional estimator -- the average treatment effect (ATE) under the source distribution -- but shrinks its magnitude toward zero. This shrinkage reflects a precautionary adjustment for distributional uncertainty. Importantly, the extent of shrinkage depends on the degree of treatment effect heterogeneity. When treatment effects are homogeneous, we see a delayed shrinkage. Namely, the robust predictor coincides with the naive ATE within a certain neighborhood of the source distribution, implying no adjustment for small distributional shifts. Only when the target distribution is sufficiently different from the source distribution does the best-predicted treatment effect begin to shrink toward zero. By contrast, under heterogeneous treatment effects, even small deviations from the source distribution induce immediate shrinkage. These patterns align with the intuition that heterogeneity amplifies sensitivity to distributional changes.

In Section \ref{sec:estimation}, we propose bound estimators based on an M-estimation formulation of the dual problem and establish its consistency and asymptotic distribution. To conduct inference on the partially identified minimax parameter, we develop a two-step procedure that combines ideas from the partial identification literature (see \cite{imbens2004confidence} and \cite{stoye2009more}) with Bonferroni-type corrections. In Section \ref{sec:choice}, we discuss practical considerations for selecting the radius of the Wasserstein ball, which governs the degree of robustness. Monte Carlo simulations and synthetic data illustrate the finite-sample performance of our method in Section \ref{sec:simulation}. Section \ref{sec:conclude} concludes.

\subsection{Related Literature} 
Our paper is related to two strands of literature. The first is on transfer learning and external validity. A seminal contribution is \cite{hotz2005predicting}, which provides a framework for extrapolating causal effects across populations. In addition to standard identification assumptions for internal validity, they impose a form of locational unconfoundedness, under which differences across locations arise solely from shifts in covariates, while the conditional distribution of the potential outcomes remains invariant. Under this assumption, treatment effects in the target population can be recovered by reweighting conditional average treatment effects using the target covariate distribution. This approach -- often referred to as covariate shifts -- has been extended in subsequent work, including \cite{spini2021robustness}, \cite{huang2023leveraging}, \cite{jin2024tailored}, \cite{huang2024sensitivity}, and \cite{menzel2023transfer}.
A growing body of evidence, however, suggests that such assumptions are restrictive in practice. As emphasized by \cite{allcott2015site} and \cite{jin2025beyond}, unobserved differences across locations may invalidate conditional exchangeability. In contrast to this literature, we allow for distributional shifts in potential outcomes, rather than restricting attention to covariate shifts alone.

Recent work has begun to relax the location unconfoundedness assumption by allowing for richer forms of distributional change. For example, \cite{guo2024statistical} and \cite{zhang2024minimax} consider settings in which the conditional distribution of outcomes in the target population lies within a convex combination of conditional distributions observed across multiple sites. Similarly, \cite{jeong2024out} model site-level heterogeneity as random perturbations around a common target distribution. These approaches leverage multisite data or partial information about the target population to reduce uncertainty about the target distribution. By contrast, our setting is intentionally more limited: we consider a single observed sample and allow for minimal or no information about the target population, a scenario that is common when extrapolating to new environments or future periods. This distinction leads us to adopt a different approach to modeling uncertainty.
We use a Wasserstein neighborhood instead of a linear combination of source sites. 

The second strand of literature is distributionally robust optimization (DRO). DRO has been widely studied across operations research, statistics, and machine learning; see, for example, \cite{blanchet2019quantifying}, \cite{duchi2021learning}, \cite{gao2023distributionally}, and \cite{fan2025quantifying}. More recently, DRO devices have been incorporated into econometric applications. \cite{bertsimas2022distributionally} use DRO to conduct sensitivity analysis with respect to unobserved confounding. \cite{qu2024distributionally} develop distributionally robust instrumental variables estimator that is resilient to weak or invalid instruments in finite samples. \cite{chen2024robust} apply DRO to relax rational expectations in moment restrictions. In addition, DRO has been used to study robust policy learning; see, for example, \cite{mo2021learning}, \cite{adjaho2022externally}, \cite{kido2022distributionally}, and \cite{lei2023policy}. Other work, such as \cite{christensen2023counterfactual} and \cite{gu2024wasserstein}, employs DRO to assess the sensitivity of counterfactuals to parametric assumptions about the latent variable distribution in a class of structural models. 

Our contribution differs from existing DRO-based approaches in several dimensions. First, unlike the literature on individualized policy learning, which typically evaluates the robustness of treatment rules, we focus on predicting treatment effects under distributional shift. This distinction leads to a different objective function: we minimize the worst-case mean squared error of treatment effect prediction, rather than optimizing policy performance. Second, our analysis explicitly addresses partial identification arising from the unobserved joint distribution of potential outcomes, a feature that is largely absent in existing DRO applications. Third, we provide a formal inference procedure for our proposed estimator, while inference is overlooked in most DRO literature.

%Perhaps the closest paper to ours is \cite{jeong2020assessing}, who study the worst-case subpopulation treatment eﬀect with a linear loss function and an ambiguity set consisting of perturbed distributions of covariates. Hence, they impose that the conditional distribution of potential outcomes given covariates remain unchanged over subpopulations, which we do not require. Their parameter of interest turns out to be point-identified, but ours is only partially identified. 

\section{Setup}
Suppose we have access to a sample randomly drawn from a single cross section of the source/reference population distribution. In the sample, we observe a binary treatment variable $T_i$ and a realized outcome variable $Y_i=T_iY_i(1)+(1-T_i)Y_i(0)$, where $Y_i(1)$ and $Y_i(0)$ denote a pair of potential outcomes. Let us denote the joint distribution of $(Y(1), Y(0))$ from the source population by $P$. We are nevertheless interested in making inference about the treatment effect in a target distribution $Q$ of potential outcomes $(\tilde{Y}(1), \tilde{Y}(0))$.\footnote{We use $(\tilde{Y}(1),\tilde{Y}(0))$ to denote the potential outcomes under distribution $Q$ to differentiate it from $P$.} The distribution $Q$ can be different from the source distribution $P$ because it is from a different location or a future time period. We do not observe the outcome variables from $Q$ and probably not even the covariates. Hence, the treatment effect under $Q$ is not identified. 

For instance, we collect data from a job training program, given that participation is randomly assigned, we can identify $\tau^*=\mathbb{E}_P[Y(1)-Y(0)]$ for the source distribution. However, in addition to the evaluation of the program in a specific location in the past, we are interested in examining whether the job training program can be expanded to another location or should be implemented on a long-term basis. Without actually implementing the job training program at a target site and, in particular, implausible to do so for the future period, our goal is to predict the worst-case treatment effect under $Q$ distribution using a sample from the source population. Therefore, we propose an additional step to the usual program evaluations. Using the same dataset in an empirical research study, following a typical causal analysis, we provide a formal procedure for generalizing the causal estimates under distributional shift.

In the hypothetical scenario where we could observe a sample from distribution $Q$, the solution to the minimization of the MSE, $\mathbb{E}_Q\big[(\tilde{Y}(1)-\tilde{Y}(0)-\tau)^2\big]$, turns out to be $\tau^Q=\mathbb{E}_Q\big[\tilde{Y}(1)-\tilde{Y}(0)\big]$. The ATE $\tau^Q$ can be seen as the best prediction of the individual treatment effect under the target distribution $Q$. If we had access to a sample from $Q$, the prediction of the individual treatment effect and the identification of the ATE would coincide. However, this coincidence breaks down when $Q$ is unknown. Without input from $Q$, we construct a class of distributions $\mathcal{Q}=\{Q: D(P,Q)\leq \delta^2\}$ centered around the source distribution within distance $\delta^2$. The target distribution $Q$ is considered to be contained in the ambiguity set $\mathcal{Q}$ when the neighborhood radius $\delta$ is carefully chosen.

\begin{remark}\label{remark1}
Intuitively, one would like to solve a distributionally robust optimization problem 
\begin{equation}\label{eqn:naivedro}
    \inf_\tau\sup_{Q\in \mathcal{Q}}\mathbb{E}_Q\big[(\tilde{Y}(1)-\tilde{Y}(0)-\tau)^2\big],
\end{equation}
with the Wasserstein distance defined in the following way:\footnote{We will be more specific about the definition of the Wasserstein distance when we introduce (\ref{Wasser}) below.}
\begin{equation}\label{Wasser0}
D^{1/2}(P,Q)=\inf_{\pi\in \Pi(P,Q)}\sqrt{\mathbb{E}_\pi\left[\norm{(Y(1),Y(0))-(\tilde{Y}(1),\tilde{Y}(0))}_p^2\right]}.
\end{equation}  
The quadratic loss function in (\ref{eqn:naivedro}) can be seen as a location estimation or linear regression of $Y(1)-Y(0)$ on a constant 1. For the combination of the quadratic loss and 2-Wasserstein distance, \cite{chao2023statistical} demonstrate that, for location estimation and regression with a fixed design matrix, the population mean or least squares solution under the source distribution remains the minimax optimizer, regardless of the distribution shift of outcome variables. Such a prediction is not helpful as there is no way to assess the generalizability of the treatment effect. More details are provided in Appendix \ref{sec:loss}, where we also discuss the drawbacks of alternative parametric loss functions that involve regressing the observed outcome on the treatment indicator and covariates.
\end{remark}

Based on the observation in Remark \ref{remark1}, we use a slightly different combination of loss function and ambiguity set to make the task of treatment effect prediction even harder. 
Let us consider the nonparametric quadratic loss $(\tilde{Y}(1)-\tilde{Y}(0)-w\tau)^2$ with additional weighting $w$. We modify the loss function in this way to maintain the regression specification, ensuring the problem remains tractable. We also augment the source distribution $P$ with a constant weighting of 1 and the target distribution $Q$ with an adversarial weighting $w$. The augmented distributions are denoted by $\bar{P}$ and $\bar{Q}$. We use 
\begin{equation}\label{Wasser}
D^{1/2}(\bar{P},\bar{Q})=\inf_{\pi\in \Pi(\bar{P},\bar{Q})}\sqrt{\mathbb{E}_\pi\left[\norm{(Y(1),Y(0),1)-(\tilde{Y}(1),\tilde{Y}(0),w)}_p^2\right]}
\end{equation}  
to measure the distance between the augmented source distribution and target distribution, where the cost function is the quadratic of the $L_p$ norm $\| (\tilde{Y}(1),\tilde{Y}(0),w)-(Y(1),Y(0),1) \|_p^2$ for $p\in (1,\infty]$. In the definition of Wasserstein distance, $\Pi(\bar{P},\bar{Q})$ is the set of couplings of $\bar{P}$ and $\bar{Q}$. Since the adversarial weighting $w$ is unitless, we normalize the potential outcomes by the standard deviation of the realized outcome under the source distribution, following the penalized regression literature, to make it scale-invariant. The optimizer will eventually be scaled back using the same standard deviation when reported. Consequently, the neighborhood radius $\delta$ can be assessed in the magnitude of the standard deviation of the realized outcome $Y$ when we use $L_2$ norm. We apply this normalization in both the simulation and the empirical illustration based on synthetic data below.

Since we only impose that $\bar{Q}$ belongs to $\mathcal{Q}=\{\bar{Q}: D(\bar{P},\bar{Q})\leq \delta^2\}$, to be conservative we pick the distribution within the ambiguity set $\mathcal{Q}$ that leads to the largest MSE for prediction. The corresponding distribution is considered the least favorable distribution. Such a procedure is robust in the sense that, for any $\tau$, the MSE of predicting the individual treatment effect for all distributions in $\mathcal{Q}$ will be bounded by the worst-case MSE. The adversarial weighting $w$ makes the MSE minimization problem more conservative because the adversary could rescale $\tau$ via the weighting $w$ to make the prediction error of the individual treatment effect larger, which inflates the adversarial loss. In this sense, the minimax solution $\tau^{DR}$ is considered the optimal worst-case out-of-population prediction of individual treatment effect. The prediction $\tau^{DR}$ results from a bias-variance tradeoff by minimizing the MSE and hence is different from $\tau^Q$, which is not identifiable under our framework. Relatively stable $\tau^{DR}$ along the increase of the level of robustness, $\delta$, is an indicator of generalizability of the treatment effect to new populations, even with possible distributional shift. 

\begin{remark}
    Such a reweighting idea has been commonly used in the transfer learning literature. For instance, with access to multiple sites of data, \cite{guo2024statistical} and \cite{zhang2024minimax} construct the ambiguity set as the weighted average of multisite distributions: \[\mathcal{C}(\mathbb{Q}_X)\coloneq\left\{\mathbb{T}=(\mathbb{Q}_X, \mathbb{T}_{Y|X}): \mathbb{T}_{Y|X}=\sum_{l=1}^L q_l \cdot \mathbb{P}^{(l)}_{Y|X} \text{ with } q\in\Delta^L\right\},\]
    where $\Delta^L=\{q\in \mathbb{R}^L: \sum^L_{l=1}q_l=1, \min_l q_l\geq 0\}$ denotes the $L$-dimension simplex. 
    With a single site but observation of covariates in the target population, \cite{hotz2005predicting} propose that the ATE under the target population can be identified by $\int_\mathcal{X}\tau(x)d\tilde{F}_X(x)$ under only covariate shifts, where $\tilde{F}_X(x)$ is the covariate distribution under the target population and the conditional average treatment effect $\tau(x)$ is assumed to remain unchanged across populations. Without access to covariates under the target distribution, \cite{spini2021robustness} and \cite{devaux2022quantifying} use the Kullback-Leibler divergence distance of covariates to bound the ATE under the same set of assumptions as in \cite{hotz2005predicting}.
\end{remark}

If we had a hypothetical sample from $Q$, we only need marginal distributions of potential outcomes to identify the ATE under $Q$. However, for the prediction problem we set up, the joint distribution $P$ is involved in the definition of the Wasserstein ambiguity set. This is induced by our choice of the nonparametric quadratic loss, which involves the second moment of the individual treatment effect. Even though joint distribution of $(Y(1), Y(0))$ exists in the sense that $P(y_1,y_0)=C^*(P_1(y_1), P_0(y_0))$ for some copula $C^*: [0,1]^2\mapsto [0,1]$, where $P_1$ and $P_0$ are the marginal distributions of $Y(1)$ and $Y(0)$, it can never be identified using our sample. As a result, our context presents an additional layer of complexity compared to the general transfer estimates literature. 

With that said, we know the joint distribution $P(y_1,y_0)$ must belong to the Fr\'echet class of joint distributions with marginals $P_1$, $P_0$, parameterized by the set of copulas $\mathcal{C}(P_1,P_0)$. Therefore, we can pick one distribution in $\mathcal{C}(P_1,P_0)$ that gives the smallest worst-case MSE and another one that leads to the largest worst-case MSE. These two cases are considered optimistic and pessimistic cases, respectively. 
This step is related to the partial identification of the joint distribution of potential outcomes. Visually, we can imagine there is a set $\mathcal{C}(P_1,P_0)$. Each point within the set $\mathcal{C}(P_1,P_0)$ is a joint distribution of potential outcomes. Centered around each point, there is a Wasserstein neighborhood with radius $\delta$. For each point, the Wasserstein neighborhood is defined as (\ref{Wasser}) with $P$ replaced by the copula. 
 
To assess the robustness of the treatment effect, we start with the following two objective functions: 
 \begin{equation}\label{pessi_obj}
 	\inf_\tau\sup_{C\in\mathcal{C}(P_1,P_0)}\sup_{\bar{Q}\in \mathcal{Q}}\mathbb{E}_{\bar{Q}}\big[(\tilde{Y}(1)-\tilde{Y}(0)-w\tau)^2\big] 
 \end{equation}
 and
 \begin{equation}\label{optim_obj}
 	\inf_\tau\inf_{C\in\mathcal{C}(P_1,P_0)}\sup_{\bar{Q}\in \mathcal{Q}} \mathbb{E}_{\bar{Q}}\big[(\tilde{Y}(1)-\tilde{Y}(0)-w\tau)^2\big].
 \end{equation} 
The solution to (\ref{pessi_obj}) and (\ref{optim_obj}) is denoted by $\tau_p$ and $\tau_o$ respectively, which predict the treatment effect by minimizing the worst-case MSE. Essentially, we need to find a solution to a minimax optimization problem. The middle layer of $\inf$ and $\sup$ arises due to the partial identification issue.  

\begin{remark}
     The inner maximization in (\ref{pessi_obj}) and (\ref{optim_obj}) is with respect to the distribution shift of $Q$. The middle minimization and maximization concern the partial identification of the joint distribution of $P_1$ and $P_0$, which does not involve $Q$. However, the copula $C$ plays a role in the uncertainty set $\mathcal{Q}$ since the set is centered around the joint distribution $C$. The solutions to (\ref{pessi_obj}) and (\ref{optim_obj}) provide lower and upper bounds (defined according to the sign of $\tau^*$) for the non-point-identified minimax prediction $\tau^{\texttt{DR}}$. We do not consider $\inf_{\bar{Q}\in \mathcal{Q}} \mathbb{E}_{\bar{Q}}\big[(\tilde{Y}(1)-\tilde{Y}(0)-w\tau)^2\big]$ for the inner problem. When the Wasserstein radius $\delta$ is sufficiently large, this most favorable case always has an MSE of zero and hence is not meaningful. Intuitively, such a minimization problem does not provide a robust guarantee of MSE with distribution shift.  
\end{remark}

\section{Identification}

The primal problem in (\ref{pessi_obj}) and (\ref{optim_obj}) appears to be initially difficult to solve, as it involves optimization with respect to an infinite number of distributions. Inspired by the results in, for example, \cite{blanchet2019robust} and \cite{gao2023distributionally}, we can derive a closed form of the dual problem of the inner maximization problem. 
	\begin{equation}\label{dual}
		\sup_{\bar{Q}\in \mathcal{Q}}\mathbb{E}_{\bar{Q}}\big[(\tilde{Y}(1)-\tilde{Y}(0)-w\tau)^2\big] = \left\{\sqrt{\mathbb{E}_{C}\big[(Y(1)-Y(0)-\tau)^2\big]}+\delta(2+|\tau|^q)^{1/q}\right\}^2,
	\end{equation}
	where $1/p+1/q=1$ so that $q\in[1,\infty)$.

We can see that the right-hand side of (\ref{dual}) is the quadratic of the square root of the MSE under the copula $C$ plus a penalty term with the penalization parameter being the radius of the Wasserstein neighborhood. The penalization term involves the sum of a constant two and the parameter $\tau$, which is different from the usual penalization of targeting parameters alone. The MSE under $C$ in the right-hand side of (\ref{dual}) can be further decomposed into two terms. 	
			\begin{equation}\label{decompose}
				\begin{aligned}
						&\sqrt{E_{C}\big[(Y(1)-Y(0)-\tau)^2\big]}+\delta(2+|\tau|^q)^{1/q}\\
					=&\sqrt{Var_{C}\big(Y(1)-Y(0)\big)+(\tau^*-\tau)^2}+\delta(2+|\tau|^q)^{1/q}
				\end{aligned}
		\end{equation}
Because of the observation in (\ref{decompose}), we only need to find the copula that leads to the largest variance of individual treatment effect for the pessimistic case and the copula that leads to the smallest variance for the optimistic case. 

\begin{proposition}\label{prop1}
	
	$\sqrt{E_{C}\big[(Y(1)-Y(0)-\tau)^2\big]}+\delta(2+|\tau|^q)^{1/q}$ is monotonically increasing in $V=Var_{C}\big(Y(1)-Y(0)\big)$. Hence, $\forall$ $q\in[1,\infty)$, for the pessimistic case
	\begin{equation}\label{pessi_dual}
	\sup_{C\in\mathcal{C}(P_1,P_0)}\sup_{\bar{Q}\in \mathcal{Q}}\mathbb{E}_{\bar{Q}}\big[(\tilde{Y}(1)-\tilde{Y}(0)-w\tau)^2\big]= \left\{\sqrt{V_p+(\tau^*-\tau)^2}+\delta(2+|\tau|^q)^{1/q}\right\}^2,
	\end{equation}
	where $V_p=V_U(P_1,P_0)=\sup_{C\in\mathcal{C}(P_1,P_0)}V\big(C(P_1,P_0)\big)$, and $V\big(C(P_1,P_0)\big)$ denotes the variance of individual treatment effect under copula $C(P_1,P_0)$.
	And for the optimistic case
	\begin{equation}\label{optim_dual}
		\inf_{C\in\mathcal{C}(P_1,P_0)}\sup_{\bar{Q}\in \mathcal{Q}}\mathbb{E}_{\bar{Q}}\big[(\tilde{Y}(1)-\tilde{Y}(0)-w\tau)^2\big] =\left\{ \sqrt{V_o+(\tau^*-\tau)^2}+\delta(2+|\tau|^q)^{1/q}\right\}^2,
	\end{equation}
	where $V_o=V_L(P_1,P_0)=\inf_{C\in\mathcal{C}(P_1,P_0)}V\big(C(P_1,P_0)\big)$.
	\end{proposition}
			
For the outer minimization problem, minimizing the quadratic is equivalent to minimizing the terms within the curly bracket. Compared to the bridge estimator for a linear regression in (\ref{bridge}) below (see, for instance, \cite{knight2000asymptotics}), we take the square root of the loss function, and the penalization is not purely applied to the targeting parameter. Thus, solutions to the minimization of the right-hand side of (\ref{pessi_dual}) and (\ref{optim_dual}) can be considered as a square-root bridge-type estimator.
	\begin{equation}\label{bridge}
	\argmin_\beta \norm{Y-X\beta}_2^2+\lambda \norm{\beta}_q^q,\ q>0
	\end{equation}

Define the solution to the outer minimization problem as
	\[
	f(V,\delta)=\text{argmin}_\tau \sqrt{V+(\tau^*-\tau)^2}+\delta(2+|\tau|^q)^{1/q}.
	\]
\begin{proposition}\label{prop2}
(1)	When $\delta=0$, $f(V,\delta)=\tau^*$. When $\delta>0$, $f(V,\delta)\in [\tau^*, 0]$ if $\tau^*\leq 0$ and $f(V,\delta)\in [0, \tau^*]$ if $\tau^*\geq 0$. (2) $f(V,\delta)$ is monotonically decreasing in $V$ in the sense that $|f(V_p,\delta)|\leq |f(V^P,\delta)|\leq |f(V_o,\delta)|$, where $V^P=Var_P\big(Y(1)-Y(0)\big)$. 
\end{proposition}	

\begin{remark}
    Suppose we have some priors on the joint distribution. In that case, we can either achieve point identification by picking a particular copula or have a shorter identified interval by narrowing down the copula set; see \cite{heckman1997making}. 
\end{remark}
Based on a simple observation, when there is no distribution shift, $\delta=0$ and $f(V,\delta)=\tau^*$. Namely, the prediction of the treatment effect is the ATE under the source distribution when the target population coincides with the source population. 
On the other hand,  when $\delta>0$, we allow for a shift in the distribution. We can easily see that $f(V,\delta)$ shares the same sign as $\tau^*$ but shrinks toward zero, as in any regularized estimation. By being conservative, our prediction of treatment effect under $Q$ is always no larger than $\tau^*$ in magnitude. 

Since the joint distribution $P(y_1,y_0)=C^*(P_1(y_1), P_0(y_0))$ also belongs to the copula set  $\mathcal{C}(P_1,P_0)$, $V_o\leq V^P\leq V_p$. Therefore, as long as we can find a pair of ($V_p$, $V_o$), we have found a bound for the minimax optimizer with respect to the unknown joint distribution $P$.

\subsection{Homogeneous Treatment Effect}
When the treatment effect is homogeneous, we immediately know the joint distribution of the potential outcomes from the marginal distributions. This is the case where we do not need to worry about finding a copula, and the pessimistic and the optimistic cases coincide. On the other hand, if we would like to avoid the complication of finding the smallest variance of the individual treatment effect, a naive lower bound for $V_o$ is simply zero. 

Given $Var_P\big(Y(1)-Y(0)\big)=0$, the dual objective function reduces to 
		\begin{equation}\label{obj:homo}
		\inf_\tau |\tau^*-\tau|+\delta(2+|\tau|^q)^{1/q}.
		\end{equation}
		
\begin{figure}
		\centering
		\includegraphics[width=4in]{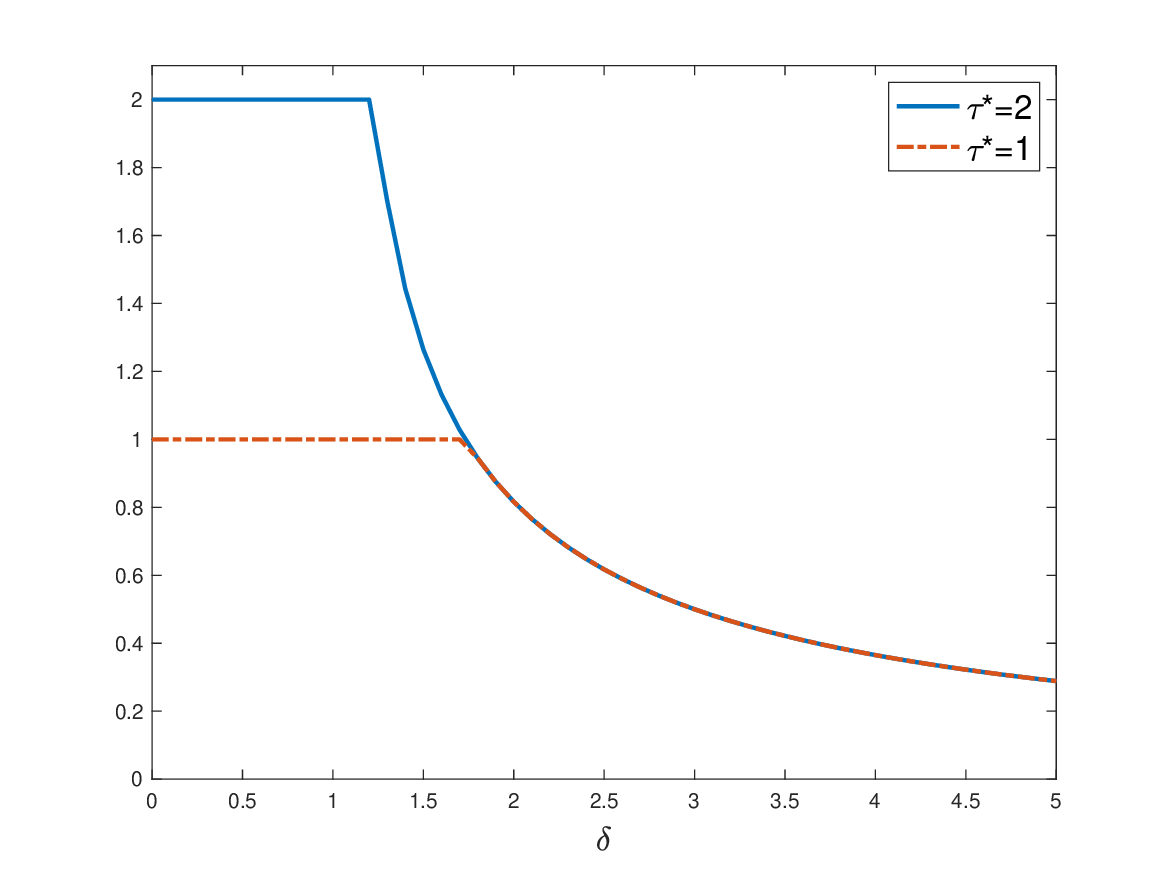} 
		\caption{Minimax Optimizer under Homogeneous Treatment Effect with $q=2$}
		\label{delayed}
	\end{figure}

	\begin{proposition}\label{prop3}
		 With homogeneous treatment effect, $f(0,\delta)=\tau^*$ for $\delta \leq \big(\frac{2}{|\tau^*|^q}+1\big)^{1-1/q} \eqqcolon \bar{\delta}$. 		
	\end{proposition}	
		Proposition \ref{prop3} implies delayed shrinkage of the minimax optimizer, which is in contrast to the typical pattern of regularized estimation but intuitive in the context of causal analysis. For any Wasserstein neighborhood radius less than or equal to $\bar{\delta}$, our prediction of the individual treatment effect under $Q$ is always $\tau^*$. Figure \ref{delayed} is a graphical illustration of delayed shrinkage, where we set $q=2$ and consider two values of $\tau^*$: $\tau^*=2$ and $\tau^*=1$. The corresponding cutoffs are $\bar{\delta}_2=1.22$ and $\bar{\delta}_1=1.73$. 
		
		We can imagine there is more generalizability of our causal estimates if the treatment effect is homogeneous. Only when the target distribution $Q$ is sufficiently different from the reference distribution $P$ by setting a relatively large $\delta$, our best prediction of the treatment effect starts to shrink toward zero. Furthermore, the boundary radius $\bar{\delta}$ decreases in $\tau^*$. This is also intuitive as it would be harder to maintain a larger treatment effect given distributional shift. 		

\subsection{Heterogeneous Treatment Effect}

\begin{figure}
	\centering
	\includegraphics[width=4in]{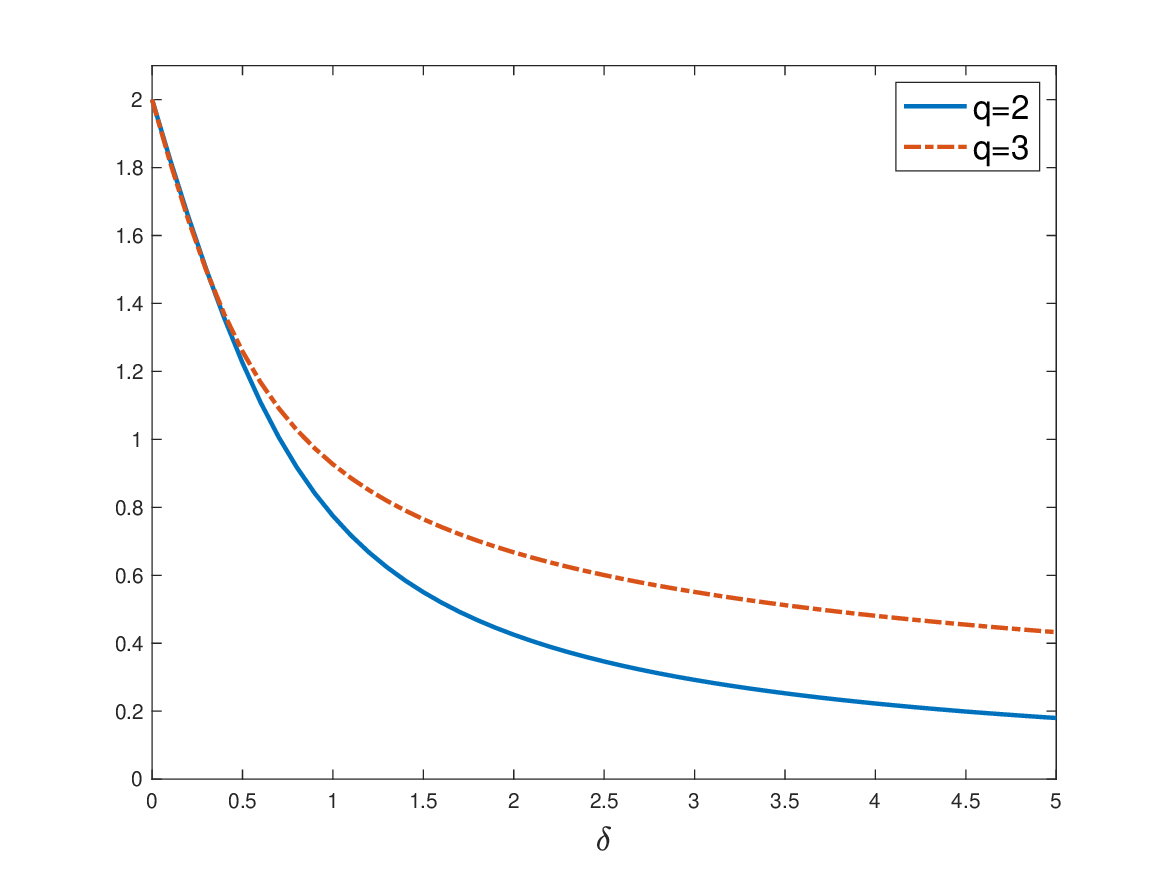} 
	\caption{Minimax Optimizer under Heterogeneous Treatment Effect}
	\label{hetero}
\end{figure}

When we switch to heterogeneous treatment effect, $Var_{C}\big(Y(1)-Y(0)\big)>0$ and $|f(V,\delta)|<|\tau^*|$ for $\delta>0$. This implies that the solution $f(V,\delta)$ shrinks toward zero immediately whenever there is a distribution shift, even for a tiny shift. Figure \ref{hetero} illustrates the immediate shrinkage when the population variance $V$ is 5 with $q=2$ and $q=3$ respectively. We can also see that shrinkage occurs to a greater extent for smaller values of $q$.
 
Given the pessimistic and optimistic cases, we need to find bounds of $Var_{C}\big(Y(1)-Y(0)\big)$. Using  Fr{\'e}chet-Hoeffding inequality, we can find sharp bounds of the variance; see, for instance, \cite{fan2010sharp}. This technique has also been used by \cite{aronow2014sharp} and \cite{imbens2021causal} to derive sharp bounds for the design-based variance-covariance matrix. 	
	Let $C^L(u,v)=\max(u+v-1,0)$ and $C^U(u,v)=\min(u,v)$. Fr{\'e}chet-Hoeffding inequality implies that 
	\[
	Cov_{C^L}(Y(1),Y(0))\leq Cov_P(Y(1),Y(0))\leq Cov_{C^U}(Y(1),Y(0)).
	\]
 As a result, the sharp upper bound of the variance of individual treatment effect is $V_p=V_U(P_1,P_0)=V\big(C^L(P_1,P_0)\big)$, where potential outcomes $Y(1)$ and $Y(0)$ are 
	perfectly negatively dependent. Similarly, the sharp lower bound of the variance is $V_o=V_L(P_1,P_0)=V\big(C^U(P_1,P_0)\big)$,  where the two potential outcomes are perfectly positively dependent. 

Ultimately, we are interested in predicting the treatment effect under $Q$ using information from the reference distribution $P$. If we can identify the joint distribution $P$, we can form our objective function as 
		\begin{equation}\label{obj_P}
				\inf_\tau\sup_{\bar{Q}\in \mathcal{Q}}\mathbb{E}_{\bar{Q}}\big[(\tilde{Y}(1)-\tilde{Y}(0)-w\tau)^2\big]. 
		\end{equation}
	Define the solution to (\ref{obj_P}) to be 
		\[
		\tau^{\texttt{DR}}\equiv f(V^P,\delta)=\text{argmin}_\tau \sqrt{Var_P\big(Y(1)-Y(0)\big)+(\tau^*-\tau)^2}+\delta(2+|\tau|^q)^{1/q}.
		\]
According to Proposition \ref{prop2}, we have found sharp bounds for $|f(V^P,\delta)|$, which is $\left(|f\left(V\big(C^L(P_1,P_0)\big),\delta\right)|, |f\left(V\big(C^U(P_1,P_0),\delta\right)|\right)$.\footnote{The sign of $f(V^P,\delta)$ and the lower and upper bounds depend on the sign of $\tau^*$.} Therefore, $f(V^P,\delta)$ can be partially identified. 

\cite{splawa1990application} proposes another set of variance bounds, which uses Cauchy–Schwarz inequality to derive the bounds for the covariance between two potential outcomes. This pair of bounds is easier to compute and is given below. 
\[
V_p^N=Var_{P_1}(Y(1))+Var_{P_0}(Y(0))+2\sqrt{Var_{P_1}(Y(1))Var_{P_0}(Y(0))}
\]
\[
V_o^N=Var_{P_1}(Y(1))+Var_{P_0}(Y(0))-2\sqrt{Var_{P_1}(Y(1))Var_{P_0}(Y(0))}
\]

\iffalse
	\begin{figure}
		\centering
		\begin{subcaptionbox}{$\tau^*=2$\label{fig:left}}[0.48\textwidth]
			{\includegraphics[width=\linewidth]{positive}}
		\end{subcaptionbox}
		\hfill
		\begin{subcaptionbox}{$\tau^*=-2$\label{fig:right}}[0.48\textwidth]
			{\includegraphics[width=\linewidth]{negative}}
		\end{subcaptionbox}
		%\caption{Two figures side by side}
		\label{fig:two-side-by-side}
	\end{figure}
\fi

\section{Estimation}\label{sec:estimation}
To proceed with estimation, we need to find the sample counterpart of the objective functions (\ref{pessi_dual}) and (\ref{optim_dual}). We first need a set of internally valid estimators.
\begin{assumption}\label{iid}
	We obtain a random sample from the source distribution, $\{Y_i,T_i\}_{i=1}^n$.  
\end{assumption}
\begin{assumption}\label{random}
	Within the source distribution, $T\inde \big(Y(1),Y(0)\big)$, $0<e=\mathbb{P}(T=1)<1$.
\end{assumption}
\begin{assumption}\label{moment}
	 $\mathbb{E}_{P_1}[Y(1)^4]<\infty$, $\mathbb{E}_{P_0}[Y(0)^4]<\infty$.
\end{assumption}
Assumptions \ref{iid}-\ref{moment} are standard assumptions in the causal inference literature. In particular, Assumption \ref{random} is the usual random assignment and overlap condition for experimental data. 

We can estimate $\tau^*$ using various estimators $\hat{\tau}^*$. For instance, we can use the difference-in-means estimator $\bar{Y}_1-\bar{Y}_0$, where $\bar{Y}_1$ and $\bar{Y}_0$ are sample averages of the treated and untreated outcomes. Or, we can use the inverse probability weighting estimator, $\frac{1}{n}\sum^n_{i=1}\frac{T_iY_i}{e}-\frac{1}{n}\sum^n_{i=1}\frac{(1-T_i)Y_i}{1-e}$.

As for $\hat{V}$, it is easy to compute $\widehat{Var}(Y(1))$ and $\widehat{Var}(Y(0))$ using the random sample of treated and control units. To estimate the sharp bounds of the covariance term, $\widehat{Cov}(Y(1),Y(0))$, we use 
	\[
\int_0^1\hat{P}^{-1}_1(u)\hat{P}^{-1}_0(u)du-\bar{Y}_1\bar{Y}_0  
\]
and 
\[
\int_0^1\hat{P}^{-1}_1(u)\hat{P}^{-1}_0(1-u)du-\bar{Y}_1\bar{Y}_0,   
\]
where	$\hat{P}_1(y)=\frac{1}{n_1}\sum^n_{i=1}T_i\mathbbm{1}\{Y_i\leq y\}$ and 
			$\hat{P}_0(y)=\frac{1}{n_0}\sum^n_{i=1}(1-T_i)\mathbbm{1}\{Y_i\leq y\}$
are empirical CDFs for the treated and untreated units and 
	$\hat{P}_1^{-1}(u)=\inf\{y:\hat{P}_1(y)\geq u\}$ and 
	$\hat{P}_0^{-1}(u)=\inf\{y:\hat{P}_0(y)\geq u\}$ are their inverse functions. 
As a result, 
\[
\hat{V}_o=\widehat{Var}(Y(1))+\widehat{Var}(Y(0))-2\left(\int_0^1\hat{P}^{-1}_1(u)\hat{P}^{-1}_0(u)du-\bar{Y}_1\bar{Y}_0  \right)
\]
and 
\[
\hat{V}_p=\widehat{Var}(Y(1))+\widehat{Var}(Y(0))-2\left(\int_0^1\hat{P}^{-1}_1(u)\hat{P}^{-1}_0(1-u)du-\bar{Y}_1\bar{Y}_0\right).
\]
	
\iffalse	
	\[
	\sum_{i=1}^P(p_i-p_{i-1})Y_{1[i]}Y_{0[i]}-\bar{Y}_1\bar{Y}_0  
	\]
		\[
	\sum_{i=1}^P(p_i-p_{i-1})Y_{1[i]}Y_{0[P+1-i]}-\bar{Y}_1\bar{Y}_0  
	\]
\fi

Estimators of the Neyman bounds are straightforward to construct based on the variance estimators of $Var_{P_1}(Y(1))$ and $Var_{P_0}(Y(0))$. They are denoted by $\hat{V}_o^N$ and $\hat{V}_p^N$:
	\[
	\hat{V}_p^N=\widehat{Var}(Y(1))+\widehat{Var}(Y(0))+2\sqrt{\widehat{Var}(Y(1))\widehat{Var}(Y(0))},
	\]
	\[
	\hat{V}_o^N=\widehat{Var}(Y(1))+\widehat{Var}(Y(0))-2\sqrt{\widehat{Var}(Y(1))\widehat{Var}(Y(0))}.
	\]

\subsection{Asymptotic Properties}\label{asym}
Based on the dual problem, the outer minimization of (\ref{pessi_dual}) and (\ref{optim_dual}) becomes an M-estimation problem. Therefore, we can apply the empirical process theory in \cite{EmpiricalProcess} to derive the asymptotic properties. Let $\tau_p=f(V_p, \delta)$ and $\tau_o=f(V_o, \delta)$. Moreover, let $\hat{\tau}_p$ and $\hat{\tau}_o$ be the solution to the sample minimization problem. 
Before showing the asymptotic properties of $(\hat{\tau}_p,\hat{\tau}_o)$, let us prove an intermediate result.

\begin{assumption}\label{quantile}
(i) (Density conditions) $P$ admits continuous density functions $f_1$ and $f_0$ for $Y(1)$ and $Y(0)$ respectively, and there exists $f_{\min} > 0$ such that:
\begin{align*}
	\inf_{u \in [0,1]} f_1(P^{-1}_1(u)) \geq f_{\min}, \quad \inf_{u \in [0,1]} f_0(P^{-1}_0(u)) \geq f_{\min}
\end{align*}

(ii) (Bounded quantiles) There exists $M < \infty$ such that:
\begin{align*}
	\sup_{u \in [0,1]} |P^{-1}_1(u)| \leq M, \quad \sup_{u \in [0,1]} |P^{-1}_0(u)| \leq M.
\end{align*}

(iii) (Bahadur remainder) For $t \in \{0,1\}$:
\begin{align*}
	\sup_{u \in [0,1]} |R_{n,t}(u)| = \sup_{u \in [0,1]} \Big| \hat{P}^{-1}_t(u) - P^{-1}_t(u) + \frac{\hat{P}_t(P^{-1}_t(u)) - u}{f_t(P^{-1}_t(u))} \Big| = o_p(n^{-1/2})
\end{align*}
where $R_{n,t}(u)$ is the Bahadur remainder term \citep{Bahadur:1966} for quantile estimators.
\end{assumption}

\begin{lemma} \label{lem:JointCLT}
(\romannumeral 1) Under Assumptions \ref{iid}-\ref{moment}, the vector $\sqrt{n}(\hat{\tau}^* - \tau^*, \hat{V}_p^N - V_p^N, \hat{V}_o^N - V_o^N)$ converges in distribution to a multivariate normal random vector with mean zero.

(\romannumeral 2) Under Assumptions \ref{iid}-\ref{quantile}, the vector $\sqrt{n}(\hat{\tau}^* - \tau^*, \hat{V}_p - V_p, \hat{V}_o - V_o)$ converges in distribution to a multivariate normal random vector with mean zero. The detailed variance-covariance matrices are given in Appendix \ref{appendixe}.
\end{lemma}

 Lemma \ref{lem:JointCLT} can be summarized as:
\begin{align} \label{eq:joint}
    \sqrt{n}(\hat{V}_p - V_p, \hat{V}_o - V_o, \hat{\tau}^\ast - \tau^*)^\intercal \conD (Z_p, Z_o, Z_\tau)^\intercal \sim N(0, \bm{\Sigma}).
\end{align}

\begin{theorem}\label{thm:consistency}
	Under the conditions in Lemma \ref{lem:JointCLT}, $\hat{\tau}_p\conP \tau_p$ and $\hat{\tau}_o\conP \tau_o$.
\end{theorem}

To describe the asymptotic distributions, we need more notation. From now on, let $V_p$ and $V_o$ be the variance bounds (either Neyman or sharp). For $b\in\{p,o\}$, let $\tau_b$ be the minimizer of $M(\tau) \coloneqq A(\tau) + \delta B(\tau)$, where
\begin{align*}
    A(\tau) = ( V_b + (\tau^* - \tau)^2 )^{1/2} \quad \text{and} \quad B(\tau) = \left(2 + |\tau|^q\right)^{1/q}.
\end{align*}
Let $\partial_{b} A(\tau)$ and $\partial_{\tau^*} A(\tau)$ denote the gradient of $A(\tau)$ with respect to $V_b$ and $\tau^*$ at $\tau$.  
In the following, both $A$ and its derivatives are evaluated at $\tau_b$:
\begin{equation}\label{eqn:D}
\begin{aligned}
	D_p = \frac{\tau^* - \tau_p}{A^2}  \begin{pmatrix}
	\partial_{b} A \\
	0 \\
	\partial_{\tau^*} A
\end{pmatrix}   +  \begin{pmatrix}
	0 \\
	0 \\
	 -A^{-1}
\end{pmatrix} ,  \,\, D_o = \frac{\tau^* - \tau_o}{A^2}  
\begin{pmatrix}
        0 \\
	\partial_{b} A \\
	\partial_{\tau^*} A
\end{pmatrix}   +  \begin{pmatrix}
	0 \\
	0 \\
	 -A^{-1}
\end{pmatrix} .
\end{aligned}
\end{equation}
For $b\in\{p,o\}$, define
   $ \bbD_b(\tau_b) = [ M''(\tau_b) ]^{-1} D_b$,
where $M''$ denotes the second-order derivative of $M(\tau)$.

\begin{theorem} \label{thm:non-zero}
Assume \eqref{eq:joint} holds. For $q\geq1$ and $b\in\{p,o\}$, if $\tau_b \neq 0$, then $M''(\tau_b)$ exists. Accordingly, we have the following convergence results:\footnote{In the main text, we consider a fixed $\delta$ in estimation. In Appendix \ref{appendixb}, we have a complete discussion of the asymptotic properties, where we set $\delta_n = \delta + \eta n^{-\gamma}$ with $\gamma\in(0,\infty]$. We recommend setting $\gamma=\infty$, which leads to a fixed $\delta$, based on our analysis.}
\begin{align*}
	\sqrt{n} \begin{pmatrix} \hat{\tau}_p - \tau_p \\
   \hat{\tau}_o - \tau_o \end{pmatrix} \conD N\left(\begin{pmatrix}0\\0\end{pmatrix},\begin{pmatrix}\bbD_p(\tau_p)^\intercal \\
  \bbD_o(\tau_o)^\intercal\end{pmatrix}\bm{\Sigma} \begin{pmatrix} \bbD_p(\tau_p) & \bbD_o(\tau_o)\end{pmatrix} \right) .
\end{align*}
\end{theorem}

The bounds estimators $\hat{\tau}_p$ and $\hat{\tau}_o$ are asymptotically jointly normal. Both $\hat{\tau}_p$ and $\hat{\tau}_o$ are subject to the same source of estimation error, but with different ``loading'' terms $\bbD_b(\tau_b)$. Hence, their asymptotic covariance can be easily calculated based on Theorem \ref{thm:non-zero}.

When $\tau_b \neq 0$, the estimation errors of $V_p$, $V_o$, and $\tau^*$ will collectively influence $\hat{\tau}_b$. However, when $\tau_b = 0$, only the estimation error of $\tau^*$ contributes to the variability of $\hat{\tau}_b$, as shown in the following theorem. 

\begin{theorem}\label{thm:normality-zero}
Assume \eqref{eq:joint} holds. For $q\geq 2$, if $\tau_b=0$, we have the following convergence result:
\begin{align*}
	\sqrt{n}\hat{\tau}_b \conD \frac{Z_\tau}{1+ 1_{\{q=2\}}\delta \sqrt{V_b/2}}.
\end{align*}
For $q\in[1,2)$, the limiting distribution is non-normal.
\end{theorem}

Theorem \ref{thm:normality-zero} is a summary of Theorem \ref{thm:normality-zero-3cases} in Appendix \ref{appendixb}, which also covers the cases of $q\in(1,2)$ and $q=1$. The asymptotic behavior of the distributionally robust estimator becomes more intricate when the corresponding $\tau_b$ is zero, even though the primary source of stochasticity comes solely from the estimation of $\tau^*$. This complexity arises due to the interplay between the penalty term and the stochastic component of the objective function, which varies depending on the value of the penalty order $q$. 

When $q \geq 2$, the objective function is locally quadratic around zero, allowing standard techniques for M-estimation to apply. Under these conditions, the estimator retains asymptotic normality, a hallmark of well-behaved quadratic penalties. Notably, when $q = 2$, the stochastic term $A(\tau)$ exhibits a functional form nearly identical to the penalty term $B(\tau)$ with respect to $\tau$. This structural similarity enables the penalty to exert a slight shrinkage effect on the asymptotic variance, thereby improving estimation efficiency compared to the unpenalized estimator $\hat{\tau}^\ast$.

\subsection{Confidence Set}

In this section, we rule out the case $p=\infty$ to construct a unified inference procedure. In the dual problem, $p=\infty$ implies $q=1$, which imposes a significant penalty on the minimax prediction in a similar spirit to a LASSO estimator as shown in \textbf{Case 2-3} of Theorem \ref{thm:normality-zero-3cases} in Appendix \ref{appendixb}.

Theorem \ref{thm:normality-zero} implies that the bound estimators are not necessarily asymptotically normal when $\tau_b=0$, which is uninteresting and also causes trouble for inference. As a result, we would like to rule out this extreme case.
We propose the following two-step inference procedure. 
In the first step, we test the null $H^P_0: \tau^*=0$. Since the solution $\tau_b=0$ holds if and only if $\tau^*=0$ is true when $p\in (1,\infty)$, testing $H^Q_0: \tau^{\texttt{DR}}=0$ is equivalent to testing $H^P_0: \tau^*=0$, which is pretty straightforward. Intuitively, if we do not find any statistically significant internally valid treatment effect based on the source data, we typically would not bother assessing the external validity of our findings. On the other hand, if the null $H^P_0$ is rejected at a small significance level, we would like to see how robust our findings are under distributional shift. This is our goal in the second step.

Since $\tau^{\texttt{DR}}$ is never zero when $\tau^*\neq 0$, zero would not be a meaningful hypothesized value in the second step. Instead, policymakers might have a breakdown point in mind, for instance, the cost to implement the policy. A reasonable hypothesis would be whether the confidence interval of our minimax predictor contains the breakdown point. This is a nonstandard inference problem since $\tau^{\texttt{DR}}$ is only partially identified. Fortunately, we have shown in Section \ref{asym} that the upper and lower bound estimators $\hat{\tau}_o$ and $\hat{\tau}_p$ are asymptotically jointly normal. As a result, we can apply the approach in \cite{imbens2004confidence} (IM hereafter) and \cite{stoye2009more} to construct a confidence interval for $\tau^{\texttt{DR}}$. 

Nevertheless, such a two-step procedure comes with a caveat. We only proceed with the second step if $H^P_0$ is rejected in the first step, which introduces pre-testing bias. To solve this problem, we modify a Bonferroni-type correction approach in the literature for nonstandard inference; see, for instance, \cite{staiger1997instrumental}, \cite{romano2014practical}, \cite{mccloskey2017bonferroni}, and \cite{guo2024statistical}.

Let the size of the test be $\alpha$. In the first step, we construct a $1-\beta$ confidence interval for $\tau^*$, $I^*_n(1-\beta)$, where $\beta\in [0,\alpha]$ is some small value. Based on the value of $\tau^*=t$, $\forall\ t\in I^*_n(1-\beta)$, we construct the second-step confidence interval based on equation (6) in \cite{imbens2004confidence} with confidence level $1-\alpha+\beta$, $I_n^{[t]}(1-\alpha+\beta)$. In practice, one can create a fine grid of $I^*_n(1-\beta)$ in the first step and compute the corresponding confidence intervals in the second step. Lastly, we take a union of the second-step confidence intervals, $\mathcal{I}_n=\cup_{t\in I^*_n(1-\beta)}I_n^{[t]}(1-\alpha+\beta)$. Each interval in the second step accounts for the uncertainty of variance bound estimation only since we fix $\tau^*=t$, and the union step accounts for the uncertainty of $\hat{\tau}^*$. The inclusion of $\beta$ in the confidence level $1-\alpha+\beta$ accounts for the possibility that $\tau^*$ may not lie in $I_n^*(1-\beta)$.

To show the coverage of the two-step inference procedure, we first strengthen the pointwise convergence result in Lemma \ref{lem:JointCLT} to uniform convergence.
\begin{lemma}\label{lem:uniform_converge}
Let $\mathcal{P}$ be a family of the underlying distributions $P$, for $j=0,1$ $\tau_j = \mathbb{E}_P[Y(j)]$, $\sigma_j^2 = \mathbb{E}_P[ (Y(j) - \tau_j)^2 ]$. Suppose the assumptions in Lemma \ref{lem:JointCLT} hold. 

(\romannumeral 1) Assume that $ \sup_{P \in \mathcal{P}} \mathbb{E}_P\left[|Y(j)|^{6} \right] <\infty$ and $0 < \underline{\sigma}_j^2 \leq \sigma_j^2 \leq  \overline{\sigma}_j^2 < \infty$, $0<\eta\leq|\sigma_1^2 - \sigma_0^2|$, $0<\underline{e}\leq \mathbb{E}(T) \leq \overline{e} <1$ for all $P \in \mathcal{P}$. Then the convergence of $\sqrt{n}(\hat{\tau}^* - \tau^*, \hat{V}_p^N - V_p^N, \hat{V}_o^N - V_o^N)$ is uniform in $P\in\mathcal{P}$. The variance estimator $\hat{\bm{\Sigma}}$ based on the influence functions in Appendix \ref{appendixe} is also uniformly consistent in $P \in \mathcal{P}$.

(\romannumeral 2) In addition to the conditions in (\romannumeral 1), suppose that Assumption \ref{quantile} holds uniformly for all $P \in \mathcal{P}$. Then the convergence of $\sqrt{n}(\hat{\tau}^* - \tau^*, \hat{V}_p - V_p, \hat{V}_o - V_o)$ is uniform in $P\in\mathcal{P}$. The corresponding variance estimator $\hat{\bm{\Sigma}}$ is also uniformly consistent in $P \in \mathcal{P}$.
\end{lemma}

\begin{remark}
    Using standard argument, we can show that the upper and lower bound estimators in Theorem \ref{thm:non-zero} are asymptotically jointly normal uniformly in $P\in\mathcal{P}$ for $\mathcal{P}$ defined in Lemma \ref{lem:uniform_converge} such that $\tau^*\neq 0$. Uniform joint normality also holds if we condition on the estimate of $\tau^*$, i.e., the third element of $D_p$ and $D_o$ defined in (\ref{eqn:D}) is forced to be zero. Uniformly consistent estimator for the variance-covariance matrix in Theorem \ref{thm:non-zero} can also be easily constructed.
\end{remark}

\begin{theorem}\label{thm:inference}
Under Theorem \ref{thm:non-zero} and Lemma \ref{lem:uniform_converge}, for $\mathcal{I}_n$ defined in the above two-step inference procedure
\begin{align*}
\lim_{n\to\infty}\inf_{\tau^{\texttt{DR}}\in[\tau_p,\tau_o]}\inf_{P\in\mathcal{P}}\mathbb{P}\left(\tau^{\texttt{DR}}\in \mathcal{I}_n \right)\geq 1-\alpha,
\end{align*}
 where $\mathcal{P}$ is the family of distributions defined in Lemma \ref{lem:uniform_converge} such that $\tau^*\neq 0$.
\end{theorem}

\section{Parameter Choice}\label{sec:choice}
\subsection{Choice of Radius}\label{sec:radius}
An open question that we have not yet addressed is how to determine the radius of the Wassertein neighborhood. In the DRO literature, a data-driven approach has been proposed. The radius $\delta$ is chosen to be decreasing in sample size; see \cite{blanchet2022confidence} and \cite{lin2022distributionally}. This approach assumes an i.i.d. sample from the unknown target distribution, which does not comply with our setting. Also, in the limit $\delta$ approaches zero, which corresponds to the case without distributional shifts. We are instead interested in the generalizability of our causal estimates, given distributional shift, even when the sample size is large.

With the connection to the regularized estimation literature via the dual problem, $\delta$ can be considered as the penalization parameter. Following this literature, one might be tempted to find the optimal $\delta$ through cross-validation. This approach typically picks the penalization parameter as the one that minimizes the prediction error in subsamples. Since the target distribution $Q$ is unknown, we do not have a good criterion to assess the performance of different $\delta$. 

The nature of $\delta$ resembles that of the sensitivity parameter in the literature on sensitivity analysis that deals with the potential failure of the unconfoundedness assumption. Instead of the concern about internal validity in the sensitivity analysis literature, we assume internal validity but examine the robustness of the internal causal estimates under distributional shift. There is no single best choice for the sensitivity parameter. The general idea is to find some benchmark. In the sensitivity analysis literature, if the sensitivity parameter that nullifies the results is larger than a reasonable benchmark, then the causal findings are considered to be insensitive to the unobserved confounders. We follow the same spirit in finding benchmarks to help with the economic interpretation of $\delta$.

There are many ways one can form benchmarks. Below, we present a few possibilities. 
Even though in the target distribution $Q$ we do not observe $Y(1)$ since no treatment has been implemented yet, we might still be able to observe $Y(0)$. Hence, one can compute the Wasserstein distance for $Y(0)$ between $P$ and $Q$ distributions. In robust prediction, $\delta$ can be set to multiples of the Wasserstein distance for $Y(0)$ to approximate the true distance between $P$ and $Q$. Without access to the data from a target distribution, we can use the heterogeneity of $P$ as a benchmark for the distributional shift from $P$ to $Q$. If we observe covariates, we can split the sample based on these covariates and compute the Wasserstein distance of the potential outcomes across the resulting subsamples. 

For example, in the analysis of job training programs, it has been demonstrated that the pre-intervention employment record is one of the most important predictors of heterogeneous treatment effects; see \cite{hotz2005predicting} and \cite{gupta2023s}. As a result, we split the data into two subsamples, one previously employed and another previously unemployed. For the $L_2$ norm, the square root of the sum of the squared 2-Wasserstein distances of the marginal distributions serves as a lower bound for the Wasserstein distance of the joint distributions of potential outcomes. As a preview, for the job training program data used in Section \ref{sec:job} below, such a lower bound based on the $L_2$ norm cost function is \$1,154 between the two subsamples with or without a previous employment record, which is about 0.23 standard deviation of the post-treatment earnings. 
        
In practice, we recommend using a spectrum of $\delta$ as a stress test. Even though our robust prediction of the treatment effect can never be exactly zero if $q>1$, we can set the minimum level of the treatment effect that can offset the cost as the threshold. The radius $\delta$ leading to a prediction equal to the threshold would be an interesting cutoff, which is considered a breakdown point. We can use one or multiple benchmarking approaches proposed above to assess whether this $\delta$ is considered too small. If so, then our internal estimates might not be robust to distributional shift. In other words, there is not much external validity to our causal findings.   
	
\subsection{Choice of Norm}\label{norm}
For our general theory, we allow for any $q\in[1,\infty)$ in the definition of the Wasserstein neighborhood. All of our theoretical results hold, regardless of the value of $q$. Therefore, another loose end is how to choose $q$. The behavior of the minimax optimizer follows the same pattern as long as $q> 1$. The only difference is that $f(V,\delta)$ shrinks toward zero more slowly with the increase of $\delta$ as $q$ becomes larger. Moreover, $f(V,\delta)$ never reaches zero when $q>1$. On the other hand, $f(V,\delta)$ can be exactly zero when $q=1$ if $\delta$ is sufficiently large, behaving like a Lasso estimator. %However, when the treatment effect is homogeneous, our robust prediction will be either $\tau^*$ or zero if $q=1$, which might not be the most interesting case. 

	\begin{figure}
		\centering
		\begin{subcaptionbox}{Heterogeneous Treatment Effect\label{fig:left}}[0.48\textwidth]
			{\includegraphics[width=\linewidth]{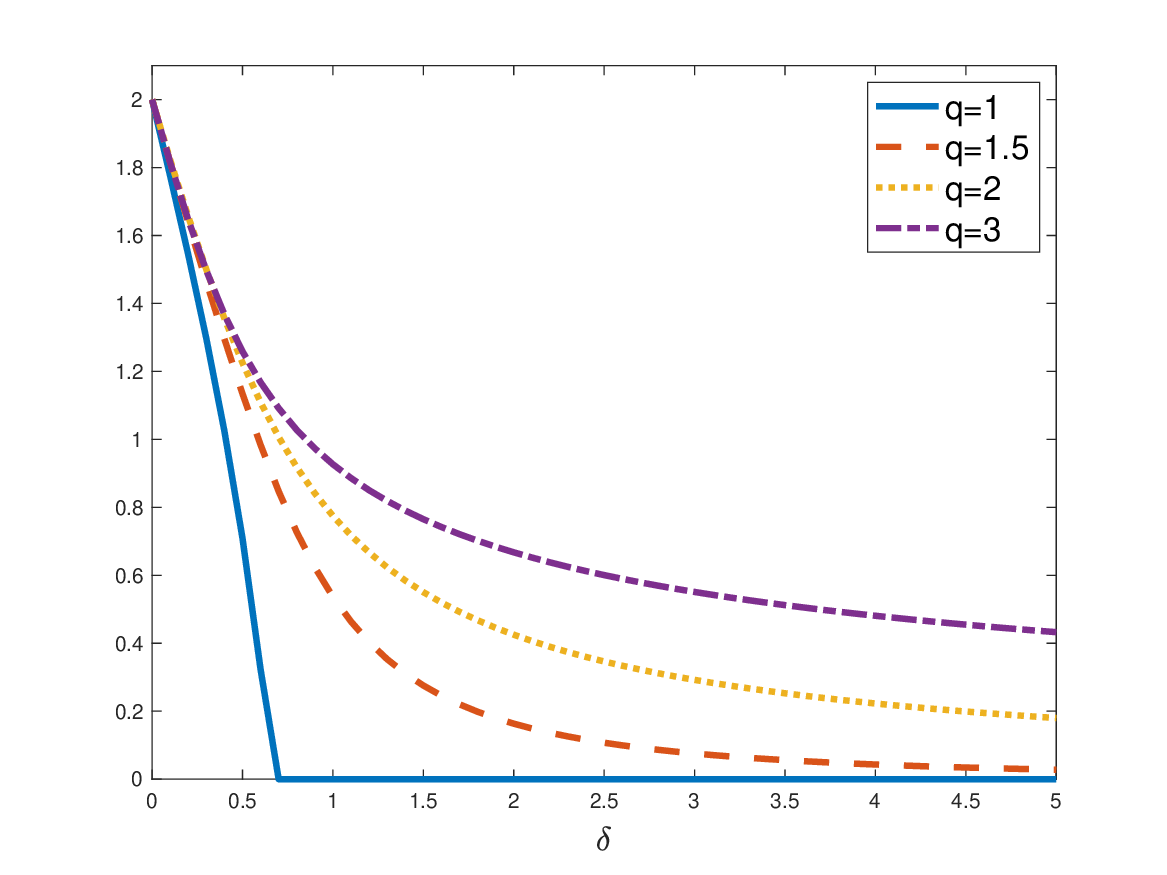}}
		\end{subcaptionbox}
		\hfill
		\begin{subcaptionbox}{Homogeneous Treatment Effect\label{fig:right}}[0.48\textwidth]
			{\includegraphics[width=\linewidth]{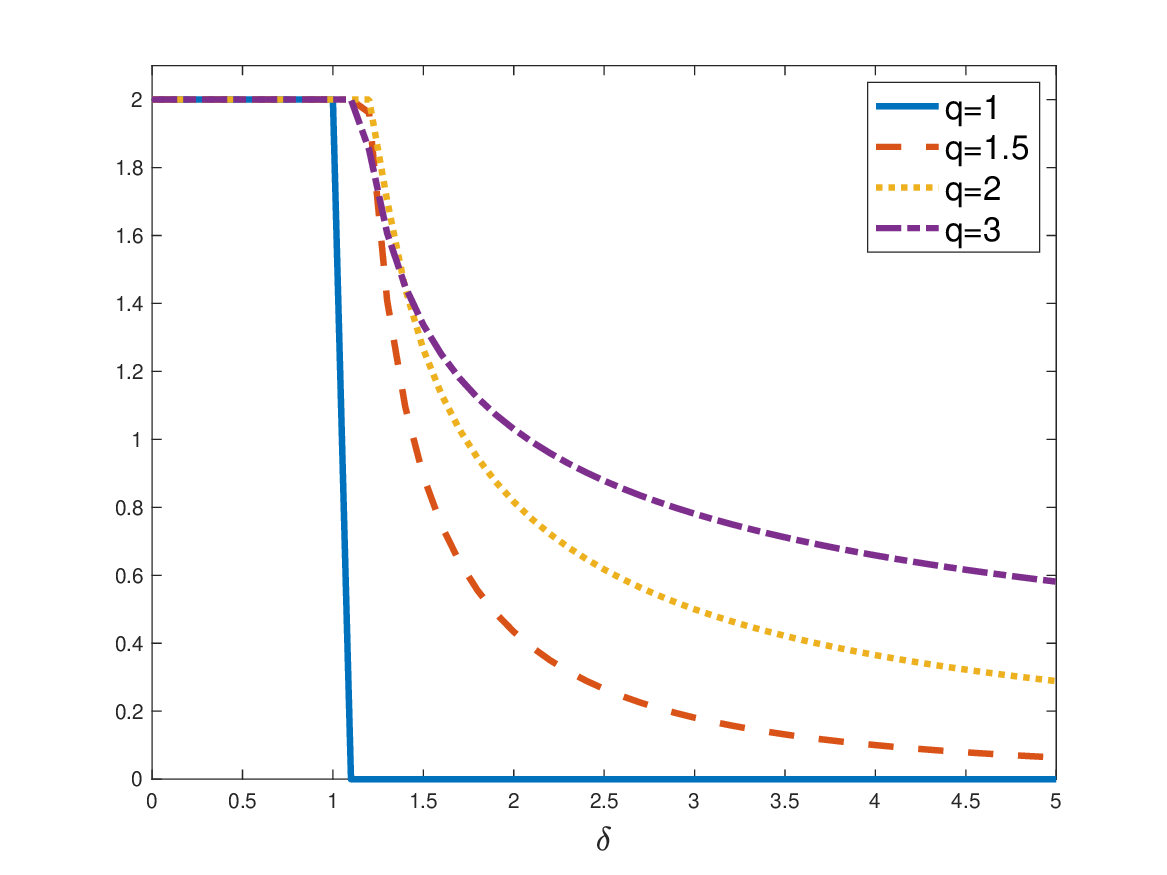}}
		\end{subcaptionbox}
		\caption{Robust Prediction for Different $q$}
		\label{fig: norm}
	\end{figure}

Figure \ref{fig: norm} plots the predicted treatment effect when $\tau^*=2$ for different values of $q$. Figure \ref{fig:left} shows the behavior of the minimax optimizer under heterogeneous treatment effect with $V=5$, and Figure \ref{fig:right} shows the case for homogeneous treatment effect with $V=0$. 

Parallel with the regularized estimation literature, multiple values of $q$ have been proposed, such as $q=1$ for Lasso and $q=2$ for Ridge. There is no single answer for the optimal $q$. In practice, we recommend using $q=2$ because of its tractability and clear interpretation. With the $L_2$ norm, which leads to $q=2$, $\|(Y(1),Y(0),1)-(\tilde{Y}(1),\tilde{Y}(0),w)\|_2^2=\|(Y(1),Y(0))-(\tilde{Y}(1),\tilde{Y}(0))\|_2^2+(1-w)^2$. Based on the worst-case distribution $\tilde{Q}$, we can quantify the distribution shift resulting from $(Y(1),Y(0))$, which turns out to be $\frac{2}{2+\tau^2}\delta^2$. Thus, when $\tau$ is small, the distribution shift is primarily driven by the change in potential outcomes.

\section{Simulation and Empirical Illustration}\label{sec:simulation}

\subsection{Simulation}

We study the finite sample performance of our two-step confidence intervals in simulation exercises. Potential outcomes $(Y(1),Y(0))$ are drawn from a bivariate normal distribution but truncated to $[-6, 6]^2$, 
\[
N\Bigg(\begin{pmatrix}\mu_1\\ \mu_0\end{pmatrix},\begin{pmatrix}
  \sigma_1^2 &\rho \sigma_1\sigma_0\\ \rho\sigma_1\sigma_0& \sigma_0^2  \end{pmatrix}\Bigg). 
\]
We set $\rho=0.7$, $\sigma_1=2$, $\sigma_0=1$, $\mu_1=\sigma_1$, and $\mu_0=0.2\sigma_0$, unless otherwise noted. Treatments are randomly assigned with probability 0.3.

We consider six cases: (\romannumeral 1) $p=2$, $\delta=0.1$; (\romannumeral 2) $p=2$, $\delta=1$; (\romannumeral 3) $p=2$, $\delta=1$, $\sigma_1=2$, $\sigma_0=0.01$; (\romannumeral 4) $p=2$, $\delta=0.1$, $\mu_1=0.2\sigma_1$, $\mu_0=0.1\sigma_0$; (\romannumeral 5) $p=1.5$, $\delta=0.1$; (\romannumeral 6) $p=3$, $\delta=0.1$. The first two cases serve as baselines with a small radius and a relatively large radius. The third case resembles near-point identification, where the sharp bounds of the variance of individual treatment effect after normalization are [1.92, 1.96]. Point identification can pose a threat to valid uniform inference under partial identification; see, for instance, \cite{imbens2004confidence}. Case (\romannumeral 4) examines the scenario where the variance of the outcome is significantly larger than the average treatment effect, making it challenging to estimate the ATE under the source distribution $P$ precisely. The first four cases use the $L_2$ norm for the Wasserstein distance. In contrast, the last two cases change the $L_p$ norm in the cost function of the Wasserstein neighborhood but otherwise remain the same as the baseline cases. We use the plug-in variance estimator $\hat{\bm{\Sigma}}$ based on the influence functions in Appendix \ref{appendixe}.

Table \ref{tab:6case} reports the coverage rate of the IM confidence intervals (CIs) with or without Bonferroni correction across 2,000 replications, the average CIs, and the average length ratio of the two-step CIs over the non-corrected CIs. To proceed with our proposed two-step CIs, replications with first-step CIs containing zero are dropped. Whenever this occurs, the reported results are averages across the remaining replications. We report the results with both sharp variance bounds and Neyman variance bounds. We set $\alpha=0.05$ and $\beta=0.045$. In our simulations, the two-step CIs are less conservative when $\beta$ is closer to $\alpha$, but there is not much improvement when $\beta$ is larger than 0.045.

In a finite sample with 500 observations, the non-corrected IM confidence intervals (CIs) exhibit slight under-coverage in cases (\romannumeral 3) and (\romannumeral 4). Nonetheless, the two-step CIs consistently achieve the nominal coverage rate, as expected. When the sample size increases to 1,000, the coverage rate of the non-corrected IM CIs exceeds 0.95 in case (\romannumeral 4). However, in case (\romannumeral 3), the coverage rate of the non-corrected CIs remains below 0.92 even with 2,000 observations per sample. These results remain qualitatively unchanged even with 5,000 replications. When both the non-corrected and two-step CIs attain the nominal coverage rate, the two-step CIs are 16–25\% wider than the non-corrected CIs, and their coverage rate can approach one as the neighborhood radius increases. The performance of the CIs is stable across different choices of $L_p$ norms and is consistent between sharp and Neyman variance bounds.  

\begin{table}
  \centering
  \caption{Performance of 95\% Confidence Intervals}
  \begin{threeparttable}
    \begin{tabular}{lcccccc}
    \toprule
   
          & $\tau^{\texttt{DR}}$ & \multicolumn{2}{c}{Coverage} & \multicolumn{2}{c}{CI} & Length Ratio \\  
          \cmidrule(lr){3-4} \cmidrule(lr){5-6}
          &       & IM    & IM\_Bonf & IM    & IM\_Bonf &  \\
          \midrule
            & \multicolumn{6}{c}{Sharp} \\
     \cmidrule{2-7}
   Case (\romannumeral 1) & 1.648 & 0.964 & 0.988 & (1.298, 1.943) & (1.246, 2.004) & 1.177 \\
    Case (\romannumeral 2) & 1.049 & 0.996 & 1.000 & (0.629, 1.418) & (0.577, 1.564) & 1.252 \\
    Case (\romannumeral 3) & 0.957 & 0.899 & 0.978 & (0.788, 1.120) & (0.740, 1.191) & 1.362 \\
    Case (\romannumeral 4) & 0.277 & 0.927 & 0.954 & (0.127, 0.733) & (0.104, 0.769) & 1.098 \\
    Case (\romannumeral 5) & 1.659 & 0.953 & 0.981 & (1.333, 1.954) & (1.289, 2.011) & 1.165 \\
    Case (\romannumeral 6) & 1.634 & 0.955 & 0.986 & (1.262, 1.932) & (1.202, 1.999) & 1.191 \\
   
    \midrule
          & \multicolumn{6}{c}{Neyman} \\
     \cmidrule{2-7}
   Case (\romannumeral 1) & 1.648 & 0.966 & 0.988 & (1.296, 1.945) & (1.245, 2.007) & 1.176 \\
    Case (\romannumeral 2) & 1.049 & 0.997 & 1.000 & (0.628, 1.433) & (0.576,  1.581) & 1.252 \\
    Case (\romannumeral 3) & 0.957 & 0.906 & 0.980 & (0.788, 1.121) & (0.740, 1.191) & 1.359 \\
    Case (\romannumeral 4) & 0.277 & 0.933 & 0.954 & (0.126, 0.735) & (0.104, 0.770) & 1.095 \\
    Case (\romannumeral 5) & 1.659 & 0.955 & 0.981 & (1.333, 1.956) & (1.289, 2.013) & 1.163 \\
    Case (\romannumeral 6) & 1.634 & 0.956 & 0.987 & (1.261, 1.935) & (1.201, 2.001) & 1.190 \\
 
    \bottomrule
    \end{tabular}%
    \begin{tablenotes}
    \footnotesize
\item[1]  In the first column, $\tau^{\texttt{DR}}$ is the non-point-identified minimax optimizer in the population. Coverage rate is with respect to $\tau^{\texttt{DR}}$.
\item[2] IM stands for Imbens and Manski confidence interval, and IM\_Bonf stands for the Imbens and Manski confidence interval with Bonferroni correction.
\item [3] The confidence intervals are averaged over 2,000 replications, and the length ratio represents the length of the two-step confidence intervals compared to the IM confidence intervals averaged across 2,000 replications (or the remaining replications where replications with first-step CIs containing zero are dropped).
\item [4] The top and bottom panels report results for the sharp and Neyman variance bounds, respectively.
\end{tablenotes}
    \end{threeparttable}
  \label{tab:6case}%
\end{table}%

\subsection{Empirical Simulation}\label{sec:job}
	% LaLonde(1986), Dehejia and Wahba(1999,2002) 
	
We illustrate our prediction method in the context of a job training program. During the mid-1970s, the National Supported Work Demonstration program randomly assigned qualified applicants to training positions. Pioneered by \cite{lalonde1986evaluating} and followed by \cite{dehejia1999causal} and many others, this dataset has been extensively studied to evaluate the performance of different causal estimators. The outcome variable is earnings in 1978 for men, and the treatment variable is participation in the job training program.  

Instead of the original experimental data, we use the artificial data of size 1,000,000 generated by (Wasserstein) Generative Adversarial Networks in \cite{athey2024using} as our population. There are 445 observations in the experimental sample used in \cite{dehejia1999causal}, comprising 185 men who were treated and 260 men who were untreated. We generate 2,000 samples randomly drawn from our artificial population, while maintaining the fixed ratio of treatment and control units. The treatment subsample size is  $n_1=n*185/(185+260)$ and the control subsample size is $n_0=n*260/(185+260)$. We try different sample sizes $n$.

The population ATE is 1,333 dollars. Because the artificial population comes with counterfactuals, we can compute the population variance of individual treatment effect, $Var_P\left(Y(1)-Y(0)\right)$. Table \ref{variance} reports the population variance, the sharp bounds, and the Neyman bounds of the variance. The Neyman bounds are wider than the sharp bounds, as expected. Both lower bounds are pretty close to zero.  

			\begin{table}
				\caption{Variance of Individual Treatment Effect}
		\centering
        \begin{threeparttable}
		\begin{tabular}{ccccc}
			\toprule
			Var & Sharp\_l & Sharp\_u & Neyman\_l & Neyman\_u \\
			\midrule
			45.76  & 1.15 & 86.38 & 1.05 & 98.47\\
			\bottomrule
		\end{tabular}
        \begin{tablenotes}
    
        \footnotesize
            \item [1] This table reports the population variance of individual treatment effect as well as the upper and lower bounds of the variance of individual treatment effect based on sharp bounds or Neyman bounds. 
            \item [2] Wages are measured in thousands of dollars in this table. 
        \end{tablenotes}
          \end{threeparttable}
        \label{variance}
	\end{table}

\begin{figure}[h]
		\centering
		\includegraphics[width=4in]{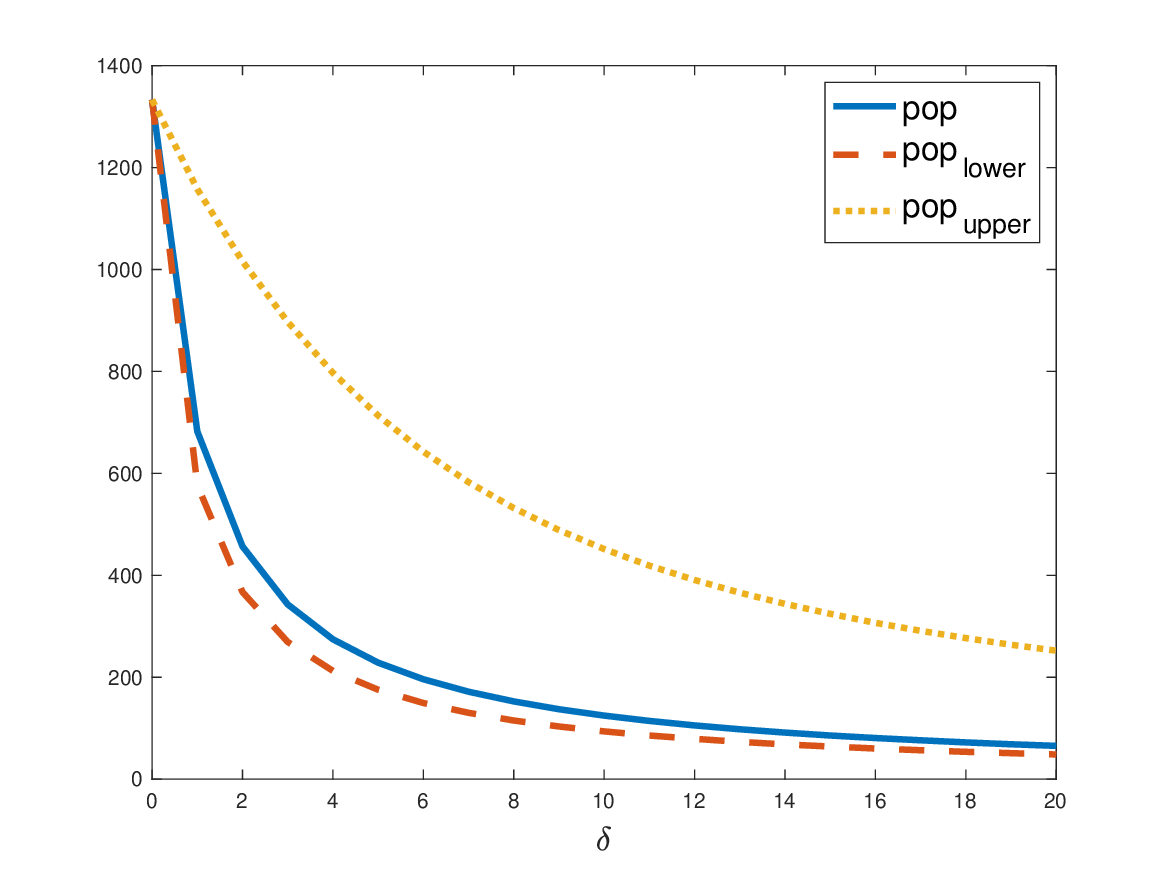} 
		\caption{Population Prediction $\tau^{\texttt{DR}}$, $\tau_p$, and $\tau_o$}
		\label{fig: pop}
	\end{figure}

We first compute the population prediction with respect to the joint distribution $P$, the perfectly positively dependent copula, and the perfectly negatively dependent copula, respectively. Figure \ref{fig: pop} depicts the predictions for a range of $\delta$ with $q=2$. As expected, $\tau_p\leq \tau^{\texttt{DR}}\leq \tau_o$.

\begin{figure}[H]
		\centering
		\begin{subcaptionbox}{$q=2$, $n=445$}[0.48\textwidth]
			{\includegraphics[width=\linewidth]{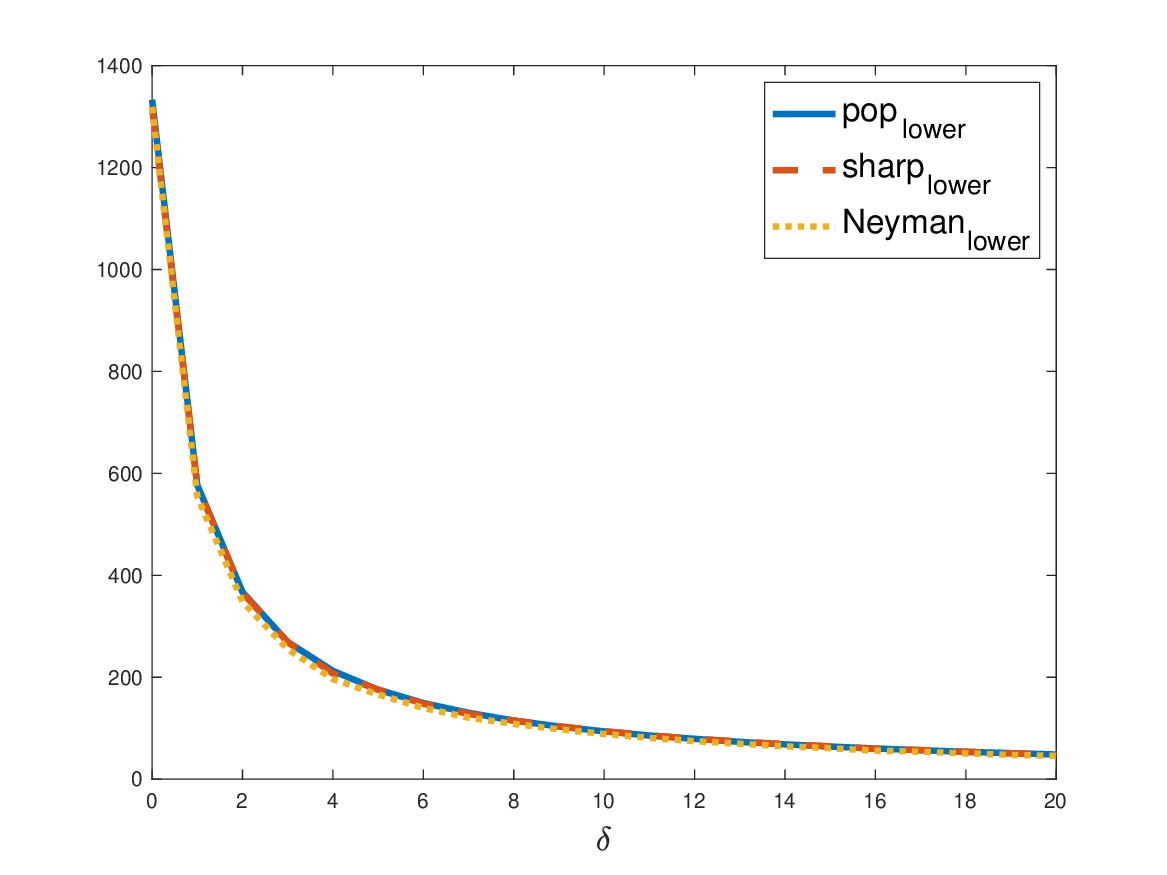}}
		\end{subcaptionbox}
		\hfill
		\begin{subcaptionbox}{$q=2$, $n=4450$}[0.48\textwidth]
			{\includegraphics[width=\linewidth]{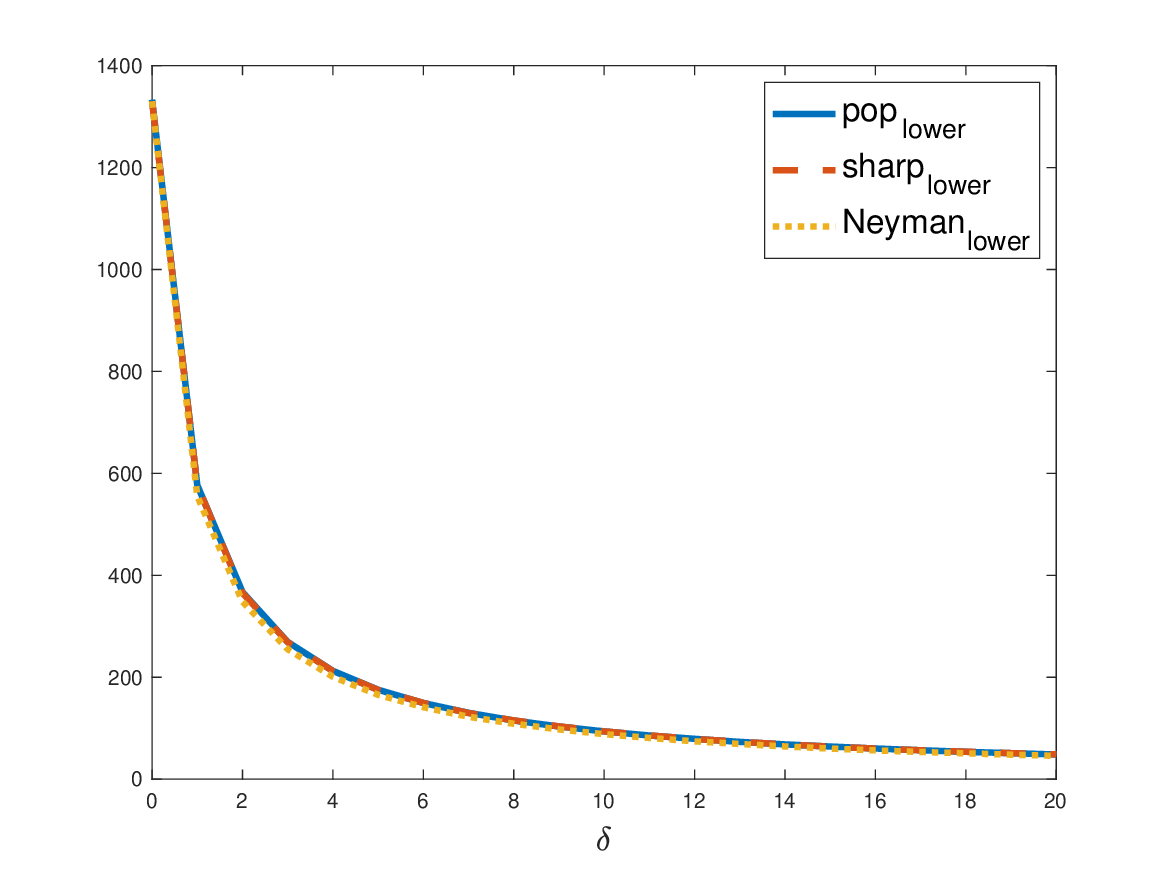}}
		\end{subcaptionbox}
	 \caption{Lower Bound of the Best Prediction}
		\label{fig: lower}
	\end{figure}

\begin{figure}[H]
		\centering
		\begin{subcaptionbox}{$q=2$, $n=445$}[0.48\textwidth]
			{\includegraphics[width=\linewidth]{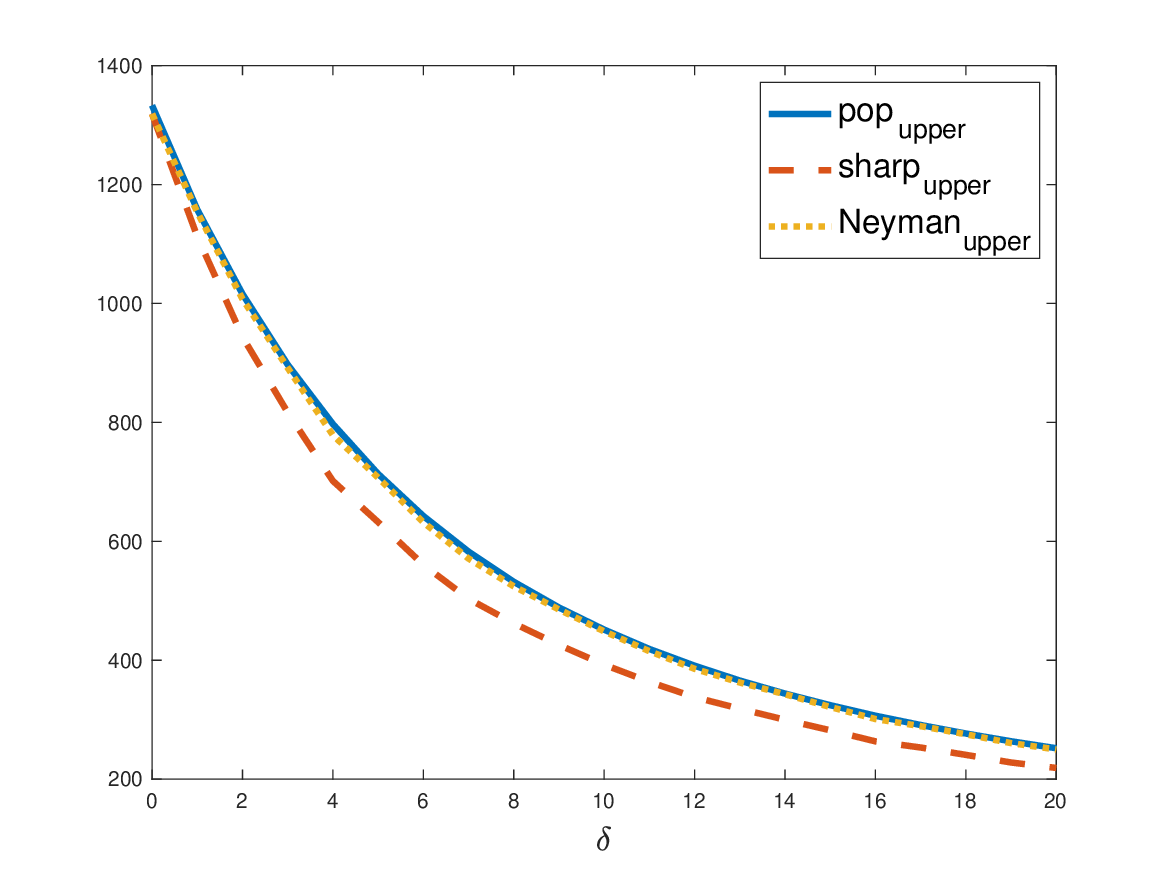}}
		\end{subcaptionbox}
		\hfill
		\begin{subcaptionbox}{$q=2$, $n=4450$}[0.48\textwidth]
			{\includegraphics[width=\linewidth]{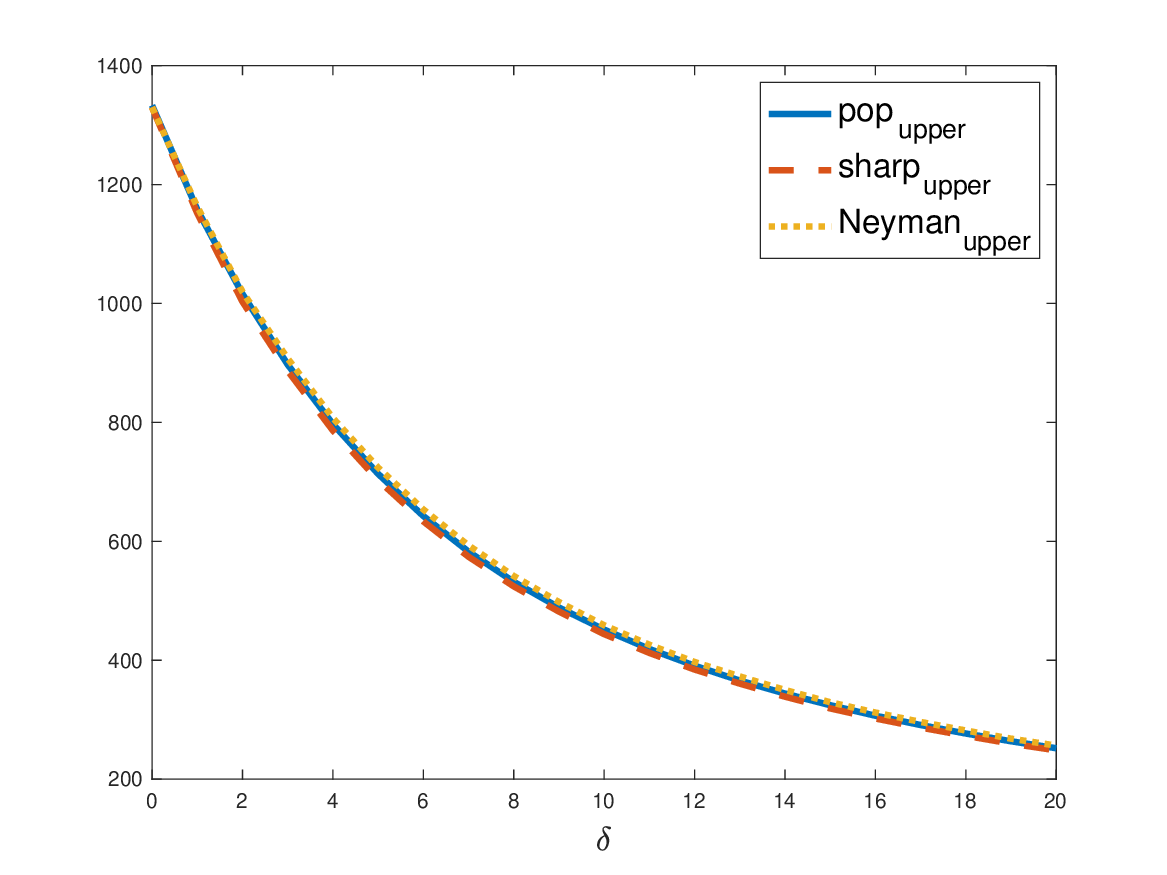}}
		\end{subcaptionbox}
		\caption{Upper Bound of the Best Prediction}
		\label{fig: upper}
	\end{figure}

Next, we compare the average prediction across 2,000 samples with the population prediction. The bound parameters are obtained based on sharp variance bounds using population data. We examine the upper and lower bounds of the prediction in the sample, respectively, using either Neyman variance bounds or sharp variance bounds. For the left panel of Figures \ref{fig: lower} and \ref{fig: upper}, the sample size is 445. The sample predictions are close to the population prediction for the lower bound. However, there are noticeable gaps for the upper bound estimator based on the sharp variance bound. Increasing the sample size by tenfold in the right panels leads to sample predictions aligning much more closely with the population prediction.

    \begin{figure}[H]
        \centering
        \includegraphics[width=4in]{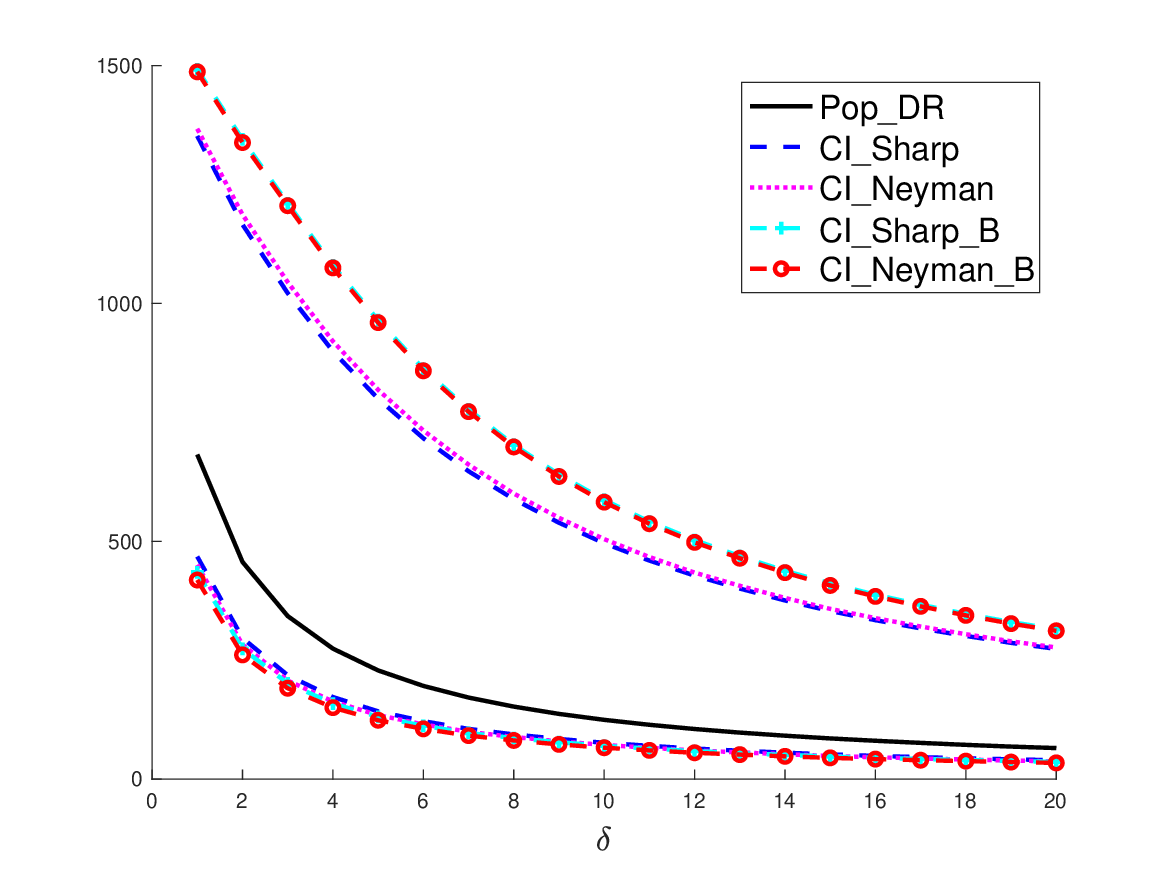}
        \caption{Comparison of 95\% Confidence Intervals}
        \label{fig:CI}
    \end{figure}

For the sample size $n=4,450$, we also examine the IM confidence intervals based on sharp variance bounds and Neyman variance bounds, with and without the Bonferroni correction. The results are plotted in Figure \ref{fig:CI}. The solid black line collects the value of $\tau^{\texttt{DR}}$ in the population corresponding to different radii of the Wasserstein neighborhood. Each dot on the upper and lower curves is the average of the confidence interval endpoints across 2,000 replications. The Neyman bound CIs are pretty similar to the sharp bound CIs. Not surprisingly, our two-step CIs are wider than the non-corrected CIs, but they are not a lot wider. 

The length ratios of the CIs with or without Bonferroni correction are reported in Table \ref{tab:lengthratio} below. The ratios are quite reasonable, ranging from 1.16 to 1.27. The length ratios based on the Neyman variance bounds are overall smaller, implying that two-step CIs based on the Neyman bounds are less conservative compared with the two-step CIs based on the sharp bounds.

In the literature, it has been recorded that the average cost of providing job training services ranges from \$953, \$919, \$430 to \$118 per trainee across various locations in the US in the early 1980s; see \cite{hotz2005predicting}. For the job training program on which our artificial population data is based, \cite{lalonde1986evaluating} records that the program cost is at least \$2,700 per trainee. 
The average cost per trainee across the four locations is \$605. Based on the first-order condition of the outer minimization problem, the lower bound $\tau_p=605$ when $\delta=0.92$ using the sharp bound of the variance. Using the benchmark \$1,154 calculated in Section \ref{sec:radius} based on the heterogeneity between two subsamples, which is equivalent to 0.23 standard deviation of the realized outcome, this $\delta$ is sizable, indicating some robustness of the treatment effect against distributional shift.

\begin{table}[htbp]
  \centering
  \caption{CI Length Ratio}
  \begin{threeparttable}
    \begin{tabular}{lcccccccccc}
    \toprule
     $\delta$ & 1     & 3     & 5     & 7     & 9     & 11    & 13    & 15    & 17    & 19 \\
    \midrule
     Sharp & 1.20  & 1.26  & 1.27  & 1.26  & 1.24  & 1.22  & 1.21  & 1.21  & 1.20  & 1.19 \\
    Neyman & 1.17  & 1.21  & 1.22  & 1.21  & 1.20  & 1.19  & 1.18  & 1.17  & 1.17  & 1.16 \\
    \bottomrule
    \end{tabular}%
    \begin{tablenotes}
        \footnotesize
        \item [1] This table reports the length ratio of our two-step CIs compared to the IM CIs averaged over 2,000 replications.
        \item [2] ``Sharp" stands for the predictions based on the sharp bounds of the variance of individual treatment effect and ``Neyman" stands for the predictions based on the Neyman variance bounds.  
    \end{tablenotes}
    \end{threeparttable}
  \label{tab:lengthratio}%
\end{table}%

\section{Conclusion}\label{sec:conclude}

We propose a method for out-of-population prediction of treatment effect with only retrospective data. Although our robust prediction is partially identified, we provide a confidence set for the prediction through a two-step procedure. 

In the current paper, we consider only distributional shifts in potential outcomes and do not include covariates. In practice, however, covariates play an important role in observational data. Extending our framework to incorporate covariates is an important direction for future research. Ideally, we would allow for distributional shifts both in covariates and in the conditional distribution of potential outcomes. 

We focus on a single cross section in this paper. However, panel data have been used extensively in empirical works. A popular method for identifying causal effects is the difference-in-differences approach, which utilizes panel or pooled cross-sectional data. There are typically multiple periods post treatment. It would be interesting to generalize our method to short panel data.

\bibliography{external}

\appendix
\numberwithin{figure}{section}
\numberwithin{table}{section}
\numberwithin{equation}{section}
\numberwithin{assumption}{section}

\section{Proof of the Identification Results}

\textbf{Proof of Proposition \ref{prop1}:} 

The proof closely follows that of Proposition 2 in \cite{blanchet2019robust}. By setting $\bar{\beta}=(1,-1,-\tau)$ and $\bar{X}=(Y(1),Y(0),w)$, we get (\ref{dual}) directly. 

\bigskip
\noindent
\textbf{Proof of Proposition \ref{prop2}:} 

(1) The first-order partial derivative of the dual problem $\sqrt{V+(\tau^*-\tau)^2}+\delta(2+|\tau|^q)^{1/q}$ with respect to $\tau$ is 

\begin{equation}\label{eqn:partial1}
\frac{\partial}{\partial \tau} 
\left[
\sqrt{V+(\tau^*-\tau)^2}
+\delta(2+|\tau|^q)^{1/q}
\right]
=
\frac{\tau-\tau^*}{\sqrt{V+(\tau^*-\tau)^2}}
+\delta\, |\tau|^{q-1}\operatorname{sign}(\tau)\,(2+|\tau|^q)^{\frac{1}{q}-1}.
\end{equation}

The derivative (\ref{eqn:partial1}) evaluated at $\tau^*$ is $\delta\, |\tau^*|^{q-1}\operatorname{sign}(\tau^*)\,(2+|\tau^*|^q)^{\frac{1}{q}-1}$, which has the same sign as $\tau^*$. On the other hand, (\ref{eqn:partial1}) evaluated at zero is $-\frac{\tau^*}{\sqrt{V+(\tau^*)^2}}$, which has the opposite sign of $\tau^*$. 

When $\tau\neq 0$, the second-order derivative is 
\[
\frac{V}{\big(V+(\tau^*-\tau)^2\big)^{3/2}}
+2\delta (q-1)\,|\tau|^{\,q-2}\big(2+|\tau|^q\big)^{\frac{1}{q}-2},
\]
which is positive since $q\geq 1$. Therefore, the dual problem is a convex function. As a result, the solution to its first-order condition is bounded between zero and $\tau^*$ and will be zero if $\tau^*=0$.

(2) Set (\ref{eqn:partial1}) to zero and then apply the implicit function theorem. The derivative of $\tau$ with respect to $V$ is 
\[
\frac{d\tau}{dV}
=
\frac{\tau-\tau^*}
{
2\left(V
+
2\delta (q-1)\,|\tau|^{q-2}\big(2+|\tau|^q\big)^{\frac{1}{q}-2}
\big(V+(\tau^*-\tau)^2\big)^{3/2}\right)}.
\]
Therefore, $\tau$ is monotonic in $V$. If $\tau^*>0$, $\frac{d\tau}{dV}<0$ and hence $0<f(V_p,\delta)<f(V^P, \delta)<f(V_o,\delta)$. If $\tau^*<0$, $\frac{d\tau}{dV}>0$ and hence $0>f(V_p,\delta)>f(V^P, \delta)>f(V_o,\delta)$. 

\bigskip
\noindent
\textbf{Proof of Proposition \ref{prop3}:} 

We know if $\tau^*=0$, then $\tau=0$. Let us consider the more interesting case of $\tau^*\neq 0$. 
The first-order derivative of (\ref{obj:homo}) is 
\begin{equation}
  \operatorname{sign}(\tau-\tau^*) + \delta\, |\tau|^{q-1}\operatorname{sign}(\tau)\,(2+|\tau|^q)^{\frac{1}{q}-1}.
\end{equation}
The second-order derivative is 
\[
2\delta (q-1)\,|\tau|^{\,q-2}\big(2+|\tau|^q\big)^{\frac{1}{q}-2},
\]
which is nonnegative. 
Therefore, $\tau^*>0$ is a global minimizer if and only if 
\[
-1+\delta\, (\tau^*)^{q-1}(2+(\tau^*)^q)^{\frac{1}{q}-1}\leq 0.
\]
On the other hand, $\tau^*<0$ is a global minimizer if and only if 
\[
1-\delta\, (-\tau^*)^{q-1}(2+(-\tau^*)^q)^{\frac{1}{q}-1}\geq 0.
\]
As a result, $\delta \leq \left(\frac{2}{|\tau^*|^q}+1\right)^{1-1/q}$.

\section{Proof of the Asymptotic Results}\label{appendixb}

To simplify the notation in the appendix, all expectations are taken over the source distribution $P$ unless stated otherwise.

\bigskip
\noindent
\textbf{Proof of Theorem \ref{thm:consistency}:}

Denote by $V_b$ whichever bound for the variance term (optimistic or pessimistic, Neyman or sharp). Define the sample and population objective functions:
\begin{align*}
M_n(\tau) = \Big( \hat{V}_b + (\hat{\tau}^\ast - \tau)^2 \Big)^{1/2} + \delta_n \left(2 + |\tau|^q\right)^{1/q},
\end{align*}
\begin{align*}
M(\tau) = \Big( V_b + (\tau^* - \tau)^2 \Big)^{1/2} + \delta (2 + |\tau|^q)^{1/q},
\end{align*}
where $\delta_n \rightarrow \delta \geq 0$ and $\hat{V}_b$ is a consistent estimator of $V_b$.

We aim to show that for $b\in\{p,o\}$:
\begin{align*}
\hat{\tau}_b = \operatorname{argmin}_\tau M_n(\tau) \conP \tau_b = \operatorname{argmin}_\tau M(\tau).
\end{align*}

\begin{proof}

\textbf{(i) Uniqueness of the minimizer.} It is easy to see that $\left(2 + |\tau|^q\right)^{1/q}$ is convex for $q \geq 1$ (as the $L_q$ norm is convex). Besides, the following 
\begin{align*}
	\Big( V_b + (\tau^* - \tau)^2 \Big)^{1/2}
\end{align*}
is also a convex function of $\tau$. Consequently, the population objective function $M(\tau)$ is convex in $\tau$. Therefore, it has a unique minimizer. It is easy to verify that this minimizer is finite. Similarly, for each sample, the minimizer of $M_n(\tau)$ is also finite. Hence, the estimator $\hat{\tau}_n$ is bounded in probability. 

\textbf{(ii) Consistency.} Let
\begin{align*}
	A(\tau) = (V_b + (\tau^* - \tau)^2 )^{1/2} \quad \text{and} \quad B(\tau) = \left(2 + |\tau|^q\right)^{1/q}.
\end{align*}
We then have $M_n(\tau) - M(\tau) = A_n(\tau) - A(\tau) + (\delta_n - \delta) B(\tau)$.
By Lemma \ref{lem:JointCLT} and the continuous mapping theorem, we have
\begin{align*}
	M_n(\tau) \conP M(\tau), \; \text{ for each fixed $\tau$}.
\end{align*}

Let $\calK\subset \bbR$ be a compact set. It is easy to see that the class $\{ A(\tau): \tau \in \calK \}$ is Lipschitz in $\tau$. Then the above convergence in probability holds uniformly in $\calK$.

\textbf{(iii) Uniform tightness.} The consistency of $\hat{\tau}^\ast$ implies that it is bounded in probability. For any $\tau$, it is easy to see that
\begin{align*}
	M_n(\tau) \geq |\tau - \hat{\tau}^\ast| +  \delta_n |\tau| \geq (\delta_n+1) |\tau| - |\hat{\tau}^\ast|.
\end{align*}
This ensures the empirical objective function grows unbounded as $|\tau| \rightarrow \infty$, forcing $\hat{\tau}_b$ to stay finite. For any $R>0$, we have
\begin{align*}
	\mathbb{P}(|\hat{\tau}_b| > R) &\leq \mathbb{P}\big( \inf_{|\tau| > R} M_n(\tau) \leq M_n(0) \big) \leq \mathbb{P}\big( (\delta_n+1) R - |\hat{\tau}^\ast| \leq M_n(0) \big) \\
	&= \mathbb{P}\big( M_n(0) + |\hat{\tau}^\ast| \geq (\delta_n+1) R\big) \leq \frac{\bbE[M_n(0) + |\hat{\tau}^\ast|  ]}{(\delta_n + 1) R}
\end{align*}
Since $M_n(0) = O_p(1)$ and $\hat{\tau}^\ast = O_p(1)$, the right-hand side can be made smaller than any $\epsilon>0$ uniformly in $n$. This shows that the sequence $\{\hat{\tau}_b\}_n$ is uniformly tight.

Given the above three results, Corollary 3.2.3 of \cite{EmpiricalProcess} implies that $\hat{\tau}_b \conP \tau_b$.

\end{proof}

\bigskip
\noindent
\textbf{Proof of Theorems \ref{thm:non-zero} and \ref{thm:normality-zero}:}

Suppose that $\delta_n - \delta = \eta n^{-\gamma}$. For the results in the main text, we set $\gamma=\infty$. In the appendix, we provide a detailed discussion on the asymptotic properties with all values of $\gamma$. As a result, we illustrate how other rates of $\gamma$, possibly induced by data-driven choices of $\delta_n$, are not recommended. Recall from Lemma \ref{lem:JointCLT}
\begin{equation}\label{eqn:joint-recall}
    \sqrt{n}(\hat{V}_p - V_p, \hat{V}_o - V_o, \hat{\tau}^\ast - \tau^*)^\intercal \conD (Z_p, Z_o, Z_\tau)^\intercal \sim N(0, \bm{\Sigma}).
\end{equation}

\subsection*{Case 1: $\tau_b \neq 0$ and $q\geq1$}
Let
$\bbB(\tau_b) = - \eta [ M''(\tau_b) ]^{-1}  B^\prime(\tau_b)$,
where $B^\prime$ denotes the first-order derivative of $B$ and $M''$ denotes the second-order derivative of $M$.
We state the complete theorem below.
\begin{theorem} \label{thm:non-zero-3cases}
Assume \eqref{eqn:joint-recall} holds. For $q\geq1$ and $b\in\{p,o\}$, if $\tau_b \neq 0$, then $M''(\tau_b)$ exists. Accordingly, we have the following convergence results:
\begin{itemize}
\item \textbf{Case 1-1:} $\gamma \in (1/2, \infty]$
\begin{align*}
	\sqrt{n} \begin{pmatrix} \hat{\tau}_p - \tau_p \\
   \hat{\tau}_o - \tau_o \end{pmatrix} \conD N\left(\begin{pmatrix}0\\0\end{pmatrix},\begin{pmatrix}\bbD_p(\tau_p)^\intercal \\
  \bbD_o(\tau_o)^\intercal\end{pmatrix}\bm{\Sigma} \begin{pmatrix} \bbD_p(\tau_p) & \bbD_o(\tau_o)\end{pmatrix} \right) .
\end{align*}

\item \textbf{Case 1-2:} $\gamma = 1/2$

\begin{align*}
	\sqrt{n} \begin{pmatrix} \hat{\tau}_p - \tau_p \\
   \hat{\tau}_o - \tau_o \end{pmatrix} \conD N\left(\begin{pmatrix}\bbB(\tau_p)\\
   \bbB(\tau_o)\end{pmatrix},\begin{pmatrix}\bbD_p(\tau_p)^\intercal \\
  \bbD_o(\tau_o)^\intercal\end{pmatrix}\bm{\Sigma} \begin{pmatrix} \bbD_p(\tau_p) & \bbD_o(\tau_o)\end{pmatrix} \right).
\end{align*}

\item \textbf{Case 1-3:} $\gamma \in (0,1/2)$
\begin{align*}
	n^\gamma \big( \hat{\tau}_b - \tau_b \big) \conP \bbB(\tau_b) .
\end{align*}
\end{itemize}
\end{theorem}

\begin{proof}
If $\tau_b \neq 0$, then $M(\tau)$ is twice differentiable at a neighborhood of $\tau_b$ for $q\geq1$. The first-order derivative is given by $M'(\tau) =  A'(\tau) + \delta B'(\tau)$, where
\begin{gather*}
A'(\tau) = \frac{-(\tau^* - \tau)}{(V_b + (\tau^* - \tau)^2)^{1/2}}, \\
B'(\tau) = |\tau|^{q-1} \cdot \text{sign}(\tau) \cdot (2 + |\tau|^q )^{1/q - 1} = \tau |\tau|^{q-2} (2 + |\tau|^q )^{1/q - 1}.
\end{gather*}
Similarly, we have $M_n'(\tau) =  A_n'(\tau) + \delta_n B'(\tau)$, where $A_n'(\tau)$ is obtained with $V_b$ and $\tau^*$ in $A'(\tau)$ replaced by $\hat{V}_b$ and $\hat{\tau}^\ast$, respectively.

Second-order derivatives are given by:
\begin{gather*}
A''(\tau) = \frac{V_b}{\left(V_b + (\tau^* - \tau)^2\right)^{3/2}}, \\
B''(\tau) = 2(q-1) (2 + |\tau|^q)^{\frac{1}{q} - 2} \cdot |\tau|^{q - 2}.  %(1 - q)(2 + |\tau|^q)^{\frac{1}{q} - 2} \cdot |\tau|^{2q - 2} + (q - 1)(2 + |\tau|^q)^{\frac{1}{q} - 1} \cdot |\tau|^{q - 2} 
\end{gather*}
Note that when $q=1$, we have $B''(\tau) = 0$ at $\tau\neq0$, which is in line with the above expression.

The following result readily follows from standard M-estimation theory,
\begin{align*}
\sqrt{n}(\hat{\tau}_b - \tau_b) = - [ M''(\tau_b) ]^{-1} \cdot \sqrt{n} M_n'(\tau_b) + o_p(1).
\end{align*}
Since $M'(\tau_b) = 0$, we have $\delta B'(\tau_b) = - A'(\tau_b)$. It then follows that
\begin{align*}
	M_n'(\tau_b) &= \frac{-(\hat{\tau}^* - \tau_b)}{(\hat{V}_b + (\hat{\tau}^* - \tau_b)^2)^{1/2}} + \frac{(\tau^* - \tau_b)}{(V_b + (\tau^* - \tau_b)^2)^{1/2}} + (\delta_n - \delta) B'(\tau_b) \\
	&= \frac{(\tau^* - \hat{\tau}^*)}{A_n(\tau_b)} + (\tau^* - \tau_b) \Big( \frac{1}{A(\tau_b)} -  \frac{1}{A_n(\tau_b)} \Big) + (\delta_n - \delta) B'(\tau_b) 
\end{align*}

\subsubsection*{Case 1-1: $\gamma>1/2$}

In this case, we have $(\delta_n - \delta) B(\tau) = o( n^{-1/2} )$. Accordingly, we get
\begin{align*}
	\sqrt{n} M_n'(\tau_b) &= \frac{\sqrt{n}(\tau^* - \hat{\tau}^*)}{A(\tau_b)} + \sqrt{n} \big( A_n(\tau_b) - A(\tau_b) \big) \frac{\tau^* - \tau_b}{A(\tau_b)^2} +  o_p(1).
\end{align*}
If $\tau^* \neq \tau_b$ (when $\delta \neq 0$), the second term will have a non-negligible impact. Let $\partial_{b} A(\tau)$ and $\partial_{\tau^*} A(\tau)$ denote the gradient of $A(\tau)$ with respect to $V_b$ and $\tau^*$ at $\tau$. Then we have
\begin{align*}
	\sqrt{n} M_n'(\tau_b) \conD N(0, D_b^\intercal\Sigma D_b),
\end{align*}
where, take $b=p$ as an example, 
\begin{align*}
	D_p = \frac{\tau^* - \tau_b}{A(\tau_b)^2}  \left( \begin{matrix}
	1 & 0	 \\
	0 & 0 \\
	0 & 1
\end{matrix} \right)  \Big(\partial_{b} A(\tau_b),  \partial_{\tau^*} A(\tau_b)\Big)^\intercal
 +  \Big( 0, 0, -\frac{1}{A(\tau_b)} \Big)^\intercal.
\end{align*}
$D_o$ can be defined similarly.
This in turn implies that
\begin{align*}
\sqrt{n}(\hat{\tau}_b - \tau_b) \conD N(0, [ M''(\tau_b) ]^{-1} D_b^\intercal\Sigma D_b [ M''(\tau_b) ]^{-1}).
\end{align*}

\subsubsection*{Case 1-2: $\gamma=1/2$}
In this case, we obtain the following:
\begin{align*}
	\sqrt{n} M_n'(\tau_b) &= \frac{\sqrt{n}(\tau^* - \hat{\tau}^*)}{A(\tau_b)} + \sqrt{n} \big( A_n(\tau_b) - A(\tau_b) \big) \frac{\tau^* - \tau_b}{A(\tau_b)^2} + \eta B^\prime(\tau_b) + o_p(1).
\end{align*}
Since $\eta B^\prime(\tau_b) \neq 0$, we will have an asymptotic bias term:
\begin{align*}
	\sqrt{n} M_n'(\tau_b) \conD N(\eta B^\prime(\tau_b), D_b^\intercal\Sigma D_b).
\end{align*}
We then have
\begin{align*}
\sqrt{n}(\hat{\tau}_b - \tau_b) \conD N( \eta [ M''(\tau_b) ]^{-1}B^\prime(\tau_b), [ M''(\tau_b) ]^{-1} D_b^\intercal\Sigma D_b [ M''(\tau_b) ]^{-1}).
\end{align*}

\subsubsection*{Case 1-3: $\gamma<1/2$}
It is easy to see that 
\begin{align*}
	n^{\gamma} M_n'(\tau_b) \xlongrightarrow{\bbP} \eta  B'(\tau_b).
\end{align*}
The right-hand side is non-random. Hence, we would not recommend setting $\gamma<1/2$.
\end{proof}

Any data-driven selection of $\delta_n$ at rate $\gamma \in (0, 1/2]$ leads to complex asymptotics, including the emergence of a non-negligible bias $\bbB(\tau_b)$ that may dominate the limiting distribution. Unlike standard regularization settings, uncertainty in $\delta_n$ critically affects the asymptotic behavior of our estimators. Methods such as cross-validation are less informative here due to the unknown target population. Given the complications, we recommend fixing $\delta_n \equiv \delta$ ($\gamma = \infty$).

\subsection*{Case 2: $\tau_b=0$ and $q\geq1$}
The complete theorem is given below.
\begin{theorem}\label{thm:normality-zero-3cases}
Assume \eqref{eqn:joint-recall} holds. For $q\geq1$, if $\tau_b=0$, we have the following convergence results:
\begin{itemize}
\item \textbf{Case 2-1:} $q\geq2$
\begin{align*}
	\sqrt{n}\hat{\tau}_b \conD \frac{Z_\tau}{1+ 1_{\{q=2\}}\delta \sqrt{V_b/2}} .
\end{align*}
\item \textbf{Case 2-2:} $1<q<2$
\begin{align*}
n^{\alpha} \hat{\tau}_b \conD \underset{h}{\mathrm{argmin}} \left\{ -\frac{Z_\tau}{\sqrt{V_b}} h + \kappa |h|^q \right\} = \mathrm{sign}(Z_\tau) \left( \frac{|Z_\tau|}{\kappa q \sqrt{V_b}} \right)^{\frac{1}{q-1}}.
\end{align*}
If $\delta\neq0$, we have $\kappa = 2^{1/q - 1} \delta/q$ and $\alpha=1/(2q-2)$. If $\delta=0$ and $\gamma < 1-q/2$, we have $\kappa = 2^{1/q - 1} \eta/q$ and $\alpha=(1-2\gamma)/(2q-2)$. 
\item \textbf{Case 2-3:} $q=1$

If $\delta\neq0$, we get $\hat{\tau}_b = 0$ almost surely. If $\delta=0$ and $\gamma=1/2$, we obtain
\begin{align*}
\sqrt{n} \hat{\tau}_b \conD \underset{h}{\mathrm{argmin}} \left\{ \frac{h^2}{2V_b} -\frac{Z_\tau}{\sqrt{V_b}} h + \eta |h| \right\} = \text{sign}(Z_\tau) \big( |Z_\tau| - \eta \sqrt{V_b} \big)_{+},
\end{align*}
where $f(x)_+ \coloneqq \max( f(x), 0 )$.
\end{itemize}
\end{theorem}

\begin{proof}
When $q>1$, the penalty term remains continuously differentiable at $\tau=0$. Then $M(\tau)$ being minimized at $\tau_b=0$ implies that
\begin{align*}
	0 = \frac{-(\tau^* - 0)}{(V_b + (\tau^* - 0)^2)^{1/2}} + \delta B'(0) = \frac{-(\tau^* - 0)}{(V_b + (\tau^* - 0)^2)^{1/2}}.
\end{align*}
We readily see that $\tau^* = 0$.

\subsubsection*{Case 2-1: $q\geq 2$}

It can be verified that 
\begin{align*}
	B''(0) = \lim_{\tau \rightarrow 0} B''(\tau) = \begin{dcases}
	0	& \text{ if } q>2\\
	2^{-1/2} & \text{ if } q=2
\end{dcases}.
\end{align*}
So the second-order derivative $M''(\tau)$ exists at $\tau=0$. Hence, we can follow a similar analysis as above. 

More specifically, since $B'(0)=0$ and $\tau^* = \tau_b=0$, we have the following when $q>2$:
\begin{align*}
	M_n'(\tau_b) = - \frac{\hat{\tau}^*}{(\hat{V}_b + (\hat{\tau}^* - \tau_b)^2)^{1/2}}, \quad M_n^{\prime\prime}(\tau_b) = \frac{\hat{V}_b}{(\hat{V}_b + (\hat{\tau}^* - \tau_b)^2)^{3/2}}.
\end{align*}
It is then easy to see that
\begin{align*}
	\sqrt{n} \hat{\tau}_b = - [ M''(\tau_b) ]^{-1} \cdot \sqrt{n} M_n'(\tau_b) + o_p(1) = \sqrt{n} \hat{\tau}^\ast + o_p(1) \conD Z_\tau \sim \mathcal{N}(0, \sigma^2_\tau).
\end{align*}

When $q=2$, there is an additional term in $M_n^{\prime\prime}$:
\begin{align*}
	M_n^{\prime\prime}(\tau_b) = \frac{\hat{V}_b}{(\hat{V}_b + (\hat{\tau}^* - \tau_b)^2)^{3/2}} + \delta_n 2^{-1/2}.
\end{align*}
It then follows that
\begin{align*}
	\sqrt{n}\hat{\tau}_b  = \frac{\sqrt{n} \hat{\tau}^\ast}{1 + \hat{V}_b^{1/2} \delta_n 2^{-1/2}} +o_p(1) \conD \frac{Z_\tau}{1+ \delta \sqrt{V_b/2}}.
\end{align*}

\subsubsection*{Case 2-2: $1<q<2$}

For small \(|\tau|\), we expand $A_n(\tau)$ around \(\tau = 0\):
\begin{align*}
	& A_n(\tau) - A_n(0) = \frac{ A_n(\tau)^2 - A_n(0)^2 }{ A_n(\tau) + A_n(0) } = \frac{\tau (\tau - 2\hat{\tau}^\ast)}{A_n(\tau) + A_n(0)} = \frac{\tau (\tau - 2\hat{\tau}^\ast)}{A_n(\tau) + \hat{V}_b^{1/2}} 
\end{align*}
As for the penalty term, its second-order derivative is not well-defined. Hence, standard Taylor expansions (quadratic approximations) break down. Instead, we will treat $|\tau|^q$ as a whole and draw an expansion of $(1+x)^{1/q}$ around $x=0$:
\begin{align*}
B(\tau) - B(0) = 2^{1/q} \left(1 + \frac{|\tau|^q}{2}\right)^{1/q} - 2^{1/q} = 2^{1/q} \Big( \frac{1}{q} \cdot \frac{|\tau|^q}{2} + o(|\tau|^q) \Big). 
\end{align*}

Define the re-scaled parameter:
\begin{align*}
\tau = n^{-\alpha} h, \quad \alpha > 0,
\end{align*}
and the localized criterion function:
\begin{align*}
\tilde{M}_n(h) = n^{\beta} \left[M_n\left(n^{-\alpha} h\right) - M_n(0)\right], \quad \beta > 0.
\end{align*}
It is then easy to see that $\tilde{M}_n(h)$ is minimized at $h=n^{\alpha} (\hat{\tau}_b - \tau_b)$. The two rates $\alpha$ and $\beta$ will be determined to ensure that the re-scaled stochastic and penalty terms in $\tilde{M}_n(h)$ are both of order $O_p(1)$ so that they are balanced. 

The idea is to use Theorem 3.2.2 of \cite{EmpiricalProcess} to establish \( n^{\alpha} \hat{\tau}_b \conD \underset{h}{\mathrm{argmin}}\, \tilde{M}(h) \). More specifically, we must prove:
\begin{itemize}
  \item \textbf{Criterion Convergence}: \( \tilde{M}_n(h) \conD \tilde{M}(h) \) uniformly on a compact set.
  \item \textbf{Unique Argmin}: \( \tilde{M}(h) \) has a unique minimizer.
    \item \textbf{Tightness}: \( \{n^{\alpha} \hat{\tau}_b\} \) does not diverge.
\end{itemize}

First, consider a compact subsection $\calK\in\bbR$. We first show that $\tilde{M}_n(h) \conD \tilde{M}(h)$ uniformly in $\calK$.

The leading component of the stochastic term is:
\begin{align*}
n^{\beta} \cdot \frac{\tau^2/2 - \hat{\tau}^* \tau}{A(0)} = n^{\beta} \cdot \frac{n^{-2\alpha}h^2/2 - \hat{\tau}^* n^{-\alpha} h}{V_b^{1/2}} = \frac{h^2}{2V_b^{1/2}} \cdot n^{\beta-2\alpha} -\frac{Z_\tau h}{V_b^{1/2}} \cdot n^{\beta - \alpha - \frac{1}{2}},
\end{align*}
where $\sqrt{n} \hat{\tau}^* \conD Z_\tau. $

The penalty term scales as:
\begin{align*}
\delta_n |\tau|^q \cdot n^{\beta} = (\delta + \eta n^{-\gamma}) |n^{-\alpha} h|^q \cdot n^{\beta} = \delta n^{\beta-\alpha q} |h|^q + \eta n^{\beta - \gamma - \alpha q} |h|^q.
\end{align*}

When $\delta\neq 0$, the penalty term is of order $O_p(n^{\beta - \alpha q})$ for tight $h$. To make this term $O_p(1)$, we have $\beta = \alpha q$. Since $q\in (1,2)$, the quadratic component of the stochastic term, which has order $n^{\alpha(q-2)}$ will vanish. The balance can only be achieved when $\beta = \alpha + 1/2$. We can easily solve that
\begin{align*}
	\alpha = \frac{1}{2(q-1)} \quad \text{and} \quad \beta = \frac{q}{2(q-1)}.
\end{align*}

When $\delta = 0$, the penalty term becomes $O_p(n^{\beta - \alpha q - \gamma})$. Then we get $\beta = \gamma + \alpha q$. On the other hand, we get $\beta = \alpha + 1/2$ from the stochastic term. Putting all together, we obtain
\begin{align*}
	\alpha = \frac{1-2\gamma}{2(q-1)} \quad \text{and} \quad \beta = \frac{q-2\gamma}{2(q-1)}.
\end{align*}
Then we need $\gamma < 1/2$ to ensure $\alpha > 0$ and $\gamma < 1-q/2$ to make the quadratic term smaller. Putting together, we have $\gamma < 1-q/2$.

Hence, the limit process is:
\begin{align*}
\tilde{M}(h) = -\frac{Z_\tau h}{\sqrt{V_b}} + \kappa |h|^q, \quad \text{ where } \kappa = \begin{dcases}
	\frac{\delta}{q} \cdot 2^{1/q - 1}	& \text{if } \delta\neq 0 \\
	\frac{\eta}{q} \cdot 2^{1/q - 1}	& \text{if } \delta =0 \text{ and } \gamma<1/2.
\end{dcases}
\end{align*}
As \( |h| \to \infty \), the penalty dominates:
\begin{align*}
\lim_{|h| \to \infty} \tilde{M}(h) \geq \kappa |h|^q - |Z_\tau| |h| / \sqrt{V_b} \to \infty.
\end{align*}
Thus, \( \underset{h}{\mathrm{argmin}}\, \tilde{M}(h) \) is bounded.

Similarly, the penalty term in \( \tilde{M}_n(h) \) dominates for large $|h|$:
\begin{align*}
\tilde{M}_n(h) \geq \frac{\kappa}{2} |h|^q \quad \text{uniformly in } n.
\end{align*}
Hence, the argmin \( \hat{h}_n = n^{\alpha} \hat{\tau}_b \) satisfies:
\begin{align*}
|\hat{h}_n| \leq K \quad \text{for some } K > 0 \text{ independent of } n.
\end{align*}
For any \( \epsilon > 0 \), choose \( K \) such that:
\begin{align*}
\mathbb{P}(|\hat{h}_n| > K) \leq \mathbb{P}\left(\inf_{|h| > K} \tilde{M}_n(h) \leq \inf_{|h| \leq K} \tilde{M}_n(h)\right).
\end{align*}
From the above choice of $\hat{h}_n$ and $K$, we have
\begin{align*}
\inf_{|h| > K} \tilde{M}_n(h) \geq \frac{\kappa}{2} K^q, \quad \inf_{|h| \leq K} \tilde{M}_n(h) \leq 0.
\end{align*}
It then readily follows that
\begin{align*}
\mathbb{P}(|\hat{h}_n| > K) \leq \mathbb{P}\left(\frac{\kappa}{2} K^q \leq 0\right) = 0 < \epsilon.
\end{align*}
This shows that $n^{\alpha} \hat{\tau}_b$ is tight. Then we can conclude that the asymptotic distribution of \(\hat{\tau}_b\) is:
\begin{align*}
n^{\alpha} \hat{\tau}_b \conD \underset{h}{\mathrm{argmin}} \left\{ -\frac{Z_\tau}{\sqrt{V_b}} h + \kappa |h|^q \right\} = \mathrm{sign}(Z_\tau) \left( \frac{|Z_\tau|}{\kappa q \sqrt{V_b}} \right)^{\frac{1}{q-1}}.
\end{align*}

\subsubsection*{Case 2-3: $q=1$}

The previous analysis of the stochastic term and penalty term remains valid in this case as well. When $\delta\neq0$, the penalty term dominates the stochastic term. Therefore, we would have $\hat{\tau}_b = 0$ almost surely.

When $\delta=0$, the only case that we can balance these two terms is when $\gamma=1/2$. Then by letting $\beta=\alpha+1/2$, we have
\begin{align*}
	\tilde{M}_n(h) = \frac{h^2}{2V_b^{1/2}} \cdot n^{1/2-\alpha} -\frac{Z_\tau h}{V_b^{1/2}} + \eta |h|.
\end{align*}
If $\alpha<1/2$ (slower convergence rate), then the quadratic term would dominate and yield that $\hat{\tau}_b = 0$ almost surely, which leads to a contradiction. On the other hand, it is intuitively not possible to have $\alpha>1/2$ when both $\delta_n$ and $\hat{\tau}^\ast$ are $O_p(n^{-1/2})$. Hence, the only possibility is $\alpha=1/2$.
\begin{align*}
\sqrt{n} \hat{\tau}_b \conD \underset{h}{\mathrm{argmin}} \left\{ \frac{h^2}{2V_b^{1/2}} -\frac{Z_\tau}{V_b^{1/2}} h + \eta |h| \right\} = \text{sign}(Z_\tau) \big( |Z_\tau| - \eta V_b^{1/2} \big)_{+},
\end{align*}
where $f(x)_+ \coloneqq \max( f(x), 0 )$.
\end{proof}

When $q\geq 2$, $\hat{\tau}_b$ is still well-behaved, even though its limiting distribution is different from that under the case of $\tau_b\neq 0$. In contrast, when $q \in (1, 2)$, the first-order derivative of the penalty term diminishes rapidly as $\tau \rightarrow 0$, but the second-order derivative diverges. This complicates the asymptotic analysis, as it introduces a complex interaction between the stochastic error and the penalty's effect. %The minimization problem’s first-order condition implies a specific rate relationship to balance these terms: for instance, $O_p(n^{-1/2}) \sim |\hat{\tau}_b|^q$ when $\delta \neq 0$. This constraint inherently slows down the convergence rate of the estimator. The situation deteriorates further when $\delta = 0$, as the effective influence of the penalty term weakens (specifically, $n^{-\gamma}|\hat{\tau}_b|^q$). To achieve a meaningful balance between the stochastic error and the reduced penalty effect, the tuning parameter $\delta_n$ must shrink at a slower-than-root-$n$ rate. This results in an even slower convergence rate for the estimator. 

Finally, when $q = 1$, the problem resembles the LASSO framework. Although the penalty lacks curvature at zero, its ``soft thresholding" property ensures sparsity. Specifically, when $\delta \neq 0$, the penalty dominates the objective function, forcing the estimator to zero almost surely. Even in the case of $\delta = 0$, a root-$n$ shrinking $\delta_n$ can recover the root-$n$ convergence rate through the ``soft thresholding" mechanism.

\bigskip
\noindent
\textbf{Proof of Theorem \ref{thm:inference}:}

Let $S=\{\tau^*\in I_n^*(1-\beta)\}$. We know $\displaystyle\lim_{n\to\infty}\inf_{P\in\mathcal{P}}\mathbb{P}(S)= 1-\beta$. 
Theorem \ref{thm:non-zero} and Lemma \ref{lem:uniform_converge} imply that  $\sqrt{n}( \hat{\tau}_p - \tau_p)$ and $ \sqrt{n} (\hat{\tau}_o - \tau_o) $ are jointly normal uniformly in $P\in\mathcal{P}$ and there are uniformly consistent variance matrix estimators. If $\tau^*>0$ and we reject $H_0: \tau^*=0$ in the first step, $S$ contains strictly positive values. Thus, $\tau_b\in[0,\tau^*]$ and $\tau_o-\tau_p<\infty$. Since $\hat{V}_p>\hat{V}_o$ ($\hat{V}_p^N>\hat{V}_o^N$) and $\hat{\tau}_b$ is monotonically decreasing in $\hat{V}_b$, $\mathbb{P}(\hat{\tau}_o\geq \hat{\tau}_p)=1$ for all $P\in\mathcal{P}$ in the second step. If $\tau^*<0$, then $\tau_b\in[\tau^*,0]$. The above analysis goes through with $\hat{\tau}_b$ monotonically increasing in $\hat{V}_b$ and $\hat{\tau}_o$ and $\hat{\tau}_p$ switching their order. 

By Lemma 3 and Proposition 1 in \cite{stoye2009more}, 
\[
\lim_{n\to\infty}\inf_{\tau^{\texttt{DR}}\in[\tau_p,\tau_o]}\inf_{P\in\mathcal{P}}\mathbb{P}\left(\tau^{\texttt{DR}}\in I_n^{[\tau^*]}(1-\alpha+\beta)\right)\geq 1-\alpha+\beta.
\]
As a result, by the Bonferroni inequality,
\begin{align*}
&\lim_{n\to\infty}\inf_{\tau^{\texttt{DR}}\in[\tau_p,\tau_o]}\inf_{P\in\mathcal{P}}\mathbb{P}\left(\tau^{\texttt{DR}}\in \mathcal{I}_n \right)\\
    \geq &\lim_{n\to\infty}\inf_{\tau^{\texttt{DR}}\in[\tau_p,\tau_o]}\inf_{P\in\mathcal{P}}\mathbb{P}\left(\tau^{\texttt{DR}}\in I_n^{[\tau^*]}(1-\alpha+\beta),S\right)\\
    \geq &(1-\beta)+(1-\alpha+\beta)-1=1-\alpha.
\end{align*}

\section{Proof of Lemma \ref{lem:JointCLT}}\label{appendixe}

\subsection{Neyman Bounds}

The asymptotic covariance matrix $\bm{\Sigma}$ for the joint convergence of $\sqrt{n}(\hat{\tau}^* - \tau^*, \hat{V}^N_p - V^N_p, \hat{V}^N_o - V^N_o)$ is given by:
\begin{align*}
\bm{\Sigma} = 
\begin{pmatrix}
\sigma_\tau^2 & \sigma^2_{\tau +} & \sigma^2_{\tau -} \\
\sigma^2_{\tau +} & \sigma_+^2 & \sigma^2_{+-} \\
\sigma^2_{\tau -} & \sigma^2_{+-} & \sigma_-^2
\end{pmatrix}
= \Cov\left(
\begin{pmatrix}
\dot{\tau} \\
\dot{V}^+ \\
\dot{V}^-
\end{pmatrix}, (\dot{\tau}, \dot{V}^+, \dot{V}^-)
\right), 
\end{align*}
where $\dot{\tau}, \dot{V}^+, \dot{V}^-$ are the influence functions for $\tau^*, V^N_p, V^N_o$, respectively.

To give expressions of the influence functions, we define
\begin{gather*}
	e = \mathbb{P}(T=1), \,\, \tau_1 = \bbE[Y(1)] = \E[Y|T=1], \,\, \tau_0 = \E[Y(0)] = \bbE[Y|T=0], \\
	\sigma^2_1 = \Var(Y(1)) = \Var(Y | T=1), \,\, \sigma^2_0 = \Var(Y(0)) = \Var(Y | T=0).
\end{gather*}
It is easy to see that
\begin{align*}
	\tau_1 = \frac{P\big[ T P_{(Y|T)}[Y] \big]}{P[T]} \quad \text{and} \quad \sigma^2_1 = \frac{P\big[ T P_{(Y|T)}[(Y-\tau_1)^2] \big]}{P[T]}.
\end{align*}
The quantities $\tau_0$ and $\sigma^2_0$ can be defined in a similar way. Following the methods proposed by \cite{Xu&Yang:2025}, we readily get
\begin{gather*}
	\dot{\tau}_1 = \frac{T(Y - \tau_1)}{e}, \,\,\,\, \dot{\tau}_0 = \frac{(1-T)(Y - \tau_0)}{1-e}, \\
	\dot{\sigma}^2_1 = \frac{T[(Y - \tau_1)^2 - \sigma_1^2]}{e}, \,\,\,\, \dot{\sigma}^2_0 = \frac{(1-T)[(Y - \tau_0)^2 - \sigma_0^2]}{1-e}.
\end{gather*}
Since $T(1-T) \equiv 0$, we readily get
\begin{align*}
	\Cov(\dot{\tau}_1,\dot{\tau}_0) = \Cov(\dot{\sigma}^2_1,\dot{\sigma}^2_0) = \Cov(\dot{\tau}_1,\dot{\sigma}^2_0) = \Cov(\dot{\tau}_0,\dot{\sigma}^2_1) = 0.
\end{align*}
Moreover, we have
\begin{gather*}
	\Cov(\dot{\tau}_1, \dot{\sigma}^2_1) = \frac{\bbE[(Y(1) - \tau_1)^3]}{e}, \quad \Cov(\dot{\tau}_0, \dot{\sigma}^2_0) = \frac{\bbE[(Y(0) - \tau_0)^3]}{1-e}, \\
	\Var(\dot{\sigma}^2_1) = \frac{\bbE[ ((Y(1) - \tau_1)^2 - \sigma^2_1)^2]}{e}, \quad \Var(\dot{\sigma}^2_0) = \frac{\bbE[((Y(0) - \tau_0)^2 - \sigma_0^2)^2]}{1-e}.
\end{gather*}

\textbf{Step 1. Root-$n$ negligibility of the remainder terms.} 
Write $\hat{\tau}_1 = \bar{Y}_1 = \frac{\sum_{i=1}^n T_i Y_i}{\sum_{i=1}^n T_i}$. Let $n_1 = \sum_{i=1}^n T_i$ and $\hat{e} = n_1/n$. Then:
\begin{align*}
\hat{\tau}_1 &= \frac{1}{n_1}\sum_{i=1}^n T_i Y_i = \frac{1}{\hat{e}} \cdot \frac{1}{n}\sum_{i=1}^n T_i Y_i = \frac{1}{\hat{e}} \Big[ \frac{1}{n}\sum_{i=1}^n T_i (Y_i - \tau_1) + \hat{e}\tau_1 \Big] \\
&= \tau_1 + \frac{1}{\hat{e}} \cdot \frac{1}{n}\sum_{i=1}^n T_i (Y_i - \tau_1) = \tau_1 + \Big(\frac{1}{e} + O_p(n^{-1/2})\Big) \cdot \frac{1}{n}\sum_{i=1}^n T_i (Y_i - \tau_1) \\
&= \tau_1 + \frac{1}{n}\sum_{i=1}^n \frac{T_i(Y_i - \tau_1)}{e} + O_p(n^{-1}) = \tau_1 + \frac{1}{n}\sum_{i=1}^n \dot{\tau}_{1,i} + o_p(n^{-1/2}),
\end{align*}
where we used $\hat{e} = e + O_p(n^{-1/2})$. Note that we just need the remainder term to be $o_p(n^{-1/2})$ to have root-$n$ negligibility.

For the variance, write:
\begin{align*}
\hat{\sigma}^2_1 - \sigma^2_1 &= \frac{1}{n_1}\sum_{i=1}^n T_i (Y_i - \hat{\tau}_1)^2 - \sigma^2_1 = \frac{1}{n_1}\sum_{i=1}^n T_i (Y_i - \tau_1 + \tau_1 - \hat{\tau}_1)^2 - \sigma^2_1 \\
&= \frac{1}{n_1}\sum_{i=1}^n T_i (Y_i - \tau_1)^2 - 2(\hat{\tau}_1 - \tau_1)\frac{1}{n_1}\sum_{i=1}^n T_i (Y_i - \tau_1) + (\hat{\tau}_1 - \tau_1)^2 - \sigma^2_1 \\
&= \Big(\frac{1}{n_1}\sum_{i=1}^n T_i (Y_i - \tau_1)^2 - \sigma^2_1 \Big) - (\hat{\tau}_1 - \tau_1)^2 \\
&= \frac{1}{n}\sum_{i=1}^n \dot{\sigma}^2_{1,i} \cdot (1 + \frac{e-\hat{e}}{\hat{e}} ) - (\hat{\tau}_1 - \tau_1)^2 = \frac{1}{n}\sum_{i=1}^n \dot{\sigma}^2_{1,i} + o_p(n^{-1/2}).
\end{align*}

The cases for $\tau_0$ and $\sigma_0^2$ follow the same argument, and hence are omitted.

\textbf{Step 2. Joint convergence of the leading terms.}  
Therefore, it is sufficient to only consider the influence functions. Under regularity conditions, all $\sqrt{n} \, \bbP_n[\dot{\tau}_1]$, $\sqrt{n} \, \bbP_n[\dot{\tau}_0]$, $\sqrt{n} \, \bbP_n[\dot{\sigma}^2_1]$, and $\sqrt{n} \, \bbP_n[\dot{\sigma}^2_0]$ converge in distribution to some normal random variables. It is then easy to establish that $\sqrt{n}(\hat{\tau}^* - \tau^*, \hat{V}^N_p - V^N_p, \hat{V}^N_o - V^N_o)$ jointly converge in distribution to $N(0, \bm{\Sigma})$. The influence functions are given by (assuming $\sigma^2_0 \neq \sigma^2_1$) $\dot{\tau} = \dot{\tau}_1 - \dot{\tau}_0$ and:
\begin{align*}
	\dot{V}^+ = \Big( 1 + \sqrt{\frac{\sigma_0^2}{\sigma_1^2}} \Big) \dot{\sigma}^2_1 + \Big( 1 + \sqrt{\frac{\sigma_1^2}{\sigma_0^2}} \Big) \dot{\sigma}^2_0 , \,\,
	\dot{V}^- = \Big( 1 - \sqrt{\frac{\sigma_0^2}{\sigma_1^2}} \Big) \dot{\sigma}^2_1 +  \Big( 1 - \sqrt{\frac{\sigma_1^2}{\sigma_0^2}} \Big) \dot{\sigma}^2_0.
\end{align*}

\textbf{Step 3. Variances and covariances.} It is then easy to verify that
\begin{align*}
	\sigma^2_{\tau} &= \Var(\dot{\tau}) = \Var(\dot{\tau}_1) + \Var(\dot{\tau}_0) = \frac{\sigma^2_1}{e} + \frac{\sigma^2_0}{1-e}, \\
	\sigma^2_{\tau +} 
		&= \Big( 1 + \sqrt{\frac{\sigma_0^2}{\sigma_1^2}} \Big) \frac{\bbE[(Y(1) - \tau_1)^3]}{e} - \Big( 1 + \sqrt{\frac{\sigma_1^2}{\sigma_0^2}} \Big) \frac{\bbE[(Y(0) - \tau_0)^3]}{1-e}, \\
	\sigma^2_{\tau -} 
		&= \Big( 1 - \sqrt{\frac{\sigma_0^2}{\sigma_1^2}} \Big) \frac{\bbE[(Y(1) - \tau_1)^3]}{e} - \Big( 1 - \sqrt{\frac{\sigma_1^2}{\sigma_0^2}} \Big) \frac{\bbE[(Y(0) - \tau_0)^3]}{1-e}, \\
	\sigma^2_{+} &= \Big( 1 + \sqrt{\frac{\sigma_0^2}{\sigma_1^2}} \Big)^2 \frac{\bbE[ ((Y(1) - \tau_1)^2 - \sigma^2_1)^2]}{e} + \Big( 1 + \sqrt{\frac{\sigma_1^2}{\sigma_0^2}} \Big)^2 \frac{\bbE[((Y(0) - \tau_0)^2 - \sigma_0^2)^2]}{1-e}, \\
	\sigma^2_{+-} &= \Big( 1 - \frac{\sigma_0^2}{\sigma_1^2} \Big) \frac{\bbE[ ((Y(1) - \tau_1)^2 - \sigma^2_1)^2]}{e} + \Big( 1 - \frac{\sigma_1^2}{\sigma_0^2} \Big) \frac{\bbE[((Y(0) - \tau_0)^2 - \sigma_0^2)^2]}{1-e}, \\
	\sigma^2_{-} &= \Big( 1 - \sqrt{\frac{\sigma_0^2}{\sigma_1^2}} \Big)^2 \frac{\bbE[ ((Y(1) - \tau_1)^2 - \sigma^2_1)^2]}{e} + \Big( 1 - \sqrt{\frac{\sigma_1^2}{\sigma_0^2}} \Big)^2 \frac{\bbE[((Y(0) - \tau_0)^2 - \sigma_0^2)^2]}{1-e}.
\end{align*}

\subsection{Sharp Bounds}

The asymptotic covariance matrix $\bm{\Sigma}$ for the joint convergence of $\sqrt{n}(\hat{\tau}^* - \tau^*, \hat{V}_o - V_o, \hat{V}_p - V_p)$ is given by:
\begin{align*}
\bm{\Sigma} = 
\begin{pmatrix}
\sigma_\tau^2 & \sigma^2_{\tau o} & \sigma^2_{\tau p} \\
\sigma^2_{\tau o} & \sigma_o^2 & \sigma^2_{op} \\
\sigma^2_{\tau p} & \sigma^2_{op} & \sigma_p^2
\end{pmatrix}
= \Cov\left(
\begin{pmatrix}
\dot{\tau} \\
\dot{V}_o \\
\dot{V}_p
\end{pmatrix}, (\dot{\tau}, \dot{V}_o, \dot{V}_p)
\right),
\end{align*}
where $\dot{\tau}, \dot{V}_o, \dot{V}_p$ are the influence functions for $\tau^*, V_o, V_p$, respectively.

To give expressions of the influence functions, we define
\begin{gather*}
	f_1(y) = \frac{d}{dy}P_1(y), \,\, f_0(y) = \frac{d}{dy}P_0(y), \,\, Q_1(u) = P_1^{-1}(u), \,\, Q_0(u) = P_0^{-1}(u) \\
	\gamma = \tau_0 \tau_1, \,\, \theta_o = \int_0^1 Q_1(u) Q_0(u) du, \,\, \theta_p = \int_0^1 Q_1(u) Q_0(1-u) du.
\end{gather*}
Then we have
\begin{align*}
	V_o = \sigma_1^2 + \sigma_0^2 - 2 (\theta_0 - \gamma) \quad\text{and}\quad V_p = \sigma_1^2 + \sigma_0^2 - 2 (\theta_p - \gamma)
\end{align*}

Since $Y(1)$ and $Y(0)$ both have bounded support, the densities $f_1(\cdot)$ and $f_0(\cdot)$ are positive on their respective supports. The influence functions for the quantiles $Q_1(u)$ and $Q_0(u)$ are
\begin{align*}
	\dot{Q}_1(u) &= -\frac{T \cdot [\mathbbm{1}\{Y(1) \leq Q_1(u)\} - u]}{e f_1(Q_1(u))}, \\
	\dot{Q}_0(u) &= -\frac{(1-T) \cdot [\mathbbm{1}\{Y(0) \leq Q_0(u)\} - u]}{(1-e) f_0(Q_0(u))}.
\end{align*}
These in turn imply
\begin{align*}
	\dot{\theta}_o &= \int_0^1 \left[ \dot{Q}_1(u)Q_0(u) + Q_1(u)\dot{Q}_0(u) \right] du, \\
	\dot{\theta}_p &= \int_0^1 \left[ \dot{Q}_1(u)Q_0(1-u) + Q_1(u)\dot{Q}_0(1-u) \right] du.
\end{align*}

Putting together, we have
\begin{align*}
	\dot{V}_o = \dot{\sigma}^2_1 +  \dot{\sigma}^2_0 - 2 (\dot{\theta}_o - \dot{\gamma}) \quad\text{and}\quad \dot{V}_p = \dot{\sigma}^2_1 +  \dot{\sigma}^2_0 - 2 (\dot{\theta}_p - \dot{\gamma}).
\end{align*} 

\textbf{Step 1. Root-$n$ negligibility of the remainder terms.} Given the results from the previous subsection, we just need to show
\begin{align*}
\hat{\theta}_o - \theta_o &= \frac{1}{n}\sum_{i=1}^n \dot{\theta}_{o,i} + o_p(n^{-1/2})
\end{align*}
to get
\begin{align*}
\hat{V}_o - V_o - \frac{1}{n}\sum_{i=1}^n \dot{V}_{o,i} = o_p(n^{-1/2}).
\end{align*}

First, using the identity $ab - cd = c(b-d) + d(a-c) + (a-c)(b-d)$, we write:
\begin{align*}
\hat{Q}_1(u)\hat{Q}_0(u) - Q_1(u)Q_0(u) &= Q_1(u)[\hat{Q}_0(u) - Q_0(u)] + Q_0(u)[\hat{Q}_1(u) - Q_1(u)] \\
&\quad + [\hat{Q}_1(u) - Q_1(u)][\hat{Q}_0(u) - Q_0(u)].
\end{align*}
It then follows that
\begin{align*}
\hat{\theta}_o - \theta_o &=\int_0^1 [\hat{Q}_1(u)\hat{Q}_0(u) - Q_1(u)Q_0(u)] du \\
&= \int_0^1 Q_1(u)[\hat{Q}_0(u) - Q_0(u)] du + \int_0^1 Q_0(u)[\hat{Q}_1(u) - Q_1(u)] du \\
&\quad + \int_0^1 [\hat{Q}_1(u) - Q_1(u)][\hat{Q}_0(u) - Q_0(u)] du \\
&= I_1 + I_2 + I_3.
\end{align*}

The structures of $I_1$ and $I_2$ are basically the same. We will focus on $I_2$ here. This term gives the main linear influence function. Using the Bahadur representation for $\hat{Q}_1(u)$:
\begin{align*}
\hat{Q}_1(u) - Q_1(u) = -\frac{\hat{P}_1(Q_1(u)) - u}{f_1(Q_1(u))} + R_{n,1}(u).
\end{align*}
Therefore,
\begin{align*}
I_2 &= \int_0^1 Q_0(u)\left[-\frac{\hat{P}_1(Q_1(u)) - u}{f_1(Q_1(u))} + R_{n,1}(u)\right] du \\
&= -\int_0^1 \frac{Q_0(u)[\hat{P}_1(Q_1(u)) - u]}{f_1(Q_1(u))} du + \int_0^1 Q_0(u)R_{n,1}(u) du.
\end{align*}

The first term is the linear part:
\begin{align*}
-\int_0^1 \frac{Q_0(u)[\hat{P}_1(Q_1(u)) - u]}{f_1(Q_1(u))} du &= -\int_0^1 \frac{Q_0(u)}{ef_1(Q_1(u))} \left[\frac{1}{n}\sum_{i=1}^n T_i\mathbbm{1}\{Y_i \leq Q_1(u)\} - u\right] du \\
&= \frac{1}{n}\sum_{i=1}^n \left[-\int_0^1 \frac{T_i Q_0(u)[\mathbbm{1}\{Y_i \leq Q_1(u)\} - u]}{ef_1(Q_1(u))} du\right].
\end{align*}

The second term is a remainder:
\begin{align*}
\Big|\int_0^1 Q_0(u)R_{n,1}(u) du\Big| \leq \sup_{u \in [0,1]} |Q_0(u)| \cdot \sup_{u \in [0,1]} |R_{n,1}(u)| \leq M \cdot o_p(n^{-1/2}) = o_p(n^{-1/2})
\end{align*}
by Assumption \ref{quantile}. 

The term $I_3$ is bounded by
\begin{align*}
|I_3| \leq \sup_{u \in [0,1]} |\hat{Q}_1(u) - Q_1(u)| \cdot \sup_{u \in [0,1]} |\hat{Q}_0(u) - Q_0(u)| = O_p(n^{-1}) = o_p(n^{-1/2}).
\end{align*}
Hence, the desired result readily follows.

\textbf{Step 2. Joint convergence of the leading terms.} 
Under Assumption \ref{quantile}, the empirical process for quantiles implies $\sqrt{n}  \, \bbP_n \dot{\theta}_o$ and $\sqrt{n} \, \bbP_n \dot{\theta}_p$ also converge in distribution to some normal random variables. It is then easy to establish that $\sqrt{n}(\hat{\tau}^* - \tau^*, \hat{V}_o - V_o, \hat{V}_p - V_p)$ jointly converge in distribution to $N(0, \bm{\Sigma})$.

\textbf{Step 3. Variances and covariances.}
As implied by the influence functions, most elements of $\bm{\Sigma}$ have rather complicated expressions. Since $T(1-T) \equiv 0$, we readily get
\begin{gather*}
	\Cov(\dot{\tau}_0,\dot{Q}_1(u)) = \Cov(\dot{\tau}_1,\dot{Q}_0(u)) = \Cov(\dot{\sigma}^2_0,\dot{Q}_1(u)) = \Cov(\dot{\sigma}^2_1,\dot{Q}_0(u)) = 0 \,\, \forall u \\
	\Cov(\dot{Q}_0(u), \dot{Q}_1(u')) = 0 \,\, \forall u, u'.
\end{gather*}
The above results, for example, imply 
\begin{align*}
\Cov(\dot{\tau}_1, \dot{\gamma}) &= \tau_0\Var(\dot{\tau}_1), \quad \Cov(\dot{\tau}_1, \dot{\theta}_o) = \int_0^1 Q_0(u)\Cov(\dot{\tau}_1, \dot{Q}_1(u)) du,\\
\Cov(\dot{\tau}_0, \dot{\gamma}) &= \tau_1\Var(\dot{\tau}_0), \quad \Cov(\dot{\tau}_0, \dot{\theta}_o) = \int_0^1 Q_1(u)\Cov(\dot{\tau}_0, \dot{Q}_0(u)) du. 
\end{align*}

The derivation of $\sigma^2_{\tau o}$ and $\sigma^2_{\tau p}$ are similar. We will focus on the first one here. 
\begin{align*}
\sigma^2_{\tau o} &= \Cov(\dot{\tau}_1, \dot{\sigma}^2_1) - \Cov(\dot{\tau}_0, \dot{\sigma}^2_0) - 2 \Cov(\dot{\tau}_1, \dot{\theta}_o) + 2 \Cov(\dot{\tau}_0, \dot{\theta}_o)\\
&\quad + 2\Cov(\dot{\tau}_1, \dot{\gamma}) - 2\Cov(\dot{\tau}_0, \dot{\gamma}) \\
	&= \frac{1}{e} \, E\big[(Y(1)-\tau_1)^3\big]
-\frac{1}{1-e}\,E\big[(Y(0)-\tau_0)^3\big] - 2 \Cov(\dot{\tau}_1, \dot{\theta}_o) + 2 \Cov(\dot{\tau}_0, \dot{\theta}_o) \\
	&\quad + 2\tau_0 \Var(\dot{\tau}_1) - 2\tau_1 \Var(\dot{\tau}_0)
\end{align*}

Since influence functions have an expectation of zero, the covariance is just the expectation of their product:
\begin{align*}
& \quad \Cov(\dottau_1, \dotQ_1(u)) = \mathbb{E}[\dot{\tau}_1 \cdot \dot{Q}_1(u)] \\
&= -\frac{1}{e^2 f_1(Q_1(u))} \bbE(T(Y(1)-\tau_1) \ind\{Y(1) \leq Q_1(u)\} )  \\
&= -\frac{1}{e f_1(Q_1(u))}\E[(Y(1)-\tau_1)\ind\{Y(1) \leq Q_1(u)\} ].
\end{align*}
This in turn implies
\begin{align*}
\Cov(\dot{\tau}_1, \dot{\theta}_o) &= -\int_0^1 \frac{Q_0(u)}{ef_1(Q_1(u))}\E[(Y(1)-\tau_1)\ind\{Y(1) \leq Q_1(u)\} ] du \\
&= -\int_{\calY} \frac{Q_0(F_1(y))}{e} \E[(Y(1)-\tau_1)\ind\{Y(1) \leq y\} ] dy,
\end{align*}
where we used the transform $y = Q_1(u)$, so $du = f_1(y)dy$. Similarly, we have
\begin{align*}
	\Cov(\dot{\tau}_0, \dot{\theta}_o) &= -\int_{\calY} \frac{Q_1(F_0(y))}{1-e} \E[(Y(0)-\tau_0)\ind\{Y(0) \leq y\} ] dy, \\
	\Cov(\dot{\tau}_1, \dot{\theta}_p) &= -\int_{\calY} \frac{Q_0(1-F_1(y))}{e} \E[(Y(1)-\tau_1)\ind\{Y(1) \leq y\} ] dy, \\
	\Cov(\dot{\tau}_0, \dot{\theta}_p) &= -\int_{\calY} \frac{Q_1(1-F_0(y))}{1-e} \E[(Y(0)-\tau_0)\ind\{Y(0) \leq y\} ] dy,
\end{align*}

There are three terms remaining: $\sigma^2_{oo}$, $\sigma^2_{op}$, and $\sigma^2_{pp}$. Their expressions can be found in similar ways. Hence, we only show to what extent we can simplify $\sigma^2_{op}$:
\begin{align*}
	\sigma^2_{op} =\,& \Var(\dot{\sigma}^2_1) + \Var(\dot{\sigma}^2_0) +  4 \tau_0^2 \Var(\dot{\tau}_1) + 4 \tau_1^2 \Var(\dot{\tau}_0) \\
		& + 4 \tau_0 \Cov(\dot{\sigma}^2_1, \dot{\tau}_1) + 4 \tau_1 \Cov(\dot{\sigma}^2_0, \dot{\tau}_0) - 4 \tau_0 \Cov(\dot{\tau}_1, \dot{\theta}_o) - 4 \tau_1 \Cov(\dot{\tau}_0, \dot{\theta}_o)\\
		& - 4 \tau_0 \Cov(\dot{\tau}_1, \dot{\theta}_p) - 4 \tau_1 \Cov(\dot{\tau}_0, \dot{\theta}_p)\\
		& - 2\Cov(\dot{\sigma}^2_1, \dot{\theta}_o) - 2\Cov(\dot{\sigma}^2_0, \dot{\theta}_o) - 2\Cov(\dot{\sigma}^2_1, \dot{\theta}_p) - 2\Cov(\dot{\sigma}^2_0, \dot{\theta}_p) \\
		& + 4 \Cov(\dot{\theta}_o, \dot{\theta}_p).
\end{align*}
The terms in the first three lines have been analyzed. We will mainly focus on the last two lines. Similar to the analysis of $\Cov(\dot{\tau}_1, \dot{\theta}_o)$, we can get
\begin{align*}
	\Cov(\dot{\sigma}^2_1, \dot{\theta}_o) &= -\int_{\calY} \frac{Q_0(F_1(y))}{e} \E\big[ [(Y(1)-\tau_1)^2 - \sigma^2_1] \ind\{Y(1) \leq y\} \big] dy, \\
	\Cov(\dot{\sigma}^2_0, \dot{\theta}_o) &= -\int_{\calY} \frac{Q_1(F_0(y))}{1-e} \E\big[ [(Y(0)-\tau_0)^2 - \sigma^2_0] \ind\{Y(0) \leq y\} \big] dy, \\
	\Cov(\dot{\sigma}^2_1, \dot{\theta}_p) &= -\int_{\calY} \frac{Q_0(1-F_1(y))}{e} \E\big[ [(Y(1)-\tau_1)^2 - \sigma^2_1] \ind\{Y(1) \leq y\} \big] dy, \\
	\Cov(\dot{\sigma}^2_0, \dot{\theta}_p) &= -\int_{\calY} \frac{Q_1(1-F_0(y))}{1-e} \E\big[ [(Y(0)-\tau_0)^2 - \sigma^2_0] \ind\{Y(0) \leq y\} \big] dy. 
\end{align*}

The last term is 
\begin{align*}
	\Cov(\dot{\theta}_o, \dot{\theta}_p) &= \int_0^1 \Big( Q_0(u) Q_0(1-u)  \bbE[ \dot{Q}_1(u)^2 ] +  Q_1(u) Q_1(1-u)  \bbE[ \dot{Q}_0(u)^2 ]  \Big) du \\
		&= \int_0^1 \Big( Q_0(u) Q_0(1-u) \frac{u(1-u)}{e [f_1(Q_1(u))]^2} \\
		&\qquad + Q_1(u) Q_1(1-u) \frac{u(1-u)}{(1-e) [f_0(Q_0(u))]^2}\Big) du.
\end{align*}

To construct consistent estimators of the above variance-covariance terms, we need the following:
\begin{itemize}
	\item $n_1 = \sum_i T_i$, $n_0 = \sum_i (1 - T_i)$, $\hat{e} = n_1/n$. %In a randomized experiment, $e$ is known.
	\item Empirical CDFs:
\begin{align*}
	\hat{P}_0(y) = \frac{1}{n_0} \sum_{i: T_i=0} \mathbf{1}\{Y_i \leq y\}  \quad \text{and} \quad \hat{P}_1(y) = \frac{1}{n_1} \sum_{i: T_i=1} \mathbf{1}\{Y_i \leq y\}.
\end{align*}
	\item Empirical quantile functions:
\begin{gather*}
	\hat{Q}_0(u) = 
\begin{cases}
Y_{0(k)}, & \text{if } u \in \left( \frac{k-1}{n_0}, \frac{k}{n_0} \right], \quad k = 1, 2, \dots, n_0 \\
Y_{0(n_0)}, & \text{if } u = 1
\end{cases} \\
\hat{Q}_1(u) = 
\begin{cases}
Y_{1(k)}, & \text{if } u \in \left( \frac{k-1}{n_1}, \frac{k}{n_1} \right], \quad k = 1, 2, \dots, n_1 \\
Y_{1(n_1)}, & \text{if } u = 1
\end{cases},
\end{gather*}
where $Y_{0(1)} \leq Y_{0(2)} \leq \cdots \leq Y_{0(n_0)}$ is the ordered control outcomes and $Y_{1(1)} \leq Y_{1(2)} \leq \cdots \leq Y_{1(n_1)}$ is the ordered treated outcomes.
\end{itemize}

Then, for example, the term $\Cov(\dot{\tau}_1, \dot{\theta}_o)$ can be estimated by
\begin{align*}
	\widehat{\Cov}(\dot{\tau}_1, \dot{\theta}_o) = - \sum_{j=1}^{m} \frac{ \hat{Q}_0(\hat{P}_1(y_j)) }{ \hat{e} } \left( \frac{1}{n_1} \sum_{i: T_i=1} (Y_i - \hat{\tau}_1) \mathbf{1}\{Y_i \leq y_j\} \right) \Delta y_j,
\end{align*}
where $\{y_j\}_{j=1}^m$ is a grid over the support of $Y(1)$ and $\Delta y_j = y_j - y_{j-1}$ (or use trapezoidal weights),

\section{Uniform Convergence}

\subsection{Uniform Joint CLT}

The following result has been proved by \cite{Bentkus:2005}.

\begin{lemma}[Multivariate Berry-Esseen Bound]  
Let $\mathbf{X}_1, \dots, \mathbf{X}_n \in \mathbb{R}^d$ be i.i.d. random vectors. Let $\mathbb{E}[\mathbf{X}_i] = \boldsymbol{\mu}$. Assume matrix $0<\lambda_{\min}(\mathbb{V}) < \lambda_{\max}(\mathbb{V}) < \infty $, where $\mathbb{V} = \mathbb{E}[(\mathbf{X}_i - \boldsymbol{\mu})(\mathbf{X}_i - \boldsymbol{\mu})^\top]$. Assume the third-order moment $\rho = \mathbb{E}\left[ \|\mathbb{V}^{-1/2}(\mathbf{X}_i - \boldsymbol{\mu})\|^3 \right] < \infty$.  
Define the normalized sum  
\begin{align*}
S_n = \frac{1}{\sqrt{n}} \sum_{i=1}^n (\mathbf{X}_i - \boldsymbol{\mu}),
\end{align*}  
and let $\mathcal{K}$ denote the class of all convex sets in $\mathbb{R}^d$. Let $\Phi_{\mathbb{V}}(C) =\mathbb{P}(\mathcal{N}(0, \mathbb{V}) \in C)$ for $C \in \mathcal{K}$. Then  
\begin{align*}
\sup_{C \in \mathcal{K}} \Big|  \mathbb{P}(S_n \in C) - \Phi_{\mathbb{V}}(C) \Big|  \leq C_d \cdot \frac{\rho}{\sqrt{n}},
\end{align*}  
where $C_d > 0$ is an absolute constant depending only on $d$. Specifically, $C_d \leq 20 d^{1/4}$.  
\end{lemma}

\subsubsection{Neyman Bounds}

In the case of Neyman bounds, we have shown that the influence functions $\dot{\tau}, \dot{V}^+, \dot{V}^-$ are linear transformation of $\dot{\tau}_1$, $\dot{\tau}_0$, $\dot{\sigma}_1^2$, and $\dot{\sigma}_0^2$, where:
\begin{align*}
\dot{\tau}_1 &= \frac{T(Y - \tau_1)}{e}, & \dot{\sigma}_1^2 &= \frac{T[(Y - \tau_1)^2 - \sigma_1^2]}{e}, \\
\dot{\tau}_0 &= \frac{(1-T)(Y - \tau_0)}{1-e}, & \dot{\sigma}_0^2 &= \frac{(1-T)[(Y - \tau_0)^2 - \sigma_0^2]}{1-e}.
\end{align*}

Under the assumption that $0 < \underline{\sigma}_j^2 \leq \sigma_j^2 \leq \overline{\sigma}_j^2 < \infty$ and $0<\eta\leq|\sigma_1^2 - \sigma_0^2|$ for all $P \in \mathcal{P}$, all the coefficients of this linear transformation are uniformly bounded in $\mathcal{P}$. Therefore, it is sufficient to just focus on these four influence functions, which together make up $\mathbf{X}$ in this case. 

When $0 < \underline{\sigma}_j^2 \leq \sigma(j)^2  \leq  \overline{\sigma}_j^2 < \infty$ uniformly in $P\in\mathcal{P}$, it can be shown that the eigenvalues of $\mathbb{V}$ are uniformly bounded over $\mathcal{P}$: $0 < \lambda_{\min} \leq \lambda_{\min}(\mathbb{V}) \leq \lambda_{\max}(\mathbb{V}) \leq \lambda_{\max} < \infty$. Therefore, $\|\mathbb{V}^{-1/2}\|_2 \leq \lambda_{\min}^{-1/2}$, and:
\begin{align*}
\|\mathbb{V}^{-1/2}\mathbf{X}_i\|^3 \leq \lambda_{\min}^{-3/2} \|\mathbf{X}_i\|^3.
\end{align*}

Since $\|\mathbf{X}_i\|^3 \leq C_d \sum_{j=1}^d |X_{i,j}|^3$ for some constant $C_d$ depending only on dimension $d$, it suffices to show that the third moments of each component of $\mathbf{X}_i$ are uniformly bounded over $\mathcal{P}$.

For $\dot{\tau}_1$:
\begin{align*}
\mathbb{E}[|\dot{\tau}_1|^3] = \mathbb{E}\left[\Big|\frac{T(Y - \tau_1)}{e}\Big|^3\right] \leq \frac{1}{e^3} \mathbb{E}[|Y(1) - \tau_1|^3].
\end{align*}
By the bounded moment condition and Jensen's inequality, $\sup_{P \in \mathcal{P}} \mathbb{E}_P[|Y(1)|^3] < \infty$. Since $e \geq \underline{e} > 0$ uniformly, and $|\tau_1|^3 \leq \mathbb{E}[|Y(1)|^3]$ by Jensen's inequality, we have:
\begin{align*}
\sup_{P \in \mathcal{P}} \mathbb{E}_P[|Y - \tau_1|^3] \leq 8\left(\sup_{P \in \mathcal{P}} \mathbb{E}_P[|Y(1)|^3] + \sup_{P \in \mathcal{P}} |\tau_1|^3\right) < \infty.
\end{align*}
Thus, $\sup_{P \in \mathcal{P}} \mathbb{E}_P[|\dot{\tau}_1|^3] < \infty$. The same argument applies to $\dot{\tau}_0$.

For $\dot{\sigma}_1^2$:
\begin{align*}
\mathbb{E}[|\dot{\sigma}_1^2|^3] = \mathbb{E}\left[\Big|\frac{T[(Y - \tau_1)^2 - \sigma_1^2]}{e}\Big|^3\right] \leq \frac{1}{e^3} \mathbb{E}[|(Y - \tau_1)^2 - \sigma_1^2|^3].
\end{align*}
Using the inequality $|a-b|^3 \leq 4(|a|^3 + |b|^3)$:
\begin{align*}
\mathbb{E}[|(Y - \tau_1)^2 - \sigma_1^2|^3] \leq 4\left(\mathbb{E}[|Y - \tau_1|^6] + |\sigma_1^2|^3\right).
\end{align*}
Since $\sigma_1^2 \leq \overline{\sigma}_1^2 < \infty$ uniformly, $|\sigma_1^2|^3$ is uniformly bounded. Since $\sup_{P \in \mathcal{P}} \mathbb{E}_P[|Y(1)|^{6}] < \infty$, we have $\sup_{P \in \mathcal{P}} \mathbb{E}_P[|Y - \tau_1|^6] < \infty$. Therefore, $\sup_{P \in \mathcal{P}} \mathbb{E}_P[|\dot{\sigma}_1^2|^3] < \infty$. The same argument applies to $\dot{\sigma}_0^2$.

Putting all the above together,  we have
\begin{align*}
\sup_{P \in \mathcal{P}} \mathbb{E}_P[\|\mathbf{X}_i\|^3] < \infty.
\end{align*}
Since $\|\mathbb{V}^{-1/2}\mathbf{X}_i\|^3 \leq \lambda_{\min}^{-3/2} \|\mathbf{X}_i\|^3$ and $\lambda_{\min} > 0$ uniformly, we have:
\begin{align*}
\sup_{P \in \mathcal{P}} \rho = \sup_{P \in \mathcal{P}} \mathbb{E}\left[ \|\mathbb{V}^{-1/2}\mathbf{X}_i\|^3 \right] \leq \lambda_{\min}^{-3/2} \sup_{P \in \mathcal{P}} \mathbb{E}_P[\|\mathbf{X}_i\|^3] < \infty.
\end{align*}
Let $\overline{\rho}$ denote this uniform bound. Then by the multivariate Berry-Esseen bound:
\begin{align*}
\sup_{P \in \mathcal{P}}\sup_{C \in \mathcal{K}} \Big|  \mathbb{P}(S_n \in C) - \Phi_{\mathbb{V}}(C) \Big|  \leq C_d \cdot \frac{\overline{\rho}}{\sqrt{n}}.
\end{align*}

\subsubsection{Sharp Bounds}

For the sharp bounds, Assumption \ref{quantile} (iii) suggests that the asymptotics of the quantiles are completely determined by the influence functions $\dot{Q}_1$ and $\dot{Q}_0$. The influence functions $\dot{V}_{o}$ and $\dot{V}_p$ are linear transformations of $\dot{\tau}_1$, $\dot{\tau}_0$, $\dot{\sigma}_1^2$, $\dot{\sigma}_0^2$, $\dot{\theta}_o$, and $\dot{\theta}_p$, which together make up $\mathbf{X}$ in this case. If both $f_1(P_1^{-1}(u))$ and $f(P_0^{-1}(u))$ are bounded below from 0 and above from $\infty$ uniformly in $u \in [0,1]$ and $P \in \mathcal{P}$, the two additional terms $\dot{\theta}_o$ and $\dot{\theta}_p$ also have uniformly bounded third-order moment. The eigenvalues of their variance-covariances are also uniformly bounded below from 0 and above from $\infty$. Following a similar argument. The convergence in distribution is also uniform in $\mathcal{P}$ in the sharp bound case.

\subsection{Uniform Consistent Variance Estimator}

For any sequence of statistics $S_n$, we write $S_n \xlongrightarrow{u.p.} S$ if and only if
\begin{align*}
	\limsup_{n\to\infty} \sup_{P\in\mathcal P} \mathbb{P}( \|S_n - S\| \geq \epsilon ) = 0.
\end{align*}

\begin{lemma}[Uniform Stochastic Boundedness and Convergence]
\label{lem:CD}
Let $\mathcal{P}$ be a family of probability measures. Suppose that:
\begin{enumerate}
    \item The sequence $\{C_n\}_{n=1}^\infty$ is \textbf{stochastically bounded uniformly in $P\in\mathcal{P}$}, i.e., for every $\epsilon > 0$, there exists $M_\epsilon < \infty$ such that
    \begin{align*}
    \sup_{P \in \mathcal{P}} \mathbb{P}\left(|C_n| > M_\epsilon\right) < \epsilon \quad \text{for all } n \in \mathbb{N}.
    \end{align*}
    
    \item The sequence $\{D_n\}_{n=1}^\infty$ converges to zero \textbf{in probability uniformly} over $\mathcal{P}$, denoted $D_n \xlongrightarrow{u.p.} 0$, i.e., for every $\delta > 0$,
    \begin{align*}
    \lim_{n \to \infty} \sup_{P \in \mathcal{P}} \mathbb{P}\left(|D_n| > \delta\right) = 0.
    \end{align*}
\end{enumerate}
Then the product sequence $\{C_n D_n\}_{n=1}^\infty$ also converges to zero uniformly in probability over $\mathcal{P}$:
\begin{align*}
C_n D_n \xlongrightarrow{u.p.} 0.
\end{align*}
\end{lemma}

\begin{proof}
Let $\epsilon > 0$ and $\eta > 0$ be arbitrary. We need to show that there exists $N \in \mathbb{N}$ such that for all $n \geq N$,
\begin{align*}
\sup_{P \in \mathcal{P}} \mathbb{P}\left(|C_n D_n| > \epsilon\right) < \eta.
\end{align*}

Fix an arbitrary $M > 0$. For any probability measure $P \in \mathcal{P}$, we decompose the probability using the union bound:
\begin{align*}
\mathbb{P}\left(|C_n D_n| > \epsilon\right) 
&= \mathbb{P}\left(|C_n D_n| > \epsilon,\ |C_n| \leq M\right) + \mathbb{P}\left(|C_n D_n| > \epsilon,\ |C_n| > M\right) \\
&\leq \mathbb{P}\left(|C_n D_n| > \epsilon,\ |C_n| \leq M\right) + \mathbb{P}\left(|C_n| > M\right).
\end{align*}
When $|C_n| \leq M$ and $|C_n D_n| > \epsilon$, it follows that $|D_n| > \epsilon/M$. Therefore,
\begin{align*}
\mathbb{P}\left(|C_n D_n| > \epsilon,\ |C_n| \leq M\right) \leq \mathbb{P}\left(|D_n| > \epsilon/M\right).
\end{align*}
Combining these inequalities, we obtain:
\begin{equation}
\label{eq:decomposition}
\mathbb{P}\left(|C_n D_n| > \epsilon\right) \leq \mathbb{P}\left(|C_n| > M\right) + \mathbb{P}\left(|D_n| > \epsilon/M\right).
\end{equation}

By the uniform stochastic boundedness of $\{C_n\}$ (Condition 1), for the given $\eta > 0$, there exists $M_0 < \infty$ such that
\begin{equation}
\label{eq:boundedness_Cn}
\sup_{P \in \mathcal{P}} \mathbb{P}\left(|C_n| > M_0\right) < \eta/2 \quad \text{for all } n \in \mathbb{N}.
\end{equation}

By the uniform convergence in probability of $\{D_n\}$ to zero (Condition 2), for the fixed $\delta = \epsilon/M_0 > 0$, there exists $N \in \mathbb{N}$ such that for all $n \geq N$,
\begin{equation}
\label{eq:convergence_Dn}
\sup_{P \in \mathcal{P}} \mathbb{P}\left(|D_n| > \epsilon/M_0\right) < \eta/2.
\end{equation}

Substituting $M = M_0$ into equation \eqref{eq:decomposition} and taking the supremum over $P \in \mathcal{P}$, we obtain for all $n \geq N$:
\begin{align*}
\sup_{P \in \mathcal{P}} \mathbb{P}\left(|C_n D_n| > \epsilon\right) 
&\leq \sup_{P \in \mathcal{P}} \mathbb{P}\left(|C_n| > M_0\right) + \sup_{P \in \mathcal{P}} \mathbb{P}\left(|D_n| > \epsilon/M_0\right) \\
&< \eta/2 + \eta/2 = \eta,
\end{align*}
where the inequality follows from equations \eqref{eq:boundedness_Cn} and \eqref{eq:convergence_Dn}.

Since $\epsilon > 0$ and $\eta > 0$ were arbitrary, this establishes that $C_n D_n \xlongrightarrow{u.p.} 0$, completing the proof.
\end{proof}

\subsubsection{Neyman Bounds}

Let us consider the case of Neyman bounds first. To have easy reference to the assumptions, we restate the key ones below.

\begin{assumption}[Uniform Moment Conditions]
\label{ass:moment}
There exists $\delta > 0$ such that:
\begin{align*}
	\lim_{M \to \infty} \sup_{P \in \mathcal{P}} \mathbb{E}_P\left[|Y(1)|^{4+\delta} \mathbf{1}\{|Y(1)|^{4+\delta} > M\}\right] = 0, \\
	\lim_{M \to \infty} \sup_{P \in \mathcal{P}} \mathbb{E}_P\left[|Y(0)|^{4+\delta} \mathbf{1}\{|Y(0)|^{4+\delta} > M\}\right] = 0.
\end{align*}
\end{assumption}
Assumption \ref{ass:moment} can be implied by the bounded 6th moments in Lemma \ref{lem:uniform_converge}. 
\begin{assumption}[Bounded Treatment Probability]
\label{ass:treatment}
There exist constants $0 < e_{\min} < e_{\max} < 1$ such that $e_{\min} \leq e \leq e_{\max}$.
\end{assumption}

\begin{assumption}[Variance Bounds]
\label{ass:variance}
There exist constants $0 < \sigma^2_{\min} < \sigma^2_{\max} < \infty$ such that for all $P \in \mathcal{P}$:
\begin{align*}
\sigma^2_{\min} \leq \sigma^2_0, \sigma^2_1 \leq \sigma^2_{\max}.
\end{align*}
\end{assumption}

\begin{assumption}[Uniform Separation]
\label{ass:separation}
There exists $\eta > 0$ such that for all $P \in \mathcal{P}$:
\begin{align*}
|\sigma^2_1 - \sigma^2_0| \geq \eta.
\end{align*}
\end{assumption}

Under Assumptions \ref{ass:moment}-\ref{ass:separation}, we have:
\begin{align*}
\sup_{P \in \mathcal{P}} \mathbb{P} \big( \|\hat{\bm{\Sigma}} - \bm{\Sigma}\| > \epsilon \big) \rightarrow 0 \text{ as } n \to \infty,
\end{align*}
where $\|\cdot\|$ denotes the spectral norm. For simplicity, denote this type of uniform consistency by $\hat{\bm{\Sigma}} \xlongrightarrow{u.p.} \bm{\Sigma}$.

\begin{proof}
We establish uniform consistency of the variance-covariance matrix estimator by showing uniform consistency of its diagonal elements. The off-diagonal elements follow by similar arguments using the Cauchy-Schwarz inequality and uniform consistency of the diagonal moments.

\textbf{Step 1: Uniform consistency of basic moments.} By Assumptions \ref{ass:moment}-\ref{ass:separation} and standard arguments:
\begin{equation} \label{eq:basic_moments}
	\hat{e} \xlongrightarrow{u.p.} e, \quad \hat{\tau}_1 \xlongrightarrow{u.p.} \tau_1, \quad \hat{\tau}_0 \xlongrightarrow{u.p.} \tau_0,
\end{equation}
\begin{equation} \label{eq:variances}
	\hat{\sigma}^2_1 \xlongrightarrow{u.p.} \sigma^2_1, \quad \hat{\sigma}^2_0 \xlongrightarrow{u.p.} \sigma^2_0.
\end{equation}

Since $T_i \sim \text{Bernoulli}(e)$ and by Assumption \ref{ass:treatment}, $e \in [e_{\min}, e_{\max}]$ uniformly over $\mathcal{P}$, we have:
\begin{align*}
	\text{Var}_P(\hat{e}) = \frac{e(1-e)}{n} \leq \frac{e_{\max}(1-e_{\min})}{n} \leq \frac{C_1}{n}
\end{align*}
for some constant $C_1 < \infty$ independent of $P$. Then by Chebyshev's inequality:
\begin{align*}
	\sup_{P \in \mathcal{P}} \mathbb{P}\left(|\hat{e} - e| > \epsilon\right) \leq \sup_{P \in \mathcal{P}} \frac{\text{Var}_P(\hat{e})}{\epsilon^2} \leq \frac{C_1}{n\epsilon^2} \to 0
\end{align*}
as $n \to \infty$, which gives $\hat{e} \xlongrightarrow{u.p.} e$.

The analysis of $\hat{\tau}_0$ and $\hat{\tau}_1$ are essentially the same. We focus on $\hat{\tau}_1$ to save space. Define $n_1 = \sum_{i=1}^n T_i$ and write:
$$\hat{\tau}_1 - \tau_1 = \frac{\sum_{i=1}^n T_i (Y_i - \tau_1)}{n_1} = \frac{\frac{1}{n}\sum_{i=1}^n T_i (Y_i - \tau_1)}{\hat{e}}$$
It is easy to see
\begin{align*}
	|\hat{\tau}_1 - \tau_1| = \Big| \frac{\sum_{i=1}^n T_i (Y_i - \tau_1)}{n_1} \Big| = \Big|\frac{\frac{1}{n}\sum_{i=1}^n T_i (Y_i - \tau_1)}{e} \Big| \cdot \Big|\frac{e}{\hat{e}} \Big|.
\end{align*}
Assumption \ref{ass:treatment} and the uniform consistency of $\hat{e}$ imply that $\sup_{P \in \mathcal{P}} \mathbb{P} [ | e/\hat{e} |>M_\epsilon ] < \epsilon$. Then according to Lemma \ref{lem:CD}, it is sufficient to show
\begin{align*}
	\frac{1}{ne}\sum_{i=1}^n T_i (Y_i - \tau_1)  \xlongrightarrow{u.p.} 0.
\end{align*}

Let $Z_i = T_i(Y_i - \tau_1)/e$. Then $\mathbb{E}_P[Z_i] = 0$ and:
\begin{align*}
	\mathbb{E}_P \Big[\frac{T_i^2 (Y_i - \tau_1)^2}{e^2}\Big] = \frac{1}{e^2} \mathbb{E}_P[T_i (Y_i - \tau_1)^2] = \frac{\sigma_1^2}{e} \leq \frac{\sigma_{\max}^2}{e_{\min}}
\end{align*}
By Assumption \ref{ass:moment} and the fact that uniform integrability of 4th moments implies uniform boundedness of 2nd moments, we have:
\begin{align*}
	\sup_{P \in \mathcal{P}}  \mathbb{P}\Big( \Big| \frac{1}{n} \sum_{i=1}^n Z_i \Big| > \epsilon \Big) \leq \frac{1}{n} \frac{\sup_{P \in \mathcal{P}} \mathbb{E}_P[|Z_i|^2]}{\epsilon^2} \leq \frac{1}{n} \frac{\sigma_{\max}^2}{\epsilon^2 e_{\min}}.
\end{align*}
This is the desired result, which in turn implies $\hat{\tau}_1 \xlongrightarrow{u.p.} \tau_1$ by Lemma \ref{lem:CD}.

For the variance estimators, note that:
\begin{align*}
\hat{\sigma}^2_1 - \sigma^2_1 = \Big( \frac{1}{n \hat{e}} \sum_{i=1}^n T_i (Y_i - \tau_1)^2 - \sigma^2_1 \Big) - (\hat{\tau}_1 - \tau_1)^2.
\end{align*}
The second term converge in probability to zero uniformly in $P\in\mathcal{P}$ by applying continuous mapping theorem to $(\hat{\tau}_1 - \tau_1)^2 \xlongrightarrow{u.p.} 0$.

We now establish the uniform consistency of the first term $G_n$, where
\begin{equation} \label{eq:A_n_decomposition}
G_n = \frac{1}{n\hat{e}}\sum_{i=1}^n T_i (Y_i - \tau_1)^2 - \sigma^2_1 = \Big( \frac{1}{ne}\sum_{i=1}^n T_i (Y_i - \tau_1)^2 - \sigma^2_1 \Big) \cdot \frac{e}{\hat{e}} + \sigma^2_1 \Big(\frac{e}{\hat{e}} - 1\Big).
\end{equation}

Define $G_n^1 = \frac{1}{n e}\sum_{i=1}^n T_i (Y_i - \tau_1)^2 - \sigma^2_1$ and $G_n^2 = \sigma^2_1 (e / \hat{e} - 1 )$, so that $G_n = G_n^1 \cdot e / \hat{e} + G_n^2$. The previous result implies that $G_n^2 \xlongrightarrow{u.p.} 0$. We also have $\sup_{P \in \mathcal{P}} \mathbb{P} [ | e/\hat{e} |>M_\epsilon ] < \epsilon$. So we just need to show $G_n^1 \xlongrightarrow{u.p.} 0$ by Lemma \ref{lem:CD}. In fact, for any other terms with $n_1$, we can replace it by $ne$. 

In this case, we define (we use the same notation $Z_i$ for simplicity, but its definition varies)
\begin{align*}
Z_i = \frac{T_i (Y_i - \tau_1)^2}{e}.
\end{align*}
Then $G_n^1 = \frac{1}{n}\sum_{i=1}^n Z_i - \sigma^2_1$. Under the standard assumptions, $\mathbb{E}_P[Z_i] = \sigma^2_1$, so $\mathbb{E}_P[G_n^1] = 0$. The variance of $G_n^1$ is:
\begin{align*}
\operatorname{Var}_P(G_n^1) = \operatorname{Var}_P\left(\frac{1}{n}\sum_{i=1}^n Z_i\right) = \frac{1}{n}\operatorname{Var}_P(Z_i),
\end{align*}
since the $Z_i$ are i.i.d. across $i$. We now bound $\sup_{P \in \mathcal{P}} \operatorname{Var}_P(Z_i)$. Note that:
\begin{align*}
\mathbb{E}_P[Z_i^2] = \mathbb{E}_P\left[\left(\frac{T_i (Y_i - \tau_1)^2}{e}\right)^2\right] = \frac{1}{e^2} \mathbb{E}_P\left[T_i (Y_i - \tau_1)^4\right] = \frac{1}{e} \mathbb{E}_P\left[(Y(1) - \tau_1)^4\right],
\end{align*}
where we use that $T_i$ is independent of $Y_i(1)$ and $\mathbb{E}_P[T_i] = e$. By the inequality $(a-b)^4 \leq 8a^4 + 8b^4$ and Jensen's inequality ($\tau_1^4 \leq \mathbb{E}_P[Y(1)^4]$), we have:
\begin{align*}
\mathbb{E}_P\left[(Y(1) - \tau_1)^4\right] \leq 8\mathbb{E}_P[Y(1)^4] + 8\tau_1^4 \leq 16\mathbb{E}_P[Y(1)^4].
\end{align*}
By Assumption \ref{ass:moment} (uniform $L_{4+\delta}$ integrability), there exists $M_0 < \infty$ such that $\sup_{P \in \mathcal{P}} \mathbb{E}_P\left[Y(1)^4 \mathbf{1}\{Y(1)^4 > M_0\}\right] < 1$, which implies:
\begin{align*}
\sup_{P \in \mathcal{P}} \mathbb{E}_P[Y(1)^4] \leq M_0 + 1 < \infty.
\end{align*}
By Assumption \ref{ass:treatment}, $e \geq e_{\min} > 0$ uniformly over $\mathcal{P}$. Therefore:
\begin{align*}
\sup_{P \in \mathcal{P}} \mathbb{E}_P[Z_i^2] \leq \frac{16}{e_{\min}} \sup_{P \in \mathcal{P}} \mathbb{E}_P[Y(1)^4] < \infty.
\end{align*}
Since $\operatorname{Var}_P(Z_i) \leq \mathbb{E}_P[Z_i^2]$, it follows that:
\begin{align*}
\sup_{P \in \mathcal{P}} \operatorname{Var}_P(G_n^1) \leq \frac{1}{n} \sup_{P \in \mathcal{P}} \mathbb{E}_P[Z_i^2] \leq \frac{C}{n}
\end{align*}
for some constant $C < \infty$ independent of $P$ and $n$. By Chebyshev's inequality, for any $\epsilon > 0$:
\begin{align*}
\mathbb{P}\left(|G_n^1| > \epsilon\right) \leq \frac{\operatorname{Var}_P(G_n^1)}{\epsilon^2} \leq \frac{C}{n\epsilon^2}.
\end{align*}
Taking the supremum over $P \in \mathcal{P}$:
\begin{align*}
\sup_{P \in \mathcal{P}} \mathbb{P}\left(|G_n^1| > \epsilon\right) \leq \frac{C}{n\epsilon^2} \xrightarrow{n \to \infty} 0.
\end{align*}
Thus, $G_n^1 \xlongrightarrow{u.p.} 0$.

Putting all the pieces together, we have
\begin{equation}
\label{eq:C_n_consistency}
	\hat{\sigma}^2_1 \xlongrightarrow{u.p.} \sigma^2_1.
\end{equation}
A similar argument gives $\hat{\sigma}^2_0 \xlongrightarrow{u.p.} \sigma^2_0$.

\textbf{Step 2: Uniform consistency of fourth central moments.} Define:
\begin{align*}
\hat{\mu}_{4,1} = \frac{1}{n_1}\sum_{i=1}^n T_i (Y_i - \hat{\tau}_1)^4, \quad \hat{\mu}_{4,0} = \frac{1}{n_0}\sum_{i=1}^n (1-T_i) (Y_i - \hat{\tau}_0)^4,
\end{align*}
where $n_1 = \sum_{i=1}^n T_i$ and $n_0 = n - n_1$. 

As shown in Lemma \ref{lem:CD} and the previous step, it is sufficient to prove the result with $n_1=n\hat{e}$ replaced by $ne$. Then we only need to show the following two terms converge to zero uniformly in probability over $\mathcal{P}$:
\begin{align*}
G_n^1
    := \frac{1}{n e}\sum_{i=1}^n T_i (Y_i-\tau_1)^4
        \;-\; \mu_{4,1},
    \qquad
G_n^2
    := \frac{1}{n e}\sum_{i=1}^n
        T_i \big[(Y_i-\hat{\tau}_1)^4 - (Y_i-\tau_1)^4 \big].
\end{align*}

\textbf{Control of $G_n^1$.}
Under Assumption~\ref{ass:moment}, the class
$\{(Y(1)-\tau_1)^4 : P\in\mathcal P\}$ is uniformly integrable,
because $\sup_{P\in\mathcal P} \mathbb{E}_P[|Y(1)|^{4+\delta}]<\infty$.
Fix $M>0$ and write
\begin{align*}
Z_i := T_i (Y_i-\tau_1)^4, \qquad
Z_i^{(M)} := T_i (Y_i-\tau_1)^4\mathbf{1}\{|Y_i|\le M\}.
\end{align*}
Then
\begin{align*}
\frac{1}{ne}\sum_{i=1}^n Z_i
    = \frac{1}{ne}\sum_{i=1}^n Z_i^{(M)}
    + \frac{1}{ne}\sum_{i=1}^n (Z_i - Z_i^{(M)}).
\end{align*}
Since $Z_i^{(M)}$ are bounded, a standard LLN for triangular arrays gives
\begin{align*}
\lim_M \limsup_n \sup_{P\in\mathcal P}
\mathbb{P}\Big(
    \Big|
        \frac{1}{ne}\sum_{i=1}^n Z_i^{(M)}
        - \mathbb{E}_P[Z_1^{(M)}]/e
    \Big| > \varepsilon
\Big) = 0, \,\, \forall \epsilon>0.
\end{align*}

For the tail term, note that the summands $(Z_i - Z_i^{(M)})$ are non-negative. By Markov's inequality, for any $\varepsilon > 0$,
\begin{align*}
\mathbb{P}\left(\frac{1}{ne}\sum_{i=1}^n (Z_i - Z_i^{(M)}) > \varepsilon\right) 
\leq \frac{\mathbb{E}_P\left[\frac{1}{ne}\sum_{i=1}^n (Z_i - Z_i^{(M)})\right]}{\varepsilon}
= \frac{\mathbb{E}_P[Z_1 - Z_1^{(M)}]/e}{\varepsilon}.
\end{align*}
Since $\mathbb{E}_P[Z_1 - Z_1^{(M)}] = e \cdot \mathbb{E}_P\big[(Y(1)-\tau_1)^4\mathbf{1}\{|Y(1)| > M\}\big]$, we have
\begin{align*}
\sup_{P \in \mathcal{P}} P \Big(\frac{1}{ne}\sum_{i=1}^n (Z_i - Z_i^{(M)}) > \varepsilon \Big) 
\leq \frac{\sup_{P \in \mathcal{P}} \mathbb{E}_P\big[(Y(1)-\tau_1)^4\mathbf{1}\{|Y(1)| > M\}\big]}{\varepsilon}.
\end{align*}

By Assumption~\ref{ass:moment} and the inequality $(a-b)^4 \leq 8a^4 + 8b^4$ together with Jensen's inequality ($\tau_1^4 \leq \mathbb{E}_P[Y(1)^4]$), the class $\{(Y(1)-\tau_1)^4 : P \in \mathcal{P}\}$ is uniformly integrable. Moreover, since $\sup_{P \in \mathcal{P}} |\tau_1| < \infty$ (as $\sup_{P \in \mathcal{P}} \mathbb{E}_P[|Y(1)|^{4+\delta}] < \infty$ implies bounded first moments), there exists $C < \infty$ such that $|\tau_1| \leq C$ for all $P \in \mathcal{P}$. Hence, for $M > C$, we have $\{|Y(1)| > M\} \subseteq \{|Y(1)-\tau_1| > M - C\}$, and
\begin{align*}
\mathbb{E}_P\big[(Y(1)-\tau_1)^4\mathbf{1}\{|Y(1)| > M\}\big]
\leq \mathbb{E}_P\big[(Y(1)-\tau_1)^4\mathbf{1}\{|Y(1)-\tau_1| > M - C\}\big].
\end{align*}
By uniform integrability, the right-hand side converges to zero uniformly over $\mathcal{P}$ as $M \to \infty$. Therefore,
\begin{align*}
\lim_{M \to \infty} \limsup_{n \to \infty} \sup_{P \in \mathcal{P}} 
\mathbb{P}\left(\frac{1}{ne}\sum_{i=1}^n (Z_i - Z_i^{(M)}) > \varepsilon\right) = 0.
\end{align*}

Combining this with the bounded part and noting that $\big|\mathbb{E}_P[Z_1^{(M)}]/e - \mu_{4,1}\big| \to 0$ uniformly over $\mathcal{P}$ as $M \to \infty$ by dominated convergence and uniform integrability, we obtain $G_n^1 \xlongrightarrow{u.p.} 0$.

\textbf{Control of $G_n^2$.} 
Let $\Delta_n = \hat{\tau}_1 - \tau_1$. Using the polynomial expansion:
\begin{align*}
(y-\hat{\tau}_1)^4 - (y-\tau_1)^4 = -4(y-\tau_1)^3\Delta_n + 6(y-\tau_1)^2\Delta_n^2 - 4(y-\tau_1)\Delta_n^3 + \Delta_n^4.
\end{align*}
Therefore,
\begin{align*}
G_n^2 &= \frac{1}{ne}\sum_{i=1}^n T_i \big[(Y_i-\hat{\tau}_1)^4 - (Y_i-\tau_1)^4 \big] \\
&= -4\Delta_n \cdot \frac{1}{ne}\sum_{i=1}^n T_i (Y_i-\tau_1)^3 
   + 6\Delta_n^2 \cdot \frac{1}{ne}\sum_{i=1}^n T_i (Y_i-\tau_1)^2 \\
&\quad - 4\Delta_n^3 \cdot \frac{1}{ne}\sum_{i=1}^n T_i (Y_i-\tau_1) 
   + \Delta_n^4 \cdot \frac{1}{ne}\sum_{i=1}^n T_i \\
&=: A_n + B_n + C_n + D_n.
\end{align*}

We analyze each term separately:

\textbf{Term $A_n = -4\Delta_n \cdot \frac{1}{ne}\sum_{i=1}^n T_i (Y_i-\tau_1)^3$:} 
From Step 1 of the proof, $\Delta_n \xlongrightarrow{u.p.} 0$. Under Assumption~\ref{ass:moment}, $\sup_{P\in\mathcal{P}} \mathbb{E}_P[|Y(1)-\tau_1|^3] < \infty$ (since $3 < 4+\delta$ and uniform $L_{4+\delta}$ integrability implies uniform boundedness of lower moments). Similar to the analysis of $G_n^1$, the class $\{(Y(1)-\tau_1)^3 : P\in\mathcal{P}\}$ is uniformly integrable, which implies
\begin{align*}
\frac{1}{ne}\sum_{i=1}^n T_i (Y_i-\tau_1)^3 = O_p(1) \quad \text{uniformly in } \mathcal{P}.
\end{align*}
Therefore, by Lemma~\ref{lem:CD}, $A_n \xlongrightarrow{u.p.} 0$.

\textbf{Term $B_n = 6\Delta_n^2 \cdot \frac{1}{ne}\sum_{i=1}^n T_i (Y_i-\tau_1)^2$:}
Since $\Delta_n \xlongrightarrow{u.p.} 0$, we have $\Delta_n^2 \xlongrightarrow{u.p.} 0$. From the previous step of the proof, we know that
\begin{align*}
\frac{1}{ne}\sum_{i=1}^n T_i (Y_i-\tau_1)^2 \xlongrightarrow{u.p.} \sigma_1^2,
\end{align*}
so this term is stochastically bounded uniformly in $\mathcal{P}$. By Lemma~\ref{lem:CD}, $B_n \xlongrightarrow{u.p.} 0$.

\textbf{Term $C_n = -4\Delta_n^3 \cdot \frac{1}{ne}\sum_{i=1}^n T_i (Y_i-\tau_1)$:}
We have $\Delta_n^3 \xlongrightarrow{u.p.} 0$. We have already proved that
\begin{align*}
	\frac{1}{ne}\sum_{i=1}^n T_i (Y_i - \tau_1)  \xlongrightarrow{u.p.} 0.
\end{align*}
It is then easy to see that $C_n \xlongrightarrow{u.p.} 0$.

\textbf{Term $D_n = \Delta_n^4 \cdot \frac{1}{ne}\sum_{i=1}^n T_i$:}
We have $\Delta_n^4 \xlongrightarrow{u.p.} 0$ and $\frac{1}{ne}\sum_{i=1}^n T_i = \hat{e}/e \xlongrightarrow{u.p.} 1$ (by Assumption~\ref{ass:treatment} and the uniform consistency of $\hat{e}$). Therefore, $D_n \xlongrightarrow{u.p.} 0$.

Since each term $A_n, B_n, C_n, D_n$ converges to zero uniformly in probability over $\mathcal{P}$, we conclude that $G_n^2 \xlongrightarrow{u.p.} 0$. This completes the proof that $\hat{\mu}_{4,1} \xlongrightarrow{u.p.} \mu_{4,1}$. The proof for $\hat{\mu}_{4,0}$ follows analogously.

\textbf{Step 3: Uniform consistency of $\sigma^2_\tau$.} The estimator is:
\begin{align*}
\hat{\sigma}^2_\tau = \frac{\hat{\sigma}^2_1}{\hat{e}} + \frac{\hat{\sigma}^2_0}{1-\hat{e}}.
\end{align*}
By Assumptions \ref{ass:treatment} and \ref{ass:variance}, the function $g(e, \sigma^2_1, \sigma^2_0) = \frac{\sigma^2_1}{e} + \frac{\sigma^2_0}{1-e}$ is continuous and bounded on the compact set $\{e \in [e_{\min}, e_{\max}], \sigma^2_1, \sigma^2_0 \in [\sigma^2_{\min}, \sigma^2_{\max}]\}$. According to the continuous mapping theorem, when applied uniformly:
\begin{equation}
\label{eq:sigma_tau_consistency}
\sup_{P \in \mathcal{P}} |\hat{\sigma}^2_\tau - \sigma^2_\tau| \conP 0.
\end{equation}

\textbf{Step 4: Uniform consistency of $\sigma^2_{+}$.} The estimator is:
\begin{align*}
\hat{\sigma}^2_{+} = \Big(1 + \sqrt{\frac{\hat{\sigma}_0^2}{\hat{\sigma}_1^2}}\Big)^2 \frac{\hat{\mu}_{4,1} - (\hat{\sigma}^2_1)^2}{\hat{e}} - \Big(1 + \sqrt{\frac{\hat{\sigma}_1^2}{\hat{\sigma}_0^2}}\Big)^2 \frac{\hat{\mu}_{4,0} - (\hat{\sigma}^2_0)^2}{1-\hat{e}}.
\end{align*}

Consider the first term. Based on the assumptions and what we have proved previously, it is easy to see that
\begin{align*}
\sup_{P \in \mathcal{P}} \Big| \sqrt{\frac{\hat{\sigma}_0^2}{\hat{\sigma}_1^2}} - \sqrt{\frac{\sigma_0^2}{\sigma_1^2}}\Big|  \conP 0, \quad \sup_{P \in \mathcal{P}} \Big| \frac{1}{\hat{e}} - \frac{1}{e}\Big|  \conP 0,
\end{align*}
so we have uniform consistency of each component. Since the function
\begin{align*}
h_1(\sigma^2_0, \sigma^2_1, \mu_{4,1}, e) = \Big(1 + \frac{\sigma_0^2}{\sigma_1^2}\Big)^2 \frac{\mu_{4,1} - (\sigma^2_1)^2}{e}
\end{align*}
is continuous on the compact parameter space defined by our assumptions, by the continuous mapping theorem applied uniformly:
\begin{align*}
\sup_{P \in \mathcal{P}} \Big|\Big(1 + \frac{\hat{\sigma}_0^2}{\hat{\sigma}_1^2}\Big)^2 \frac{\hat{\mu}_{4,1} - (\hat{\sigma}^2_1)^2}{\hat{e}} - \Big( 1 + \frac{\sigma_0^2}{\sigma_1^2}\Big)^2 \frac{\E[((Y(1) - \tau_1)^2 - \sigma^2_1)^2]}{e} \Big| \conP 0.
\end{align*}
The same reasoning applies to the second term in $\sigma^2_{+}$. Therefore,
\begin{equation}
\label{eq:sigma_pp_consistency}
\sup_{P \in \mathcal{P}} |\hat{\sigma}^2_{+} - \sigma^2_{+}| \conP 0.
\end{equation}

\textbf{Step 5: Uniform consistency of $\sigma^2_{-}$.} The argument is identical to Step 4, replacing the coefficients $(1 + \sigma_0^2/\sigma_1^2)$ with $(1 - \sigma_0^2/\sigma_1^2)$, etc. By Assumption \ref{ass:separation} (uniform separation), these coefficients are uniformly bounded away from zero, ensuring continuity. Thus:
\begin{equation}
\label{eq:sigma_mm_consistency}
\sup_{P \in \mathcal{P}} |\hat{\sigma}^2_{-} - \sigma^2_{-}| \conP 0.
\end{equation}

\textbf{Step 6: Uniform consistency of off-diagonal elements.} The uniform consistency of the off-diagonal elements can be obtained by the Cauchy-Schwarz inequality and the previous results. The details are hence omitted to save space. 

Since all entries of $\hat{\bm{\Sigma}}$ converge uniformly in probability to their population counterparts, and the matrix norm is dominated by the maximum entry norm, we conclude:
\begin{align*}
\sup_{P \in \mathcal{P}} \|\hat{\bm{\Sigma}} - \bm{\Sigma}\| \rightarrow 0.
\end{align*}
\end{proof}

\subsubsection{Sharp Bounds}

Recall that $\gamma = \tau_1 \tau_0$ and
\begin{align*}
	\dot{V}_o = \dot{\sigma}^2_1 +  \dot{\sigma}^2_0 - 2 (\dot{\theta}_o - \dot{\gamma}) \quad\text{and}\quad \dot{V}_p = \dot{\sigma}^2_1 +  \dot{\sigma}^2_0 - 2 (\dot{\theta}_p - \dot{\gamma}).
\end{align*}

We have analyzed the variance estimators of $\dot{\tau}_j$, $\dot{\sigma}_j^2$, $j=0,1$ in the previous section. In view of the Cauchy-Schwartz inequality, we only need to show that the variance estimators of $\dot{\theta}_o$ and $\dot{\theta}_p$ are uniformly consistent within $\mathcal{P}$.

In addition to Assumptions \ref{ass:moment}, \ref{ass:treatment}, and \ref{ass:variance}, we also need Assumption \ref{quantile} (i) and (ii)  uniformly for $P \in \mathcal{P}$. For easy reference, we restate the assumption below.

\begin{assumption}\label{ass:quantile-uniform}
(i) (Density conditions) $P$ admits continuous density functions $f_1$ and $f_0$ for $Y(1)$ and $Y(0)$ respectively, and there exists $f_{\min} > 0$ such that:
\begin{align*}
	\inf_{P\in\mathcal{P}} \inf_{u \in [0,1]} f_1(P^{-1}_1(u)) \geq f_{\min}, \quad \inf_{P\in\mathcal{P}} \inf_{u \in [0,1]} f_0(P^{-1}_0(u)) \geq f_{\min}
\end{align*}

(ii) (Bounded quantiles) There exists $M < \infty$ such that:
\begin{align*}
	\sup_{P\in\mathcal{P}} \sup_{u \in [0,1]} |P^{-1}_1(u)| \leq M, \quad \sup_{u \in [0,1]} |P^{-1}_0(u)| \leq M.
\end{align*}

\end{assumption}

\begin{proposition}[Uniform Consistency of Variance Estimator for $\dot{\theta}_o$]
\label{prop:theta_o_variance}
Under Assumptions \ref{ass:treatment} and \ref{ass:quantile-uniform}, the variance estimator $\widehat{\Var}(\dot{\theta}_o) = \frac{1}{n}\sum_{i=1}^n \hat{\dot{\theta}}_{o,i}^2$ satisfies:
\begin{align*}
\sup_{P \in \mathcal{P}} \Big| \widehat{\Var}(\dot{\theta}_o) - \Var_P(\dot{\theta}_o) \Big| \conP 0 \quad \text{as } n \to \infty,
\end{align*}
where $\hat{\dot{\theta}}_{o,i}$ is the estimated influence function:
\begin{align*}
\hat{\dot{\theta}}_{o,i} = \int_0^1 \left[ \hat{\dot{Q}}_{1,i}(u)\hat{Q}_0(u) + \hat{Q}_1(u)\hat{\dot{Q}}_{0,i}(u) \right] du,
\end{align*}
with
\begin{align*}
\hat{\dot{Q}}_{1,i}(u) = -\frac{T_i \cdot [\mathbbm{1}\{Y_i \leq \hat{Q}_1(u)\} - u]}{\hat{e} \hat{f}_1(\hat{Q}_1(u))}, \,\,
\hat{\dot{Q}}_{0,i}(u) = -\frac{(1-T_i) \cdot [\mathbbm{1}\{Y_i \leq \hat{Q}_0(u)\} - u]}{(1-\hat{e}) \hat{f}_0(\hat{Q}_0(u))}.
\end{align*}
\end{proposition}

\begin{proof}
We need to show that under Assumptions \ref{ass:treatment} and \ref{ass:quantile-uniform}:
\begin{align*}
\sup_{P \in \mathcal{P}} \Big| \frac{1}{n}\sum_{i=1}^n \hat{\dot{\theta}}_{o,i}^2 - \Var_P(\dot{\theta}_o) \Big| \conP 0.
\end{align*}

\textbf{Step 1: Uniform boundedness of influence functions}. From Assumption \ref{ass:quantile-uniform}(ii), we have:
\begin{align*}
\sup_{P \in \mathcal{P}} \sup_{u \in [0,1]} |\hat{Q}_1(u)| \leq M + o_p(1), \quad \sup_{P \in \mathcal{P}} \sup_{u \in [0,1]} |\hat{Q}_0(u)| \leq M + o_p(1).
\end{align*}
From Assumptions \ref{ass:treatment} and \ref{ass:quantile-uniform}(i), the denominators are bounded away from zero:
\begin{align*}
\hat{e} \hat{f}_1(\hat{Q}_1(u)) \geq e_{\min} f_{\min} + o_p(1), \quad
(1-\hat{e}) \hat{f}_0(\hat{Q}_0(u)) \geq (1-e_{\max}) f_{\min} + o_p(1).
\end{align*}
Therefore, the estimated influence function components are uniformly bounded:
\begin{align*}
\sup_{P \in \mathcal{P}} \sup_{i,u} |\hat{\dot{Q}}_{1,i}(u)| \leq \frac{2}{e_{\min}f_{\min}} + o_p(1), \quad
\sup_{P \in \mathcal{P}} \sup_{i,u} |\hat{\dot{Q}}_{0,i}(u)| \leq \frac{2}{(1-e_{\max})f_{\min}} + o_p(1).
\end{align*}

This implies uniform boundedness of the integrated influence function:
\begin{align*}
\sup_{P \in \mathcal{P}} \sup_i |\hat{\dot{\theta}}_{o,i}| \leq C_1 < \infty,
\end{align*}
where $C_1 = 2M \left(\frac{1}{e_{\min}f_{\min}} + \frac{1}{(1-e_{\max})f_{\min}}\right)$. Similarly, the true influence functions $\dot{\theta}_{o,i}$ are uniformly bounded by the same constant $C_1$.

\textbf{Step 2: Uniform convergence of quantile and density estimators}. Under Assumption \ref{ass:quantile-uniform}(i)-(ii), the quantile estimators satisfy:
\begin{align*}
\lim_n \sup_{P \in \mathcal{P}} \sup_{u \in [0,1]} P( |\hat{Q}_t(u) - Q_t(u)| > \epsilon ) = 0, \quad t = 0,1.
\end{align*}

Assuming consistent density estimators (e.g., kernel density estimators with appropriate bandwidth), we have:
\begin{align*}
\sup_{P \in \mathcal{P}} \sup_{y \in [-M,M]} |\hat{f}_t(y) - f_t(y)| \conP 0, \quad t = 0,1.
\end{align*}
We have already proved that $\hat{e} \xlongrightarrow{u.p.} e$.

\textbf{Step 3: Uniform convergence of influence function components}. Consider the difference for the treated group component:
\begin{align*}
\hat{\dot{Q}}_{1,i}(u) - \dot{Q}_{1,i}(u) = -T_i \Big[ \frac{\mathbbm{1}\{Y_i \leq \hat{Q}_1(u)\} - u}{\hat{e}\hat{f}_1(\hat{Q}_1(u))} - \frac{\mathbbm{1}\{Y_i(1) \leq Q_1(u)\} - u}{e f_1(Q_1(u))} \Big].
\end{align*}

We decompose this as:
\begin{align*}
\hat{\dot{Q}}_{1,i}(u) - \dot{Q}_{1,i}(u) = A_i(u) + B_i(u),
\end{align*}
where:
\begin{align*}
A_i(u) &= -T_i \cdot \frac{\mathbbm{1}\{Y_i(1) \leq \hat{Q}_1(u)\} - \mathbbm{1}\{Y_i(1) \leq Q_1(u)\}}{\hat{e}\hat{f}_1(\hat{Q}_1(u))}, \\
B_i(u) &= -T_i \cdot (\mathbbm{1}\{Y_i(1) \leq Q_1(u)\} - u) \Big[ \frac{1}{\hat{e}\hat{f}_1(\hat{Q}_1(u))} - \frac{1}{e f_1(Q_1(u))} \Big].
\end{align*}

For $A_i(u)$, note that:
\begin{align*}
& |\mathbbm{1}\{Y_i(1) \leq \hat{Q}_1(u)\} - \mathbbm{1}\{Y_i(1) \leq Q_1(u)\}| \\
\leq\,& \mathbbm{1}\{Y_i(1) \in [\min(\hat{Q}_1(u), Q_1(u)), \max(\hat{Q}_1(u), Q_1(u))]\}.
\end{align*}
Since $f_1(y) \geq f_{\min} > 0$ on its support, we have:
\begin{align*}
\mathbb{E}[|\mathbbm{1}\{Y_i(1) \leq \hat{Q}_1(u)\} - \mathbbm{1} \{Y_i(1) \leq Q_1(u)\}|] \\
    \leq f_{\max} \mathbb{E}[|\hat{Q}_1(u) - Q_1(u)|] \leq f_{\max} O(n^{-1/2}),
\end{align*}
where $f_{\max}$ exists due to the bounded support and continuity of $f_1$. 

For $B_i(u)$, the boundedness of quantiles and densities bounded away from zero ensures:
\begin{align*}
\Big| \frac{1}{\hat{e}\hat{f}_1(\hat{Q}_1(u))} - \frac{1}{e f_1(Q_1(u))} \Big| \leq K \Big( |\hat{e} - e| + |\hat{f}_1(\hat{Q}_1(u)) - f_1(Q_1(u))| \Big)
\end{align*}
for some constant $K$.

Combining these results and using the uniform convergence from Step 2:
\begin{align*}
\lim_n \sup_{P \in \mathcal{P}} \sup_{i,u} P ( |\hat{\dot{Q}}_{1,i}(u) - \dot{Q}_{1,i}(u)| > \epsilon) = 0.
\end{align*}
A similar conclusion holds for the control group.

\textbf{Step 4: $L_1$-Convergence of the Influence Function}. Expand the difference:
\begin{align*}
\hat{\dot{\theta}}_{o,i} - \dot{\theta}_{o,i} = \int_0^1 \Big[ &\hat{\dot{Q}}_{1,i}(u)(\hat{Q}_0(u) - Q_0(u)) + (\hat{\dot{Q}}_{1,i}(u) - \dot{Q}_{1,i}(u))Q_0(u) \\
+ &(\hat{Q}_1(u) - Q_1(u))\hat{\dot{Q}}_{0,i}(u) + Q_1(u)(\hat{\dot{Q}}_{0,i}(u) - \dot{Q}_{0,i}(u)) \Big] du.
\end{align*}
By Assumption \ref{ass:quantile-uniform}(ii) and Step 1, all quantile and influence function components are uniformly bounded by a constant $C_Q$. Taking absolute values and using the triangle inequality:
\begin{align*}
|\hat{\dot{\theta}}_{o,i} - \dot{\theta}_{o,i}| \leq C_Q \int_0^1 \Big( &|\hat{Q}_0(u) - Q_0(u)| + |\hat{Q}_1(u) - Q_1(u)| \\
&+ |\hat{\dot{Q}}_{1,i}(u) - \dot{Q}_{1,i}(u)| + |\hat{\dot{Q}}_{0,i}(u) - \dot{Q}_{0,i}(u)| \Big) du.
\end{align*}
From Steps 2 and 3, each term inside the integral converges to $0$ in probability uniformly over $P \in \mathcal{P}$ and $u \in [0,1]$. Moreover, by boundedness, the integrand is dominated by an integrable constant. By the Uniform Dominated Convergence Theorem, we have:
\begin{align*}
\lim_{n \to \infty} \sup_{P \in \mathcal{P}} \frac{1}{n} \sum_{i=1}^n \mathbb{E}_P\big[ |\hat{\dot{\theta}}_{o,i} - \dot{\theta}_{o,i}| \big] = 0.
\end{align*}
%In fact, by symmetry, this expectation does not depend on $i$.

\textbf{Step 5: Uniform Convergence of the Average Squared Difference}.  
Using the identity $|a^2 - b^2| = |a-b||a+b|$ and the uniform bound $|\hat{\dot{\theta}}_{o,i}|, |\dot{\theta}_{o,i}| \leq C_1$ from Step 1:
\begin{align*}
\frac{1}{n}\sum_{i=1}^n |\hat{\dot{\theta}}_{o,i}^2 - \dot{\theta}_{o,i}^2| \leq 2C_1 \cdot \frac{1}{n}\sum_{i=1}^n |\hat{\dot{\theta}}_{o,i} - \dot{\theta}_{o,i}|.
\end{align*}
Take expectations over $P$:
\begin{align*}
\sup_{P \in \mathcal{P}} \mathbb{E}_P\left[ \frac{1}{n}\sum_{i=1}^n |\hat{\dot{\theta}}_{o,i}^2 - \dot{\theta}_{o,i}^2| \right] \leq 2C_1 \sup_{P \in \mathcal{P}} \frac{1}{n} \sum_{i=1}^n \mathbb{E}_P\big[ |\hat{\dot{\theta}}_{o,i} - \dot{\theta}_{o,i}| \big] \xrightarrow{n \to \infty} 0.
\end{align*}
By Markov's inequality, for any $\epsilon > 0$:
\begin{align*}
\sup_{P \in \mathcal{P}} P_P\left( \Big| \frac{1}{n}\sum_{i=1}^n (\hat{\dot{\theta}}_{o,i}^2 - \dot{\theta}_{o,i}^2) \Big| > \epsilon \right) \leq \frac{1}{\epsilon} \sup_{P \in \mathcal{P}} \mathbb{E}_P\left[ \frac{1}{n}\sum_{i=1}^n |\hat{\dot{\theta}}_{o,i}^2 - \dot{\theta}_{o,i}^2| \right] \to 0.
\end{align*}
Thus, $\sup_{P \in \mathcal{P}} \left| \frac{1}{n}\sum_{i=1}^n \hat{\dot{\theta}}_{o,i}^2 - \frac{1}{n}\sum_{i=1}^n \dot{\theta}_{o,i}^2 \right| \conP 0$.

\textbf{Step 6: Uniform Convergence of the Variance Estimator}.   
Decompose as in the original proof:
\begin{align*}
\sup_{P \in \mathcal{P}} \Big| \frac{1}{n}\sum_{i=1}^n \hat{\dot{\theta}}_{o,i}^2 - \Var_P(\dot{\theta}_o) \Big| \leq\,& \underbrace{\sup_{P \in \mathcal{P}} \Big| \frac{1}{n}\sum_{i=1}^n \hat{\dot{\theta}}_{o,i}^2 - \frac{1}{n}\sum_{i=1}^n \dot{\theta}_{o,i}^2 \Big|}_{\text{Step 5 } \xrightarrow{p} 0} \\
& + \underbrace{\sup_{P \in \mathcal{P}} \Big| \frac{1}{n}\sum_{i=1}^n \dot{\theta}_{o,i}^2 - \Var_P(\dot{\theta}_o) \Big|}_{\text{Term 2}}.
\end{align*}
For Term 2, note that $\dot{\theta}_{o,i}^2$ are uniformly bounded by $C_1^2$ and \text{i.i.d.} under each $P$. By the Uniform Law of Large Numbers for uniformly bounded classes (which holds trivially under our boundedness assumptions), Term 2 converges to $0$ in probability uniformly over $\mathcal{P}$. Combining both terms completes the proof. 
\end{proof}

\section{Choice of Loss Function}\label{sec:loss}

We discuss alternative choices of loss functions and show that they do not lead to meaningful distributionally robust predictions.

\begin{proposition}\label{prop:e1} (Linear regression with no covariate shifts) For $q\in(1,\infty]$, consider $l(x,y;\beta)=(y-x\beta)^2$ and the 2-Wasserstein neighborhood with cost function 
\begin{equation}\label{eq:cost1}
    \left(N_q((x,y),(u,v))\right)^2=\begin{cases}
    \norm{y-v}_q^2=|y-v|^2 & \text{if } x=u\\
    \infty & \text{if } otherwise.
\end{cases}
\end{equation}

Then
\[
\inf_{\beta\in\mathbb{R}^d} \sup_{D(P,Q)\leq \delta^2}\mathbb{E}_Q\left[l(X,Y;\beta)\right]=\inf_{\beta\in\mathbb{R}^d}\left\{\sqrt{\mathbb{E}_P\left[(Y-X\beta)^2\right]}+\delta\right\}^2.
\]
    
\end{proposition}

The proof directly follows that of Proposition 2 and Theorem 1 in \cite{blanchet2019robust} by using the cost function in (\ref{eq:cost1}). We can easily see that the minimax solution for $\beta$ remains as the parameter under the source distribution $P$. 

The loss function in (\ref{eqn:naivedro}) can be considered a special case of Proposition \ref{prop:e1} by regressing $\tilde{Y}(1)-\tilde{Y}(0)$ on a constant 1. Additionally, even if one would like to estimate the treatment effect in a parametric linear regression of the observed outcome on the treatment indicator and covariates (sometimes with the interaction between the treatment indicator and covariates included), the distributionally robust coefficient estimates would be driven solely by covariate shifts, regardless of the distributional shift in potential outcomes, which is counterintuitive. %In fact, suppose there is distribution shift in covariates but no shift in treatment assignment, the coefficients on the covariates and the interaction term will be penalized toward zero, and the coefficient on the treatment indicator, in some DGP, could increase in magnitude along with the Wasserstein radius to fit the observed outcome better. 

\begin{remark}
    If we replace the 2-Wasserstein neighborhood in Proposition \ref{prop:e1} with the 1-Wasserstein neighborhood, there is no valid minimax solution because the inner supremum is infinite.
\end{remark}

\end{document}